\documentclass[twocolumn]{./aastex63}
\usepackage[T1]{fontenc}
\usepackage{color,epsfig,graphics,graphicx,rotating,comment,amssymb,stackengine}
\usepackage{natbib,lipsum,atbegshi}
\usepackage{sidecap}
\usepackage{ulem}

\definecolor{brown}{rgb}{0.6,0.4,0.2}
\definecolor{purple}{rgb}{0.5,0,0.5}
\shortauthors{Rho et al.}

\def\one{{\,\sc i}}
\def\two{{\,\sc ii}}

\newcommand{\kms}{km\,s$^{-1}$}

\newcommand{\hei}{He\,{\sc i}}
\newcommand{\ci}{C\,{\sc i}}
\newcommand{\caii}{Ca\,{\sc ii}}
\newcommand{\feii}{Fe\,{\sc ii}}
\newcommand{\mgi}{Mg\,{\sc i}}
\newcommand{\mgii}{Mg\,{\sc ii}}
\newcommand{\nai}{Na\,{\sc i}}
\newcommand{\oi}{O\,{\sc i}}
\newcommand{\siii}{Si\,{\sc ii}}
\newcommand{\sii}{Si\,{\sc i}}
\newcommand{\si}{S\,{\sc i}}

\newcommand{\ergs}{erg\,s$^{-1}$}

\newcommand{\ltsimeq}{\raisebox{-0.6ex}{$\,\stackrel
        {\raisebox{-.2ex}{$\textstyle <$}}{\sim}\,$}}
\newcommand{\gtsimeq}{\raisebox{-0.6ex}{$\,\stackrel
        {\raisebox{-.2ex}{$\textstyle >$}}{\sim}\,$}}

\newcommand{\mic}{$\mu$m}

\shorttitle{SN Ic and Ic-BL: SN2020oi and SN2020bvc}

\begin{document}

\title{Near-Infrared and Optical Observations of Type Ic SN2020oi and
 broad-lined Ic SN2020bvc: Carbon Monoxide, Dust and High-Velocity Supernova Ejecta}
\author{J. Rho}\affil{SETI Institute, 189 Bernardo Ave., Ste. 200, Mountain View, CA 94043, USA; jrho@seti.org}
\author{A. Evans}\affil{Astrophysics Group, Keele University, Keele, Staffordshire, ST5 5BG, UK; a.evans@keele.ac.uk}
\author{T. R. Geballe}\affil{Gemini Observatory/NSF's National Optical-Infrared Astronomy Research Laboratory, 670 N. Aohoku Place, Hilo, HI, 96720, USA; tom.geballe@noirlab.edu}
\author{D. P. K. Banerjee}\affil{Physical Research Laboratory, Navrangpura, Ahmedabad, Gujarat 380009, India}
\author{P. Hoeflich}\affil{Florida State University, Tallahassee, FL 32309, USA}
\author{M. Shahbandeh}\affil{Florida State University, Tallahassee, FL 32309, USA}
\author{S. Valenti}\affil{Department of Physics, University of California, Davis, CA 95616, USA}
\author{S.-C. Yoon}\affil{Department of Physics and Astronomy, Seoul National University, Gwanak-ro 1, Gwanak-gu, Seoul, 08826, South Korea}
\author{H. Jin}\affil{Department of Physics and Astronomy, Seoul National University, Gwanak-ro 1, Gwanak-gu, Seoul, 08826, South Korea}
\author{M. Williamson}\affil{Center for Cosmology and Particle Physics, New York University, 726 Broadway, NY, NY 11201, USA}
\author{M. Modjaz}\affil{Center for Cosmology and Particle Physics, New York University, 726 Broadway, NY, NY 11201, USA}
\author{D. Hiramatsu}\affil{Las Cumbres Observatory, 6740 Cortona Drive, Suite 102, Goleta, CA 93117-5575, USA}\affil{Department of Physics, University of California, Santa Barbara, CA 93106-9530, USA}
\author{D. A. Howell}\affil{Las Cumbres Observatory, 6740 Cortona Drive, Suite 102, Goleta, CA 93117-5575, USA}\affil{Department of Physics, University of California, Santa Barbara, CA 93106-9530, USA}
\author{C. Pellegrino}\affil{Las Cumbres Observatory, 6740 Cortona Drive, Suite 102, Goleta, CA 93117-5575, USA}\affil{Department of Physics, University of California, Santa Barbara, CA 93106-9530, USA}
\author{J. Vink\'o}\affil{CSFK Konkoly Observatory, Konkoly-Thege M. ut 15-17, Budapest, 1121, Hungary}\affil{Department of Optics and Quantum Electronics, University of Szeged, D\'om t\'er 9, Szeged, 6720 Hungary}\affil{ELTE E\"otv\"os Lor\'and University, Institute of Physics, P\'azmany P\'eter s\'et\'any 1/A, Budapest, 1117, Hungary}
\author{R. Cartier}\affil{CTIO, NSF's National Optical-Infrared Astronomy Research Laboratory, Casilla 603, La Serena, Chile}
\author{J. Burke}\affil{Las Cumbres Observatory, 6740 Cortona Drive, Suite 102, Goleta, CA 93117-5575, USA}\affil{Department of Physics, University of California, Santa Barbara, CA 93106-9530, USA}
\author{C. McCully}\affil{Las Cumbres Observatory, 6740 Cortona Drive, Suite 102, Goleta, CA 93117-5575, USA}\affil{Department of Physics, University of California, Santa Barbara, CA 93106-9530, USA}
\author{H. An}\affil{Department of Astronomy and Space Science, Chungbuk National University, Cheongju, 28644, Republic of Korea}
\author{H. Cha}\affil{Department of Astronomy and Space Science, Chungbuk National University, Cheongju, 28644, Republic of Korea}
\author{T. Pritchard}\affil{Center for Cosmology and Particle Physics, New York University, 726 Broadway, NY, NY 11201, USA}
\author{X. Wang}\affil{Physics Department, Tsinghua University, Beijing, 100084, China}\affil{Beijing Planetarium, Beijing  Academy of Science and Technology, Beijing, 100089, China}
\author{J. Andrews}\affil{Steward Observatory, University of Arizona, 933 North Cherry Avenue, Tucson, AZ 85721, USA}
\author{L. Galbany}\affil{Departamento de F\'isica Te\'orica y del Cosmos, Universidad de Granada, E-18071 Granada, Spain}
\author{S. Van Dyk}\affil{Caltech/IPAC, Mailcode 100-22, Pasadena, CA 91125, USA}
\author{M. L. Graham}\affil{DiRAC  Institute,  Department  of  Astronomy,  University of Washington, Box 351580, U.W., Seattle WA 98195}
\author{S. Blinnikov}\affil{NRC‚ Kurchatov Institute - ITEP, Moscow, 117218; Sternberg Astronomical Institute (SAI) of Lomonosov Moscow State University, Moscow, 117218; Dukhov Automatics Research Institute (VNIIA), Moscow, 127055, Russia}
\author{V. Joshi}\affil{Physical Research Laboratory, Navrangpura, Ahmedabad, Gujarat 380009, India}
\author{A. P\'al}\affil{CSFK Konkoly Observatory, Konkoly-Thege M. ut 15-17, Budapest, 1121, Hungary}\affil{ELTE E\"otv\"os Lor\'and University, Institute of Physics, P\'azmany P\'eter s\'et\'any 1/A, Budapest, 1117, Hungary}\affil{ELTE E\"otv\"os Lor\"and University, Department of Astronomy, P\'azm\'any P\'eter s\'et\'any 1/A, Budapest, 1117, Hungary}
\author{L. Kriskovics}\affil{CSFK Konkoly Observatory, Konkoly-Thege M. ut 15-17, Budapest, 1121, Hungary}\affil{ELTE E\"otv\"os Lor\'and University, Institute of Physics, P\'azmany P\'eter s\'et\'any 1/A, Budapest, 1117, Hungary}
\author{A. Ordash}\affil{CSFK Konkoly Observatory, Konkoly-Thege M. ut 15-17, Budapest, 1121, Hungary}
\author{R. Szakats}\affil{CSFK Konkoly Observatory, Konkoly-Thege M. ut 15-17, Budapest, 1121, Hungary}
\author{K. Vida}\affil{CSFK Konkoly Observatory, Konkoly-Thege M. ut 15-17, Budapest, 1121, Hungary}\affil{ELTE E\"otv\"os Lor\'and University, Institute of Physics, P\'azmany P\'eter s\'et\'any 1/A, Budapest, 1117, Hungary}
\author{Z. Chen}\affil{Physics Department, Tsinghua University, Beijing, 100084, China}
\author{X. Li}\affil{Physics Department, Tsinghua University, Beijing, 100084, China}
\author{J. Zhang}\affil{Yunnan Observatories, Chinese Academy of Sciences, Kunming, 650216, China}
\author{S. Yan}\affil{Physics Department, Tsinghua University, Beijing, 100084, China}

\begin{abstract}
We present near-infrared and optical observations of the Type Ic Supernova (SN) 2020oi in the galaxy M100 and the broad-lined Type Ic SN2020bvc in UGC 9379, using Gemini, LCO, SOAR, and other ground-based telescopes. The near-IR spectrum of SN2020oi at day 63 since the explosion shows strong CO emissions and a rising K-band continuum, which is the first unambiguous dust detection from a Type Ic SN. Non-LTE CO modeling shows that CO is still optically thick, and that the lower limit to the CO mass is 10$^{-3}$ M$_\odot$. The dust temperature is 810 K, and the dust mass is $\sim$10$^{-5}$ M$_\odot$. 
We explore the possibilities that the dust is freshly formed in the ejecta, heated dust in the pre-existing circumstellar medium, and an infrared echo. The light curves of SN2020oi are consistent with a STELLA model with canonical explosion energy, 0.07 M$_\odot$ Ni mass, and 0.7 M$_\odot$ ejecta mass. A model of high explosion energy of 10$^{52}$ erg, 0.4 M$_\odot$ Ni mass, 6.5 M$_\odot$ ejecta mass with the circumstellar matter, reproduces the double-peaked light curves of SN2020bvc.  We observe temporal changes of absorption features of the IR Ca\,{\sc ii} triplet, \si\ at 1.043 $\mu$m, and \feii\ at 5169 \AA. The blue-shifted lines indicate high velocities, up to 60,000 \kms\ for SN2020bvc and 20,000 \kms\ for SN2020oi, and the expansion velocity rapidly declines before the optical maximum. We present modeled spectral signatures and diagnostics of CO and SiO molecular bands between 1.4 and 10 $\mu$m.

\end{abstract}


\section{Introduction}

The large amount of dust seen in high-$z$ galaxies implies that dust forms in the early universe \citep{bertoldi03, isaak02, laporte17}. Intermediate mass stars, thought to produce most interstellar dust in present galaxies, when they are in the AGB phase, would not have evolved to the dust-producing stage in high-$z$ galaxies because the universe was too young \citep{michalowski15, lesniewka19}. In contrast, core-collapse supernovae (ccSNe) can occur just several million years after their progenitors form, offering an explanation for the source of dust early in the universe. The mapping of the young supernova remnant (SNR) Cas A confirmed that dust and carbon monoxide (CO) form in SNe ejecta \citep{rho08, rho12}. Several SNRs, namely Cas A \citep{deLooze17}, SN 1987A \citep{matsuura15},  G54.1+0.3 \citep{rho18}, Crab Nebula \citep{gomez12}, and 3 additional pulsar-wind-nebula SNRs \citep{chawner19} have dust masses of 0.1-0.9 M$_{\odot}$, in agreement with dust formation models \citep{todini01, nozawa03, sluder18}. These results suggest that ccSNe are viable major dust factories in the early universe. However, understanding of the amount of dust destruction by reverse shocks in the SN ejecta is a subject of debate \citep{nath08, nozawa07, silvia10, micelotta16, kirchschlager19}. Estimates from a number of studies of the dust mass per SN event in recent SNe are more than two orders of magnitude lower \citep[e.g.,][]{kotak09, andrews11} than the masses measured in the young SNRs mentioned above.

SNe from core-collapse are classified as Type II, if they exhibit H lines, as Type Ib SNe (SNe Ib), if they exhibit He lines, but no H lines, and as Type Ic SNe (SNe Ic) if they do not exhibit either H or He lines \citep[e.g.,][]{filippenko97,galYam17, williamson19}. The latter types likely arise from progenitors that have lost their hydrogen envelopes (for SNe Ib) and in addition of most, if not all, of the helium envelopes (for SNe Ic). SNe Ib and Ic are usually referred as stripped-envelope ccSNe.

The spectra of an SNe Ic subgroup, SNe Ic-BL, are characterized by broad lines (BL) implying very high velocities \citep[$>$20,000 \kms\ at maximum light;][]{modjaz16} which, for some objects, may indicate higher explosion energies ($\sim$10$^{52}$ erg) than for typical SNe Ic. SNe Ic-BL are of great interest because they are the only type of SN associated with $\gamma$-ray bursts \citep[GRBs; see reviews by, e.g.,][]{woosley06, modjaz11-rev,cano17_obs_guide}. There is considerable controversy over ``hidden'' He (e.g., \citealt{dessart11,hachinger12}), especially regarding whether the lack of obvious optical He lines in SN Ic and SNe Ic-BL spectra is evidence for helium deficiency in the ejecta. This controversy applies to the entire SN Ic family, including SN-GRBs and superluminous SN Ic \citep{mazzali16}. The apparent lack of He in SNe Ic and SNe Ic-BL spectra is puzzling since most of the progenitors of SNe Ic-BL and GRBs are He-stars \citep{fryer07,yoon15}. Answering the He question is crucial for understanding the stellar progenitors of SNe Ic and SNe Ic-BL \citep{yoon17}, including those connected with GRBs. Near-IR spectra are crucial for determining the presence of He and for identifying the stellar progenitors of SNe Ic \citep{dessart12, dessart15}.

We recently obtained a sequence of near-IR spectra of the Type IIP ccSN 2017eaw, which shows the onset of CO formation and newly-formed carbon dust \citep{rho18sn, tinyanont19, szalai19}. The timing of the appearance and the evolution of CO is remarkably similar to that seen in SN~1987A and is consistent with chemically controlled dust models. Some dust formation models \citep{todini01, nozawa03} predict that dust forms between 350 to 900 days after the explosion, with carbon dust being one of the first condensates. The dust mass produced in SNe ejecta depends on progenitor mass and on the SN type \citep{todini01, sarangi13}.

In this paper we present Gemini {GNIRS} near-IR target-of-opportunity (ToO) spectroscopy of two ccSNe, the Type Ic SN2020oi and the Type Ic-BL SN2020bvc, together with near-IR spectroscopy from SOAR and IRTF, optical spectroscopy and photometry from the Las Cumbres Observatory (LCO) network, and other ground-based telescopes. The two SNe discovered within days of the explosion were rapidly observed and classified as described below. Here we present and analyze the early data from day 1 to $\sim$100 after the explosions.

\section{Observations}

{\bf SN2020oi} (ZTF20aaelulu) was discovered on 2020 January 7 by the Zwicky Transient Facility (ZTF) at $r = 17.28$ mag \citep[TNS 51926;][]{forster20}. It is located in the nearby galaxy M100 at a distance of 16.22 Mpc based on NED\footnote{http://ned.ipac.caltech.edu/} with $z$ = 0.00524. Based on an optical spectrum obtained at the Southern Astrophysical Telescope (SOAR) it was classified as Type Ic \citep[ATel \# 13393;][]{siebert20}. The SN has also been detected by Swift with UVW2=17.97$\pm$0.33, UVM2=18.16$\pm$0.34, and UVW1=17.54$\pm$0.25 \citep[TNS 2020-8;][]{ho20TNS}. In addition, radio emission has been reported \citep[ATel \# 13398, 13400, 13401, and 13448;][]{horesh20, moldon20}.

{\bf SN2020bvc} was discovered by the All Sky Automated Survey for Supernovae (ASAS-SN) on 2020 February 4 at $g$ magnitude $\sim$17 \citep{stanek20}. It is located in the galaxy UGC 09379,  at a distance of 114 Mpc, with $z$ = 0.025235 based on NED. SN2020bvc was classified as a young ccSN based on optical spectra from LCO by \cite{hiramatsu20}. It was further classified as broad-lined Ic by \cite{perley20}. Early ZTF, LT, and ASAS-SN photometry show it to have had an extremely rapid initial rise, followed by a rapid decline \citep{perley20}. Radio emission by the Very Large Array and X-ray emission has been detected by Chandra \citep{ho20xrays}. No counterpart GRB to SN2020bvc has been detected. It has been suggested by  \citet{izzo20}, that the lack of a counterpart GRB is due to it being an off-axis GRB or having a choked jet. \citet{ho20} present double-peaked light curves of SN2020bvc.

\begin{table}[!htb]
\caption{Observational Parameters of Optical and Infrared Spectroscopy}
\label{Tobs}
\begin{center}
\begin{tabular}{llllc}
\hline \hline
No. & Date &MJD &day & Instrument\\
\hline
{\bf SN2020oi}\span\omit\span\omit\span\omit \span\omit \\
No.$^a$ &20200106 & 58854.04 & 0(=t$_0$)$^b$ &... \\
1 & 20200108 & 58856  & 2 & SOAR-Goodman \\
101 &{\bf 20200109}$^c$ & 58857 & 3 & {\bf SOAR-TripleSpec} \\
2  & 20200109 & 58857 & 3 & SOAR-Goodman \\
3  & 20200111 & 58859 & 5 & Lijiang-YFOSC\\
4  &20200117 & 58865 & 11 & Xinglong-BFOSC\\
5& 20200118 & 58867 & 12 & SOAR-Goodman \\
6&20200120 & 58868 & 14 & Xinglong-BFOSC\\
102&{\bf 20200120}$^c$  &58869 & 14 & {\bf IRTF-SPEX} \\
7&20200121 &58869 &15 &LCOFTN-FLOYDS\\
8& 20200131 &58879 &25 &LCOFTN-FLOYDS\\
103 &{\bf 20200204}$^c$ & 58884 &29 &{\bf Gemini-GNIRS} \\
9 &20200205 &58884 &30 &LCOFTN-FLOYDS\\
10&20200211 & 58890 & 36 & SOAR-Goodman \\
11&20200215 & 58894 &40 &Bok-BCspec\\
12&20200218 & 58897 &43 &Xinglong-BFOSC\\
104&{\bf 20200309}$^c$  &58917& 63  &{\bf Gemini-GNIRS}\\
13&20200313 &58921 &67 &Xinglong-BFOSC\\
\hline
{\bf SN2020bvc}\span\omit \span\omit\span\omit\span\omit \\
&20200203 &58882.67 &0(=t$_0$)$^d$ &...\\
1&20200205 &58884 &1.33 &LCOFTN-FLOYDS\\
2&20200215 &58894 &10 &Bok-BCSpec\\
3&20200224 &58903 &20 &LCOFTN-FLOYDS\\
4&20200225 &58904 &21 &APO3.5m-DIS\\
101&{\bf 20200306}$^c$ &58914 & 31    &  {\bf Gemini-GNIRS}\\
5&20200321 &58929 &46 &LCOFTN-FLOYDS\\
6&20200323 &58931 &48 &LCOFTN-FLOYDS\\
7&20200402 &58941 &58 &LCOFTN-FLOYDS\\
8&20200417 &58956 &73 &LCOFTN-FLOYDS\\
\hline
\end{tabular}
\end{center}
\renewcommand{\baselinestretch}{0.8}
{\footnotesize $^{a}$ The number of observations counting optical spectroscopy starts from 1 and near-IR starts from 101
for SN2020oi and SN2020bvc, respectively.}
{\footnotesize $^{b}$The explosion date of SN2020oi is estimated to be on 2020 January 6 (MJD=58854),
taken to be the middle point between the last non-detection reported by ZTF and the first detection.}
{\footnotesize $^{c}$Near-IR observations are marked in bold.
 The GNIRS observations of SN2020oi on 20200204 approximately have J, H, and K magnitude of 14.9, 14.5, 14.3 mag, respectively.
The GNIRS observations of SN2020oi on 20200309 have J, H, and K magnitude of 16.3, 16.2, 16.1 mag, respectively. 
The GNIRS observations of SN2020bvc on 20200306 have J, H, and K magnitude of 16.7, 15.7, 15.6 mag, respectively.
}
{\footnotesize $^{d}$The explosion date is from \cite{ho20}.}
\end{table}

Near-infrared 0.8-2.5 \mic\ spectra of SN2020oi and SN2020bvc were obtained by the Gemini Near-Infrared Spectrograph (GNIRS) on the 8.1-meter Frederick C. Gillett Gemini-North telescope, for program GN-2020A-Q-211. The observing dates are listed in Table 1. GNIRS was configured in its cross-dispersed mode, using its 32 line/mm grating and a 0.45 arcsec-wide slit to provide a resolving power, $R$, of $\sim$1200 (250~km~s$^{-1}$) for the first spectrum of SN2020oi and the SN2020bvc spectrum, and a 0.675 arcsec-wide slit, to nominally provide $R = 800$ (375 km/s) for the second SN2020oi spectrum (the seeing was much better than 0.675 arcsec). The observational set-up and data reduction procedures were the same as those used for SN2017eaw as described in \cite{rho18sn}. The spectra shown here have been binned such that the separation of adjacent data points, $\Delta$$\lambda$ corresponds to $\lambda$/$\Delta$$\lambda$ $\sim$1,000. Due to the COVID-19 pandemic, Gemini-North telescope was closed from 2020 March 22. After Gemini-North re-opened we obtained a GNIRS spectrum at the position of  SN2020oi on 2020 May 31, but the SN was not detected. 

Near-IR spectra of SN2020oi were obtained with Triple-Spec v4.1, the fourth generation of the Triple-Spec instrument mounted on the near-IR Nasmyth port of the 4.1\,.m SOAR telescope, previous versions of Triple-Spec were build for the 200-inch Palomar telescope, the ARC 3.5\,m telescope, and Keck-II telescope. Triple-Spec is a cross-dispersed spectrograph featuring six spectral orders, spanning 0.8 - 2.4 $\mu$m with a nominal resolution of R~3500, composed of a fixed-format slit assembly of 1.1$''$ wide and 28$''$ long, and a 2048$\times$2048 Hawaii-2RG HgCdTe detector array. Triple-Spec at SOAR is fed by a reflection off a dichroic, which also transmits light to the guider. As a result, the response cuts off below 1.0 microns. The nominal point of 50\% reflectivity of the dichroic is around 0.95 microns.

We used the Spextool IDL package \citep{cushing04} to reduce the Triple-Spec data, we subtracted consecutive AB pairs to remove the sky and the bias level. We flat fielded the science frames dividing by the normalized master flat. We calibrated 2D science frames in wavelength by using CuHeAr Hollow Cathode comparison lamps. To correct for telluric features and to flux calibrate our SN spectrum, we observed the A0V telluric standard after the SN and at a similar airmass. Finally, we extracted the SN and the telluric star spectra from the 2D wavelength calibrated frames. After the extraction of the individual spectra, we used the xtellcorr task \citep{vacca03} included in the Spextool IDL package to perform the telluric correction and the flux calibration of the spectra of SN2020oi.

LCO $UBVgri$-band data were obtained with the Sinistro cameras on the 1m telescopes at Sutherland (South Africa), CTIO (Chile), Siding Spring (Australia), and McDonald (USA), through the Global Supernova Project. PSF fitting was performed using lcogtsnpipe{\footnote{https://github.com/svalenti/lcogtsnpipe}} \citep{valenti16}, a PyRAF-based photometric reduction pipeline. As SN2020oi occurred in the same galaxy as SN2019ehk \citep{grzegorzek19}, image subtraction was performed using as templates Sinistro images of SN 2019ehk (taken between 20191201-20191207), using PyZOGY \citep{guevel17}, an implementation in Python of the subtraction algorithm described in \cite{zackay17}. $UBV$-band data were calibrated to Vega magnitudes \citep{stetson00} using standard fields observed on the same night by the same telescope as the SN. $gri$-band data were calibrated to AB magnitudes using the Sloan Digital Sky Survey (SDSS, SDSS Collaboration 2017).

Additional photometric data were collected with the 0.8m RC80 telescope of Konkoly Observatory through Johnson $BV$ and Sloan $g r i z$ filters at Piszk\'estet{\H o} station (Hungary). Photometry was performed via image subtraction of PS1\footnote{https://ps1images.stsci.edu} template frames implemented in IRAF. Photometric zero points in each filter were tied to PS1 photometry of 5--10 local stars used as tertiary standards. $B$ and $V$ magnitudes of the local standards in the Vega-system were derived from their $g_P$, $r_P$ and $i_P$ data via the calibration of \citet{tonry12}, while the magnitudes in the $g$, $r$, $i$ and $z$ filters are given in the AB-system. We also added $gr$-band Zwicky Transient Facility \citep[ZTF;][]{bellm19} photometric data for both SN2020oi and SN2020bvc.

LCO optical spectra were taken with the FLOYDS spectrographs mounted on the 2m Faulkes Telescope North (FTN) and South at Haleakala (USA) and Siding Spring (Australia), respectively, through the Global Supernova Project. A 2$''$ slit was placed on the target at the parallactic angle \citep{filippenko82}. One-dimensional spectra were extracted, reduced, and calibrated following standard procedures using the FLOYDS pipeline\footnote{https://github.com/svalenti/FLOYDS\_pipeline} \citep{valenti14}. The wiggles in the FLOYDS spectra that appear above 7500 \AA, are due to fringes.

We also acquired optical specta with Goodman on SOAR telescope \citep{clemens04} and the Boller \& Chivens spectrograph (BCSpec) on the 2.3m Bok telescope. In addition we obtained a low-resolution spectrum with the Dual Imaging Spectrograph (DIS), mounted on the 3.5m telescope at the Apache Point Observatory. The R300 grating was central wavelength of 7500 $\AA$. The instrument was rotated to the parallactic angle and $1\times400$ second exposures were obtained. These data were reduced using standard procedures and calibrated to a standard star obtained the same night using the PyDIS package \citep{davenport18}. Finally, a few optical spectra of SN2020oi were observed using the Yunnan Faint Object Spectrograph and Camera (YFOSC) on the 2.4-meter Lijiang optical telescope (LJT) \citep{ wang19, xin20} and Beijing Faint Object Spectrograph and Camera (BFOSC) on Xinglong 2.16m telescope \citep{fan16} of the Beijing Astronomical Observatory (BAO). The observations are summarized in Table \ref{Tobs}.

\begin{figure}
\includegraphics[scale=0.6,angle=0,width=8.2truecm,height=10truecm]{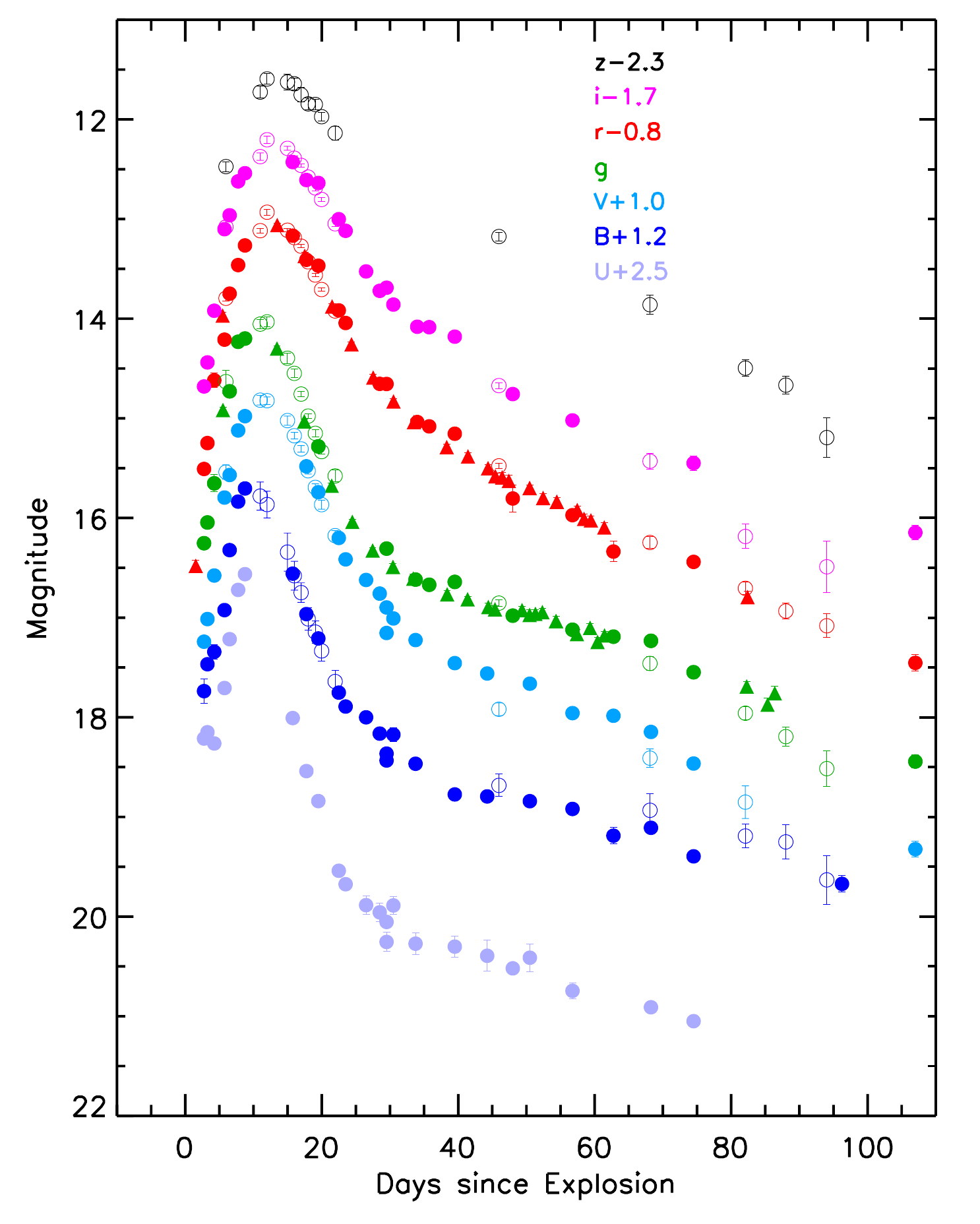}
\caption{Multi-color light curves of SN2020oi from LCO (filled circles), Konkoly (open circles)  and ZTF (filled triangles). The explosion date of 2020 January 6 (MJD=58854), taken to be the middle point between the last non-detection reported by ZTF and the first detection, is used as day 0 (t$_0$). The magnitudes are shifted as labeled for display purposes.
}
\label{Flightcurves20oi}
\end{figure}

\begin{figure}
\includegraphics[scale=0.6,angle=0,width=7.5truecm]{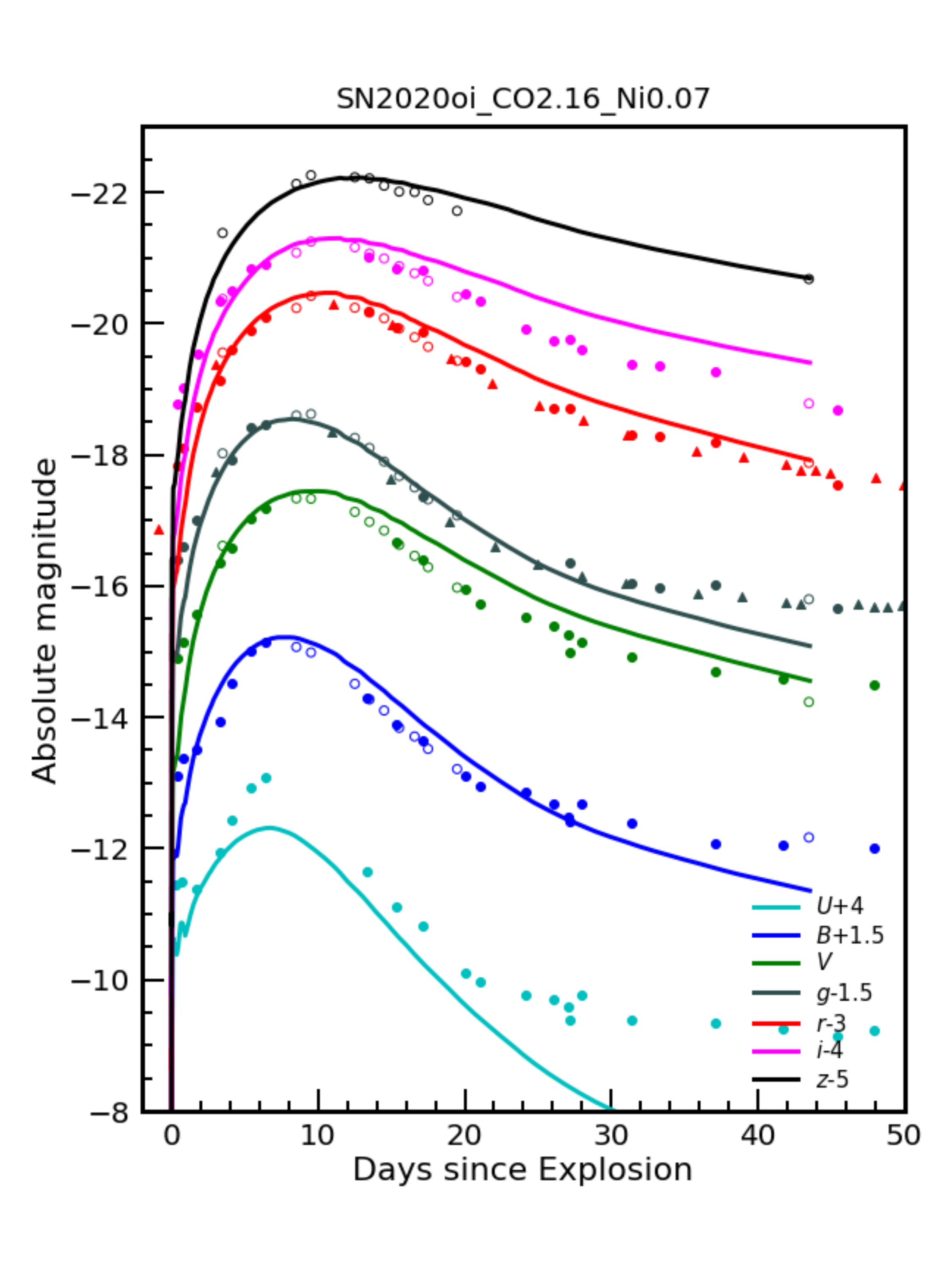}
\caption{Light curves of SN2020oi superposed on SN Ic model from a  2.16 M$_\odot$ CO star progenitor, with an explosion energy of
1$\times$10$^{51}$ erg, and a nickel mass of $0.07~M_\odot$.}
\label{FLCmodel1}
\end{figure}

\section{Results}

\subsection{Extinction}

We have examined the \nai\ D lines from the optical spectra of both SNe (see Figure \ref{Foptispec}) and do not detect distinct doublets, but notice marginal dips at Na I D1 and D2 in some of the spectra of SN2020oi. We have estimated the Galactic reddening toward the exact direction of the SN2020oi and SN2020bvc using the Galactic dust model of \cite{schlafly11}\footnote{https://irsa.ipac.caltech.edu/applications/DUST/}. The Galactic reddening is very small, $E(B-V) = 0.0227\pm 0.0002$ in the direction of the SN2020oi, and is even smaller $E(B-V) = 0.0106\pm0.0004$ in the direction of SN 2020bvc. The corrections are less than the sizes of the plotting symbols in the figures. The extinctions we derive from marginal dips in SN2020oi are consistent with the techniques described in \cite{poznanski12}. The estimated extinction of SN2020oi is comparable to those of \cite{izzo20} and \cite{ho20}, and that of SN2020bvc is comparable to that of \cite{horesh20}.

\subsection{Light Curves and Explosion Properties of SN2020oi}
\label{SLCsnoi}

\begin{table*}
\caption{Explosion and Progenitor Properties}
\label{Tprogenitor}
\begin{center}
\begin{tabular}{l|ccc|ccccc}
\hline \hline
                      & Ic &          &            &Ic-BL &&\\
                        & 2020oi  &2007gr & 1994i & 2020bvc &2020bvc & 1998bw &2006aj\\
\hline
     References$^a$         & this work & 1, 2, 3 & 4, 5, 6  & this work & Ho (7) & 8, 9, 
10, 11&  12, 13, 14 \\
C-O star (M$_{\odot}$)$^b$  & 2.16&1  &2.1& 8.26 &   & 14 &3.3 \\
Explosion Energy $E_\mathrm{exp}$ (10$^{51}$ erg)  &1  &     &     &12    &  && 2.7\\
Kinetic Energy E$_K$ (10$^{51}$ erg)    &0.6& 1-4 & 1 &10.5  & 3  &20 & 2\\
Ni mass (M$_{\odot}$)$^c$   & 0.07 &0.076 &0.07& 0.4 & 0.11 &0.4 & 0.22\\
Ejecta mass   (M$_{\odot}$) & 0.71 & 1.8 & 0.6 & 6.36 & 1.0 & 6.8 & 1.4\\
Progenitor (M$_{\odot}$)$^d$ &13 & 15  &13-15&40 - 50 & & 40  & 20\\
CSM mass (M$_{\odot}$) &$\leq$3$\times$10$^{-4}$&&&0.1 &$<$0.01 && 0.1\\
CSM $R$ (cm)  &    &&&10$^{14}$ & $>$10$^{12}$ &&3x10$^{13}$\\
\hline
\end{tabular}
\end{center}
\renewcommand{\baselinestretch}{0.8}
{\footnotesize $^a$ 1) \cite{valenti08}, 2) \cite{hunter09}; 3) \cite{mazzali10};
 4) \cite{iwamoto94}; 5) \cite{sauer06}; 6) \cite{immler02}; 
7) \cite{ho20}; 8) \cite{cano13}; 9) \cite{patat01}; 10) \cite{zhi-Yun99}; 11) \cite{nakamura01};
12) \cite{mazzali06}; 13) \cite{nakar14}; 14) \cite{waxman07}.}\\
{\footnotesize $^b$ CO-star mass is the progenitor mass at the pre-SN stage.}\\
{\footnotesize $^{c}$ The $^{56}$Ni distribution profile is a gaussian function with f$_m$=5.0 for SN2020oi, and a step function with f$_m$=0.9 for SN2020bvc \citep[][for details]{yoon19}.}\\
{\footnotesize $^{d}$ Progenitor mass here is the initial mass of the progenitor.}\\
\end{table*}

The light curves of SN2020oi are shown in Figure~\ref{Flightcurves20oi}. SN2020oi was discovered by the ZTF on 2020 January 7 (MJD=58855.54){\footnote{https://lasair.roe.ac.uk/object/ZTF20aaelulu/}} at a magnitude of $r$=17.28, and the last non-detection reported by ZTF survey is on 2020 January 4 (MJD=58852.546) to a depth of $r$=20.52 mag. Assuming that the explosion time occurred at the middle point between the discovery and the last non-detection, SN2002oi exploded on 2020 January 6 (t$_0$;  MJD = 58854.04$\pm$1.5). The light curves of SN2020oi show a gradual increase over $\sim$10 days, peaking around Jan $\sim$13--18 (depending on the wavelength), and decreasing rapidly for 25 days. Thereafter the light curves are rather flat or decrease rather more slowly ($\sim$0.3 mags) over $\sim$40 days ($\sim$0.0075 mag/day). This behavior is as expected for the $^{56}$Co to $^{56}$Fe decay with a slope of 0.0098 mag/day. 

\begin{figure}
\includegraphics[scale=0.6,angle=0,width=8.2truecm]{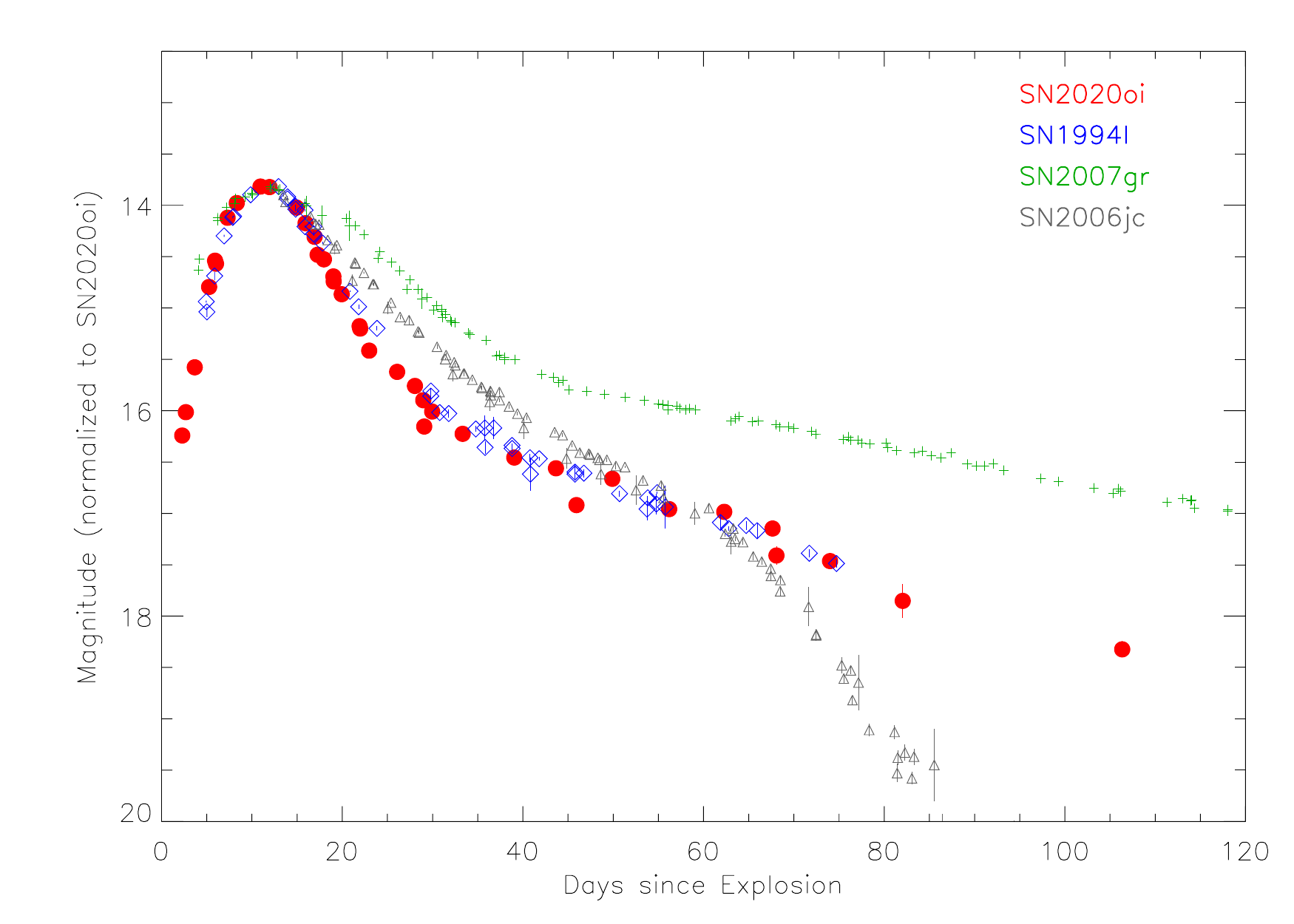}
\caption{Optical light curve of SN2020oi in $V$ band is compared with SN2007gr, SN1994I, and SN2006jc. The light curves of the three SNe are scaled to match that of SN2020oi at its peak.}
\label{FLCcomp1}
\end{figure}

After 65d, the light curves in the $g$ and $V$ bands show slightly steeper slope, which is common in SNe Ib/Ic. This more rapid decline is indicative of significant $\gamma$-ray escape due to the low mass of the ejecta \citep{clocchiatti97, sollerman98}. The peak in the light curve in the $V$ band is on 2020 January 17 (MJD$_{Vmax}$ = 58865.52$\pm$0.14) at V$_{\rm max}$ = 13.81$\pm$0.03 mag, determined using a light-curve peak finder code based on Monte Carlo simulation \citep{bianco14}. The peak in the light curve in the $B$ band is on 2020 January 15 (MJD$_{Bmax}$ = 58863.15$\pm$0.20) at B$_{\rm max}$ = 14.73$\pm$0.02 mag. The dates of $V_{\rm max}$ and $B_{\rm max}$ are 11.5 and 9.1 days after t$_0$ (Figure \ref{Flightcurves20oi}), respectively.

We have compared the observed UBV$griz$-band light curves from the LCO network, Konkoly, and ZTF with some supernova models, obtained using the one-dimensional multi-group radiation hydrodynamics code STELLA \citep{blinnikov00, blinnikov06}. The STELLA code employs a predictor-corrector high order implicit scheme for line emission and calculates the spectral energy distributions (SEDs) at each time-step. The multi-color light curves are obtained by convolving the filter response functions with the SEDs. The STELLA code also implicitly treats time-dependent equations of the angular moments of intensity averaged over a frequency bin with the variable Eddington method until the agreement with hydrodynamics is achieved at each time-step.

When a progenitor star loses its envelopes of hydrogen and helium by its interaction with a binary companion. The explosion of the carbon-oxygen (C-O) star is triggered by Fe core-collapse and leads to an explosion of Type Ic. Using the model grids in \cite{yoon19}, we find that the multicolor light curves of SN2020oi are well reproduced by the SN Ic model $``$CO2.16\_fm5.0\_E1.0" which has explosion energy of $E_\mathrm{exp} = 1.0\times10^{51}$~erg and a nickel mass of 0.07~$M_\odot$, as seen in Figure~\ref{FLCmodel1}. The progenitor of this SN model is a helium-poor C-O star of mass 2.16 $M_\odot$ corresponding initial mass of the progenitor is about 13 $M_\odot$. The adopted mass cut (where the supernova energy is injected in mass coordinate) is 1.45 $M_\odot$, assumed to be mass of the neutron star and the corresponding ejecta mass is 0.71~$M_\odot$. We assume in our model that the mass cut is the outer boundary of the iron core. The nickel is assumed to be fully mixed \citep[see Figure 1 of][for the Ni distribution of f$_m$=5.0]{yoon19} in the SN ejecta. 

We checked the light curve fitting and parameter estimates by assembling the bolometric light curve and fitting it with an Arnett-model \citep{arnett82, valenti08, chatzopoulos12}. This model family assumes homologous expansion of constant-density ejecta, powered by centrally located radioactive $^{56}$Ni, and solves the equation of radiation diffusion assuming constant opacity. The fitting resulted in physical parameters that are consistent with the STELLA model fitting above: ejecta mass in between 0.5 and 1.1 M$_\odot$ (depending on the opacity) and initial $^{56}$Ni mass of $0.07 \pm 0.01$ M$_\odot$.

\begin{figure}
\includegraphics[scale=0.6,angle=0,width=8.3truecm, height=10truecm]{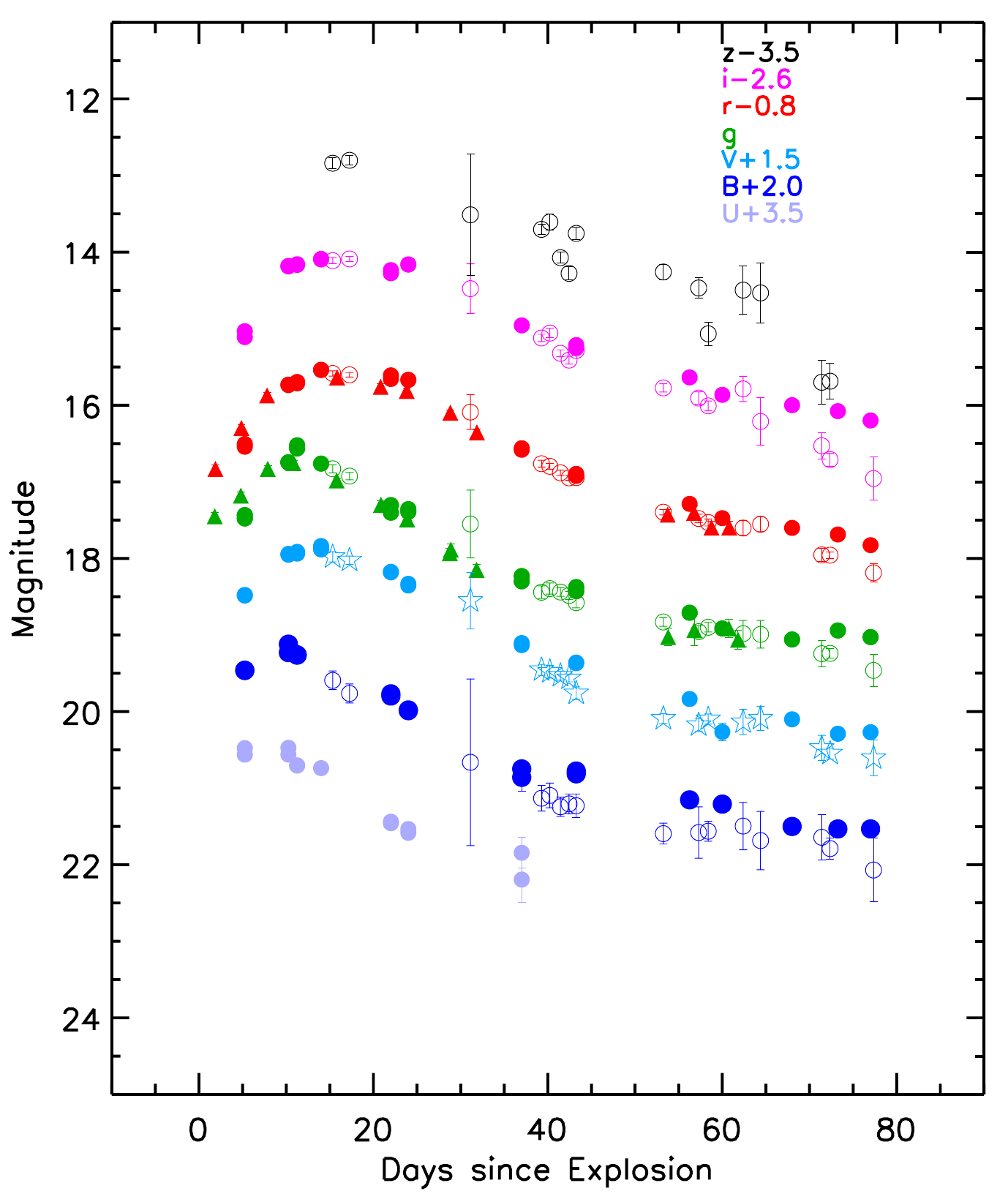}
\caption{Multi-color light curves of SN2020bvc from LCO (filled circles), Konkoly
(open circles and stars), and ZTF (triangles) with days since the
explosion. The explosion date of 2020 February 3 (MJD=58882) is between the
last non-detection and the first detection from \cite{ho20}.}
\label{Flightcurves20bvc}
\end{figure}
\begin{figure}
\includegraphics[scale=0.6,angle=0,width=8truecm]{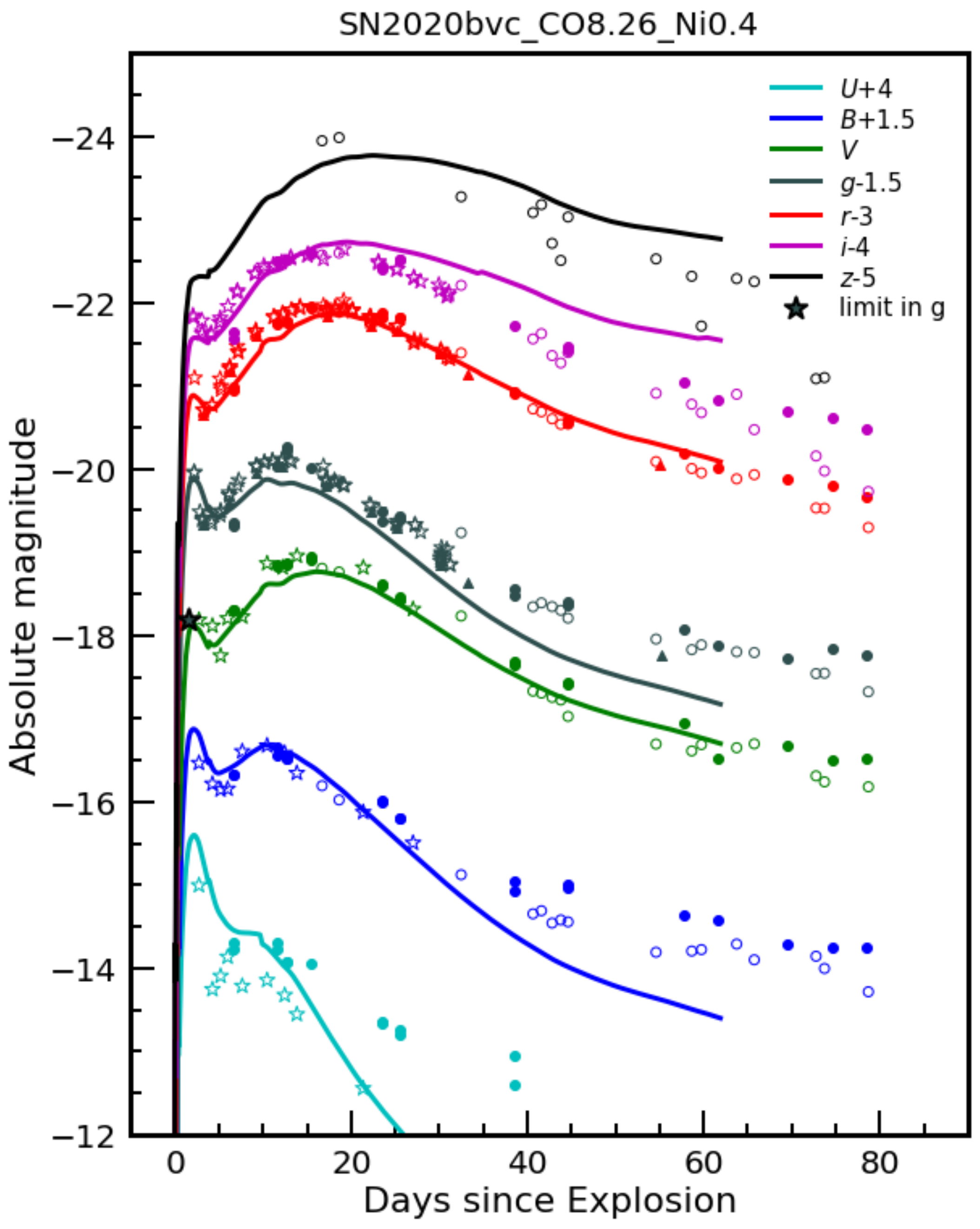}
\caption{Light curves of SN2020bvc (the double-peaked light curves from
\cite{ho20} are denoted by $``$stars") superposed on
SN Ic model  predictions from an 8.26~$M_\odot$ CO star progenitor with
explosion energy of $12\times10^{51}$~erg, nickel mass of $0.40~M_\odot$,
and CSM mass and radius of $M_\mathrm{CSM} = 0.1~M_\odot$ and
$R_\mathrm{CSM} = 10^{14}$~cm. The inconsistency of U-band photometry
between our LCO and SWIFT \citep{ho20} is due to the difference in
filters.}
\label{FLCmodel2}
\end{figure}

The light curves of SN2020oi are compared with those of other Type Ic SNe of SN2007gr, SN1994I, and SN2006jc in Figure \ref{FLCcomp1}. The photometry and spectra of other SNe are obtained from the Open SN catalog{\footnote {https://sne.space/sne/}}. The light curve of SN2020oi is remarkably similar to that of SN1994I. The two SNe, SN2007gr and SN2006jc, show slower declines after their peaks than SN2020oi.  The progenitor and explosion properties of the three Ic SNe are compared in Table \ref{Tprogenitor}. The explosion energy, Ni mass, and ejecta mass (E$_\mathrm{exp}$ $\sim$1$\times10^{51}$ erg,  M$_{\rm Ni}\sim$0.07 M$_\odot$, and M$_{\rm ej}\sim$0.71 M$_\odot$) of SN2020oi are indeed similar to those of SN1994I. In contrast, SN2007gr has a higher ejecta mass. The 2.16 M$_\odot$ CO star in SN2020oi can be made via binary interaction. The mass of circumstellar matter (CSM) could not be tightly constrained from optical curves without detection of the first peak (observed in SN2020bvc; see Figure \ref{FLCmodel2}). But the comparable model in Figure \ref{FLCmodel1} assumes the upper limit to the CSM mass of 3$\times$10$^{-4}$ M$_{\odot}$, which is still consistent with the observation. It does not contradict the recent findings based on radio observations by \cite{horesh20}.

\subsection{Light Curves and Explosion Properties of SN2020bvc}

Figure \ref{Flightcurves20bvc} shows optical light curves of SN2020bvc with LCO, Konkoly, and ZTF photometry. We use 2020 February 3.67 (MJD = 58882.67) from \cite{ho20} as day 0, which is based on the time of the ATLAS non-detection. The maximum in the $V$ band light curve occurred on 2020 February 20  (MJD$_{Vmax}$ = 58899$\pm$0.12) at $V_{\rm max}$ = 16.36$\pm$0.02 mag. The maximum in the $B$ band light curve occurred on 2020 February 12 (MJD$_{Bmax}$ = 58891.26$\pm$0.12) at $B_{\rm max}$ = 17.18$\pm$0.01 mag. The dates of $V_{\rm max}$ and $B_{\rm max}$ are 16.32 and 8.6 days after $t_0$ (Figure \ref{Flightcurves20bvc}), respectively. After reaching maxima, the light curves show gradual decreases until day $\sim$100 (the end of the coverage). The peak of SN2020bvc is shallower than that of SN2020oi by $\sim$1 magnitude.

\begin{figure*}[!ht]
\includegraphics[scale=0.6,angle=0,width=13truecm,height=17.2truecm]{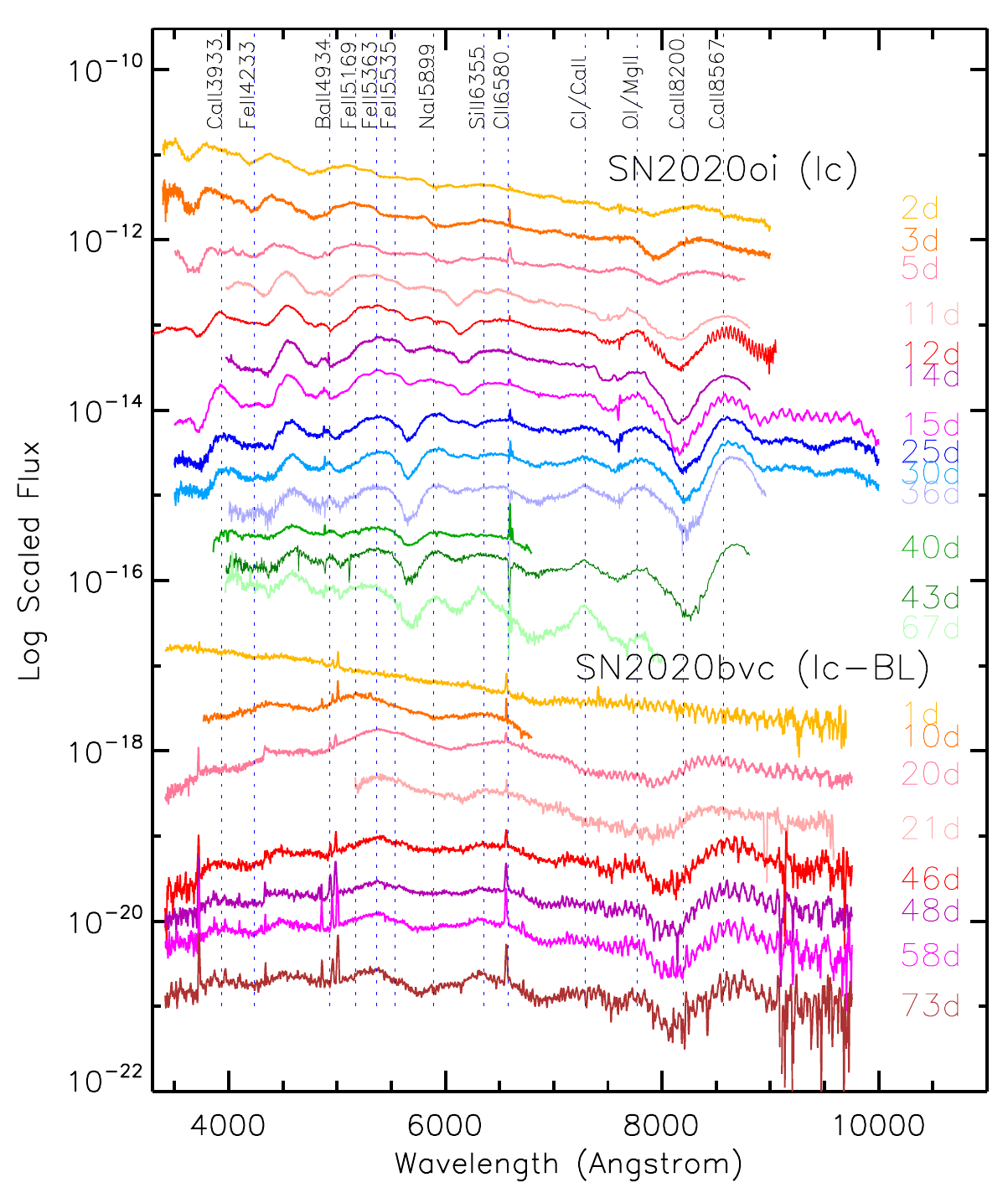}
\caption{Optical spectra of SN2020oi and SN2020bvc with the LCO network, Bok, Lijiang, Xinglong, and SOAR telescopes (see Table \ref{Tobs} for details). At top, the wavelengths of noteworthy atomic lines (at rest wavelength) are indicated (where CaII8200 infers blue-shifted \caii\ triplets.)}
\label{Foptispec}
\end{figure*}

To reproduce the light curves of SN2020bvc, we have calculated a new SN model \citep[based on][]{yoon19} with a helium-poor CO star of mass $8.26~M_\odot$, which corresponds to a progenitor 40--50~$M_\odot$ (the full grid of models will be published elsewhere). We include the double-peaked light curves from \cite{ho20}  for modeling to derive explosion parameters. We convert the light curves from \cite{ho20} in AB-system to Vega-system to compare with our LCO, Konkoly, and simulated light curves. The model is superposed on the light curve of SN2020bvc in Figure \ref{FLCmodel2}. The assumed mass cut is 1.86~$M_\odot$ and hence the ejecta mass is 6.40~$M_\odot$. We find that an explosion energy of $E_\mathrm{exp} = 12\times10^{51}$~erg and a radioactive nickel mass of 0.4 $M_\odot$ lead to a reasonably good fit for the main peak and the light curve width (Figure~\ref{FLCmodel2}). The nickel is assumed to be uniformly mixed in the inner region of the SN ejecta \citep[see Figure 1 of][for the Ni distribution of f$_m$=0.9]{yoon19}, encompassing 90\% of the ejecta mass. If this SN had a magnetar engine \citep[e.g.,][]{kasen10magnetar}, the actual nickel mass would be smaller than our estimate from the model.

The light curves of SN2020bvc in the optical are double-peaked. To explain the first peak, we assume that the progenitor was surrounded by massive circumstellar matter (CSM) having the standard $``$stellar wind density profile ($\rho$)$"$ of $\rho = \dot{M}/4\pi v_\mathrm{w} r^2$, where $\dot{M}$ is a mass-loss rate, $v_\mathrm{w}$ is a wind velocity, and $r$ is the radius of a stellar wind. Within the parameter space we explored CSM mass (i.e., $M_\mathrm{CSM}$ = 0.05, 0.1 and 0.15 $M_\odot$) and CSM radius (i.e, $R_\mathrm{CSM}$ = 10$^{13}$, 10$^{14}$ \& 10$^{15}$ $\mathrm{cm}$), we find that $M_\mathrm{CSM}$ = 0.1 $M_\odot$ and $R_\mathrm{CSM}$ = 10$^{14}~\mathrm{cm}$ ($\sim$1400 R$_\odot$) give the best fit to the first peak of the light curves in the different bands. The estimated mass and radius of the extended material is comparable to those in other Ic (e.g., super-luminous) SNe \citep{piro15}. In Section \ref{Sdoublepeak}, we compare the light curves of SN2020bvc with other Ic-BL SNe and discuss possible scenarios to explain the double-peaked light curves.

\begin{figure*}
\includegraphics[scale=0.6,angle=0,width=18.7truecm]{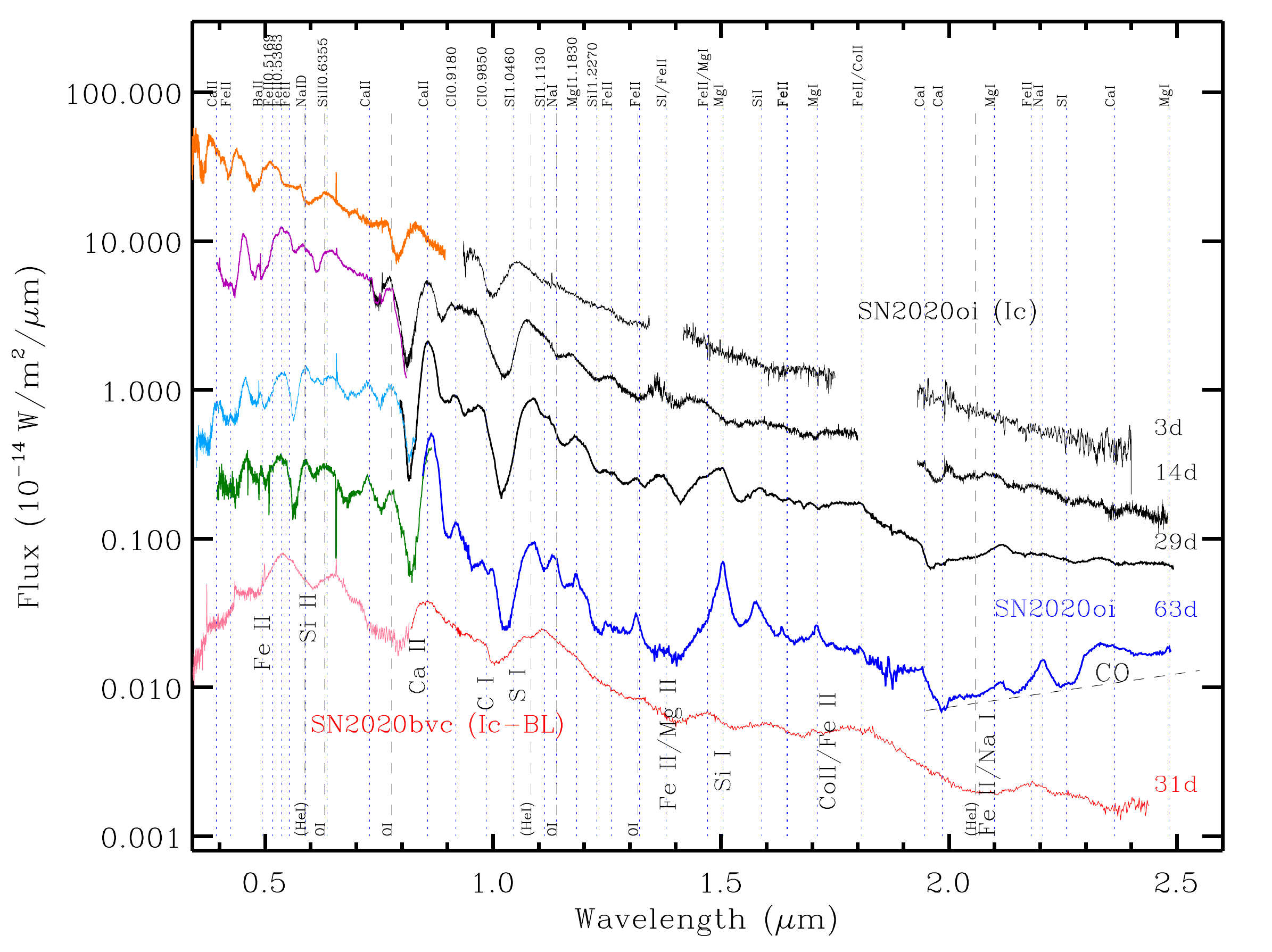}
\caption{Optical and near-infrared spectra of SN2020oi (top four) and SN2020bvc (bottom in rose and red), shifted to $z$=0. The optical spectra closest in time (see Figure \ref{Foptispec} marked in the same color) to the dates of the infrared spectra have been adjoined to the infrared spectra. At the top the wavelengths of noteworthy atomic lines (at rest wavelength) are indicated and at the bottom the broad lines observed in SN 2020bvc are identified. The dates (after explosion) at right are when the infrared spectra were obtained. 
The dashed line in the day 63 spectrum of SN2020oi (in blue) is an approximate K-band continuum, and the location of the CO emission above it is indicated.
The O I lines are marked in dashed-dotted lines with the labels at the bottom, blended with other lines. The location of He I lines (at 0.5875, 1.0830 and 2.0581\mic) are marked in grey dashed lines and indicated at the bottom.}
\label{FnearIRspec}
\end{figure*}

\subsection{Optical and Near-IR Spectroscopy of SN Ic and Ic-BL}

Figure~\ref{Foptispec} shows thirteen optical spectra of SN2020oi and eight optical spectra of SN2020bvc. Figure~\ref{FnearIRspec} shows four near-IR spectra, along with the corresponding optical data from the nearest day of SN2020oi and SN2020bvc, together with the identifications of lines (see Table \ref{Tobs}). The optical and near-IR spectra of SN2020oi span days 2 to 67 and those of SN2020bvc span days 1 to 73. The rest wavelengths of the lines (note that we use Angstroms in STP air for the optical spectra and microns in vacuum for the near-infrared spectra) are from the SN spectral models by \cite{dessart12} and synthesized spectra using SYNAPPS \citep{thomas11}, other observed spectra of SN Ic \citep{hunter09, stevance17, stevance19, gerardy02, drout16}, or Ib \citep{ergon14, jencson17}.

The spectra of both SNe are dominated by atomic lines, in absorption or in emission, and sometimes simultaneously in the form of P Cygni profiles. In some cases, where multiple lines are thought to contribute, such as the near-infrared \caii\ triplet, we refer to the combined lines as $``$features". The species making the strongest contributions are \feii, \siii, \caii, \ci, \si, and \sii\ (see Figure ~\ref{FnearIRspec}). The lines of Type Ic-BL SN2020bvc are distinctly broader than Type Ic SN2020oi. Surprisingly, despite their belonging to different subclasses of Type Ic, almost the same set of lines can be seen in each of them. The lines are labeled next to each peak; \feii\ at 4924, 5018, 5169, 5363 \AA\ \citep{kankare14}, and \siii\ at 6355 \AA. The \caii\ IR triplet shows both absorption and emission (due to its members having P Cygni profiles). We use the mean wavelength of 8567 \AA\ for the triplet, the wavelengths of whose individual lines are 8498, 8542, and 8662 \AA. The absorption line at 1.01 $\mu$m is \ci, and the nearby line at 1.046 \mic\ is \si\ with both absorption and emission in Figure \ref{FnearIRspec}. The lines of \si\ at 1.113 \mic, \feii/\mgii\ at 1.47 \mic , \sii\ at 1.589 \mic, \feii\ at 1.809 \mic, and \feii\ at 2.18 \mic\ of SN2020bvc also have counterparts in the spectra of SN2020oi. The presence of these lines is consistent with the lines in SN 2011dh, SPIRITS 15c \citep{jencson17}, and in SN2007gr \citep{hunter09}. The \feii\ at 1.64 \mic\ and \mgi\ lines at 1.5 $\mu$m are commonly detected in SN Type Ib/IIb, SPIRITS 15c \citep{jencson17} as well as in IIP SN2017eaw \citep{rho18sn}. There is little temporal change in the spectra of SN2020bvc from 20d to 73d. The spectrum at 1d is mostly black-body emission and the peak emission wavelength in the spectrum at 10d is different from those at 20 - 73d. The velocity profiles of the line features are discussed in Section \ref{Svelocity}.

Lines of He are present at 5875 \AA, 1.083 \mic, and 2.0587 \mic. At 5875 \AA, there is also \nai\ line (Figure \ref{Foptispec}), and at 1.083 $\mu$m, the \si\ line is present (Figures \ref{FnearIRspec}). The \hei\ line at 2.0587\mic\ is free of other lines However, at high velocity, the absorption of this line in SN2020oi is potentially present. The presence and velocity profiles of He lines are discussed in Sections \ref{Svelocitysnoi} and \ref{Shelium}.

The spectral signature of the CO overtone band emission at 2.3--2.5~$\mu$m with its ``sharp cut-on" on the short wavelength edge is clearly evident in the day 63 spectrum of SN2020oi; it was not present at day 29. Band-heads of CO occur at 2.294, 2.323, 2.353, 2.383, 2.414, 2.446, and 2.447 $\mu$m \citep[see Figure 2 of][]{rho18sn}. Only the cut-on due to the shortest wavelength band head is apparent. This is due to a combination of the high-velocity width and the optically thick CO gas that makes up the individual bands.

Longwards of 2.0 $\mu$m, the continuum flux density from SN2020oi increases on day 63,  whereas it was decreasing in the earlier spectrum. This is an unambiguous detection of warm dust. Detection of dust in Type Ic spectra is rare, especially at such early times.

\section{Discussion}
\subsection{CO and Dust Formation in Type Ic SN 2020oi} 
\label{SCOdust}

\subsubsection{CO Properties from LTE Model}
\label{SLTECO}
\begin{figure}
\includegraphics[scale=0.6,angle=0,width=8truecm]{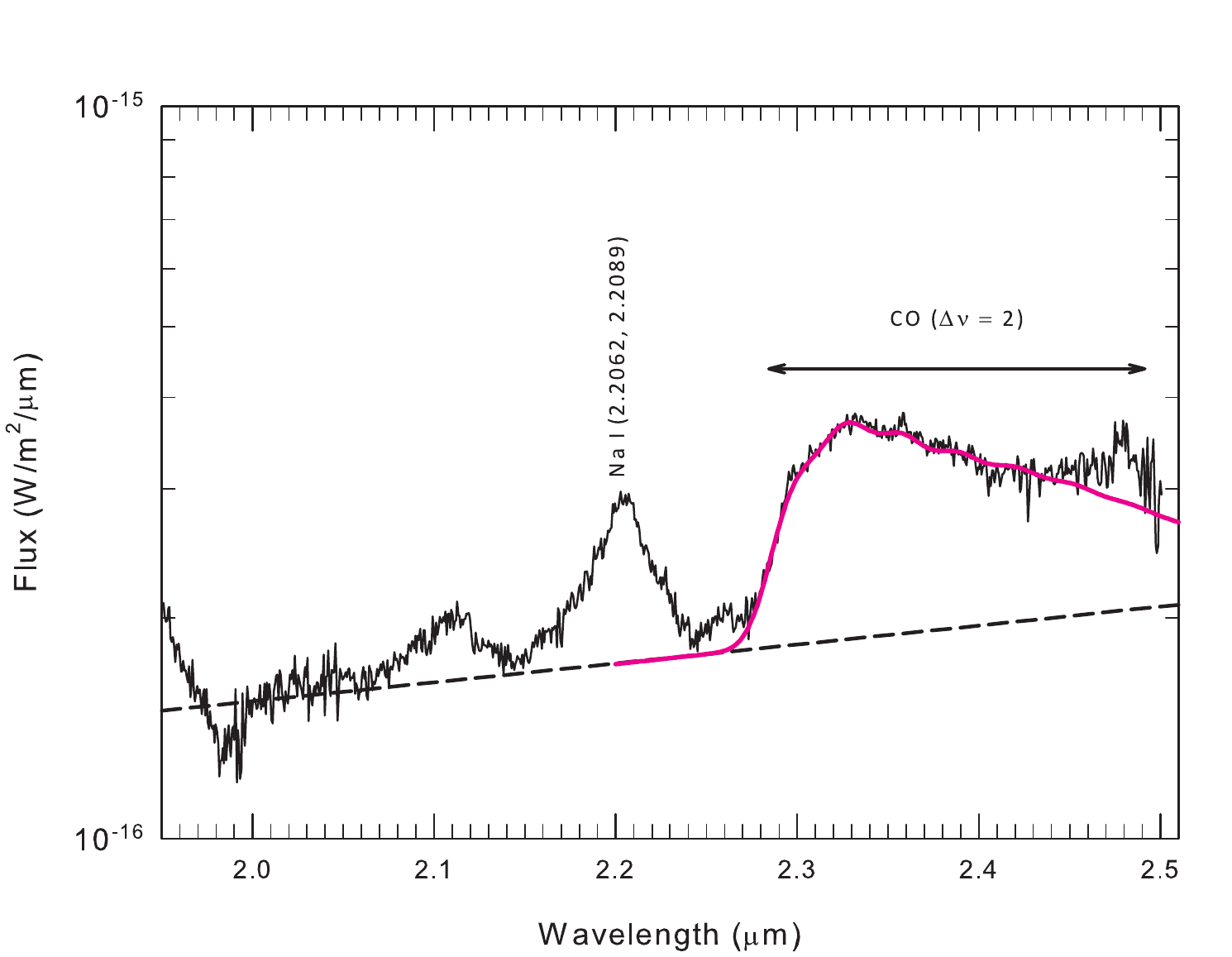}
\caption{Emission from the  CO first overtone bands  in SN2020oi on day 63 shown in black, on which is  superposed the  best LTE model fit (red line) which has the following parameters: temperature of the CO gas = 3150 K, velocity = 3700 km/s and mass = 0.8$\times$10$^{-4}$ M$_\odot$. The adopted continuum is shown by the dashed line. More details are provided in the text.}  
\label{FCOmodel}
\end{figure}

Carbon monoxide (CO) is one of the most powerful coolants in the ejecta of SNe II and is believed to be responsible in large part for cooling the ejecta to temperatures at which dust can form. It is questionable as to whether the models accurately represent the evolution of the ejecta and their mixing. Still, it is clear that measurements of CO are important tests of them, and the presence of CO is closely tied to dust formation. The onset of CO formation in SNe is best seen in the first overtone bands at 2.3-2.5 \mic, which have been detected only in a handful of ccSNe \citep[e.g.,][]{sarangi18}.

We have estimated the CO mass in SN2020oi using the LTE model developed by \cite{das09}, which assumes optically thin emission and a pure $^{12}$C$^{16}$O composition. Earlier applications of the model include SN2017eaw \citep{rho18sn}, SN2016adj \citep{banerjee18} and several novae \citep{banerjee16, joshi17}. We adopted a straight continuum (for the dust emission) passing through the two minima (2.01 -- 2.08\mic\ and 2.155 -- 2.17\mic) in the spectrum between 2.00 and 2.23 \mic. The best fit model, shown in Figure \ref{FCOmodel},  was determined by varying the CO mass, temperature, and velocity dispersion using chi-square minimization over the region 2.27 to 2.44 \mic. The region beyond 2.44 \mic\ was excluded from the fit optimization due to spectral contamination caused by a strong atomic line at $\sim$2.48 \mic, possibly due to \mgi. We estimate the following parameters for the CO gas: temperature = 3150 $\pm$ 200 K; velocity = 3700 $\pm$ 100 \kms; CO mass = (0.7--0.9)$\times$10$^{-4}$\,$(d/16.22)^2$ M$_\odot$ where $d$ is the actual distance in Mpc.

The estimated CO mass of SN2020oi is the first estimate for a Type Ic SN. The CO mass of $\sim$  10$^{-4}$ M$_\odot$ is consistent with those observed in other SNe \citep[e.g. SN1987A;][and references therein]{sarangi18} at similar epochs. Models of CO formation for SN Type IIP \citep[e.g.][]{sarangi13, sluder18} indicate that the amount of CO increases from $\ltsimeq10^{-4}$ M$_\odot$ at 100 days since the explosion to as much as 0.1 M$_\odot$ after $\gtsimeq1500$ days, irrespective of SN progenitor \citep[see e.g.][]{sarangi13}. CO was detected in Type Ib SN2016adj at day 58 after discovery and its estimated CO mass was 2$\times$10$^{-4}$ M$_\odot$ \citep{banerjee18}.

The detection of CO in SN2020oi, only $\sim$63d since the explosion ($\sim$54 and 52d after $B$ and $V$ band maximum, respectively), is one of the earliest CO detections in an SN of any type. There are only two other SNe where a similar, early  CO detection was made viz. SN 2013ge \citep{drout16} and SN2016adj \citep{banerjee18}. Detections of CO were reported in Type Ic SN2000ew and SN2007gr at day 90 and 70 (post-B maximum), respectively \citep{gerardy02, hunter09}. \cite{sarangi18}, summarizing data obtained up to 2013, found only four CO detections before day 100. In this context, the early onset of CO formation in SNe, such as SN2020oi, may present a serious challenge for theoretical models.

\subsubsection{Non-LTE CO model: Diatomic Molecule formation as a Diagnostic} \label{SnonLTECO} 

As a second approach to molecular formation, we consider spherical explosion models with the free parameters as follows: 1) the progenitor mass, 2) the initial stellar density and abundance structure, 3) the explosion energy and amount of $^{56}Ni$, and 4) possible mixing processes as key to the inherently 3D nature of the CC-SNe explosion mechanism. Parameters 1 to 3  which characterize the explosion model, can be derived from light curves and spectra.

We demonstrate in the Appendix that  the structure of the molecular bands of $CO$ and $CO^+$ provide a sensitive tool to probe the temperature, velocity, chemical composition, and that a large $CO+/CO$ indicates the exposure to hard radiation in the CO emitting region which depends on mixing.

As described below, further information on mixing comes from the time dependence of the strength of the CO feature. In CC-SNe with  massive H-rich envelopes, the onset of the CO formation is correlated with the photosphere entering the $CO$ core as discussed below. However, in SN~Ic, the CO core is exposed from the beginning, and the photosphere is hot. Thus, $^{56}Ni$ mixing may play the central role in the onset of CO formation rather than low T.

\paragraph{Mixing in Light of the Explosion Physics}
 Large and small scale mixing is not well understood, but polarization measurements of another Type Ic, SN~2002ap, and the Type IIb, SN2001ig, suggest a large scale bipolar abundance structure \citep{2003ApJ...592..457W,2007ApJ...671.1944M} which is consistent with asymmetric, axially symmetric explosion mechanisms  \citep[e.g.][]{1999ApJ...524L.107K,1999ApJ...521..179H,2017RSPTA.37560271C}. Spherical explosions form a central cavity and a dense shell supported by radiation pressure. In contrast a large scale asymmetric explosion does not form a central cavity because material is dragged down, and the void is filled. As shown in the papers cited above, the detailed structure depends on various mechanisms which include Rayleigh-Taylor instabilities of the shell, asymmetric explosions by bipolar-outflows, angular momentum inherited from the progenitor, etc., which result in different density and abundance profiles. Here, we assume mixed material fills  the inner void and use mixing of abundances of the parts of the core as a free parameter and the molecule formation as an indicator.\footnote{In the
spherical explosion model for SN2020oi, the inner cavity has a size of $\approx$ 6,000 \kms\  which is both inconsistent with the velocity indicated by the CO-feature. As we will see below, the resulting low density due large geometrical dilution causes CO formation time-scales which are inconsistent with the early CO observed in SN2020oi.}

\paragraph{Model setup and methods}
 
 We consider the questions whether and under which condition molecular formation can be understood in SNe~Ic such as SN2020oi.

 We simulate the time-dependent molecule formation via rate-equations and radiation transport as affected by the optical depth of the CO lines. We use the dynamical background based on the progenitor and the explosion parameters for SN2020oi found by the light curve-analysis of \citet{yoon17} and \citet{yoon19} with mixing of the ejecta as a free parameter. For our simulations, we use our  HYDrodynamical RAdiation code HYDRA which includes modules to provide a solution for the nuclear networks, the statistical equations needed to determine the atomic level population, the equations of state, the opacities,  the hydro in comoving hydro via the Piecewise Parabolic Method (PPM) and radiation/positron transport problems \citep{Hubeny03,Kubat09,hoeflich19}. These modules have been widely applied in non-LTE analyses of  SN1987A, and of SNe of Type II, Ibc, and Ia.

 For the background and computational efficiency, we use the same assumptions as for SN1994I \citep{2000ApJ...528..590H}, an approach  very similar to STELLA. The photon transport in molecules is solved in the co-moving frame without relativistic corrections \citep{m75,mm84} and formal integration for the emitted spectrum because the CO emission is not optically thin, and can be expected to have an effect on estimates of the amount of CO required. Our Monte-Carlo scheme is used for $\gamma-$transport \citep{hoeflich93,penney14} because the ionization may affect the time-scale of the CO and $CO^+$ formation as has been widely discussed for SN~1987A \citep{1989ESASP.290..381S,1989pafe.conf..186H,1990ApJ...358..262L, 1993MNRAS.261..535M}. We use our module for molecule formation. For details, see  \citet{1990Ap&SS.171..213S,1989HiA.....8..207S}, \citet{2000AJ....119.2968G}, and references therein. For the formation and destruction, we use the time-dependent rate equations for $C+O \rightarrow CO$,\, $C^++O \rightarrow CO^+$. For the simulations below, a typical of 50,000 to 300,000 ro-vibrational transitions have been taken into account for each $CO$, $CO^+$ and $SiO$. $SiO$ are included because the progenitor model has overlapping chemical regions with $C$, $O$, and $Si$. Depending on the conditions, $Si$ can bind $O$ which then is not available for $CO$ formation. As is common for the time-dependent networks \citep[e.g.][]{1993MNRAS.261..535M}, we include three-body association with neutral and singly-ionized C and O, radiative association and dissociation,  collisional dissociation by electrons, and charge-exchange reactions. In addition, the energy deposition and ionization by radioactive decay, $\gamma$-rays and positrons, are taken as additional ionization processes for both the atoms and molecules. Note that for SN1987A, the importance of channel via the $CO^+$ channel is still under debate \citep{1989ApJ...342..406P,1990Ap&SS.171..213S}. However, $\gamma$ -ray escape was small in SN~1987A at the time of the CO formation, whereas in a SN~Ic, we see directly the exposed core strongly formed by asymmetric explosions and conditions in which non-thermal excitation and ionization occur.

\subsubsection{Application of Non-LTE CO model to the NIR spectrum of SN2020oi}
\label{SnonLTECO2}
The discussion above revealed signatures but leaves unanswered questions on the time-dependence of CO and regarding optical depth effects at some 48d past maximum or 2 months after the explosion. Can we understand the early CO-formation needed in SN2020oi, and what amount of CO is needed?

\begin{figure}
\begin{center}$
\begin{array}{c}
\
\includegraphics[angle=360,width=0.46\textwidth,height=6.5truecm]{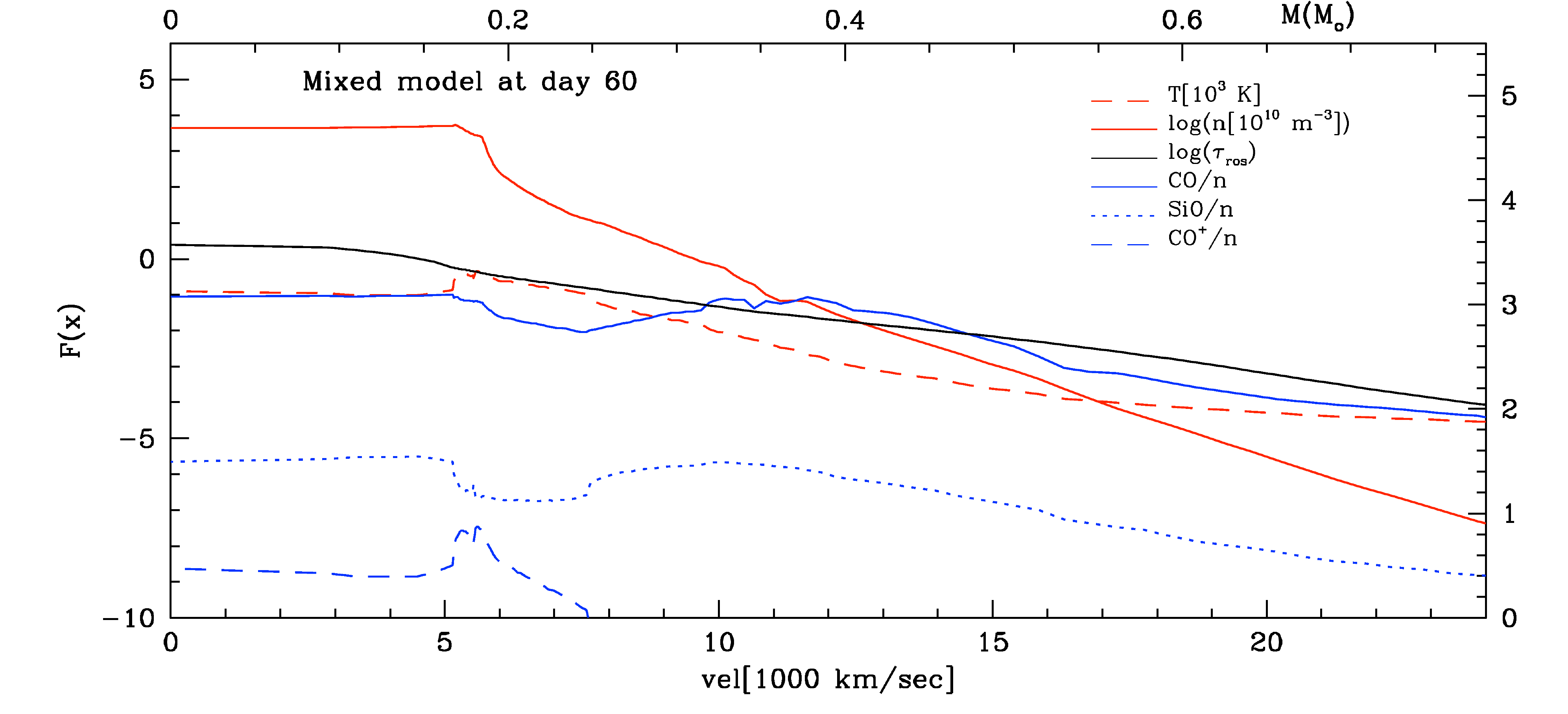}
\end{array} $
\end{center}
\caption{Structure of the mixed model at day 60. We give temperature $T$ and  particle density $n$ (right scale) and the $\tau_{ros}$, particle abundances of $CO$, $CO^+$ and $SiO$  as a functions $F(x)$ of the expansion velocity(lower scale) and mass (top scale).
The expansion velocity of 5,000-9,000 \kms\ of SN2020oi (see Section \ref{Svelocity}) infers a temperature of 3000 - 3300 K and a density of (3-5)$\times$10$^{4}$ cm$^{-3}$. 
}
\label{COST}
\end{figure}

Note that each model requires a time series of $\gamma $-ray and positron transport, and radiation transport simulations for some 1000 time-steps. Therefore, we assumed full mixing up to a certain velocity only. For the case of unmixed models, the hydrodynamical profile is used which contains a density $`$hole' as in the model. For mixed models, we assume the density hole being filled (Figure \ref{COST}). Non-thermal processes will enter by positrons and $\gamma$-rays. In all of our models, the positron component is local while the $\gamma$-ray deposition is mostly non-local. Thus, the effect on molecule formation will depend somewhat  on homogeneous or large scale mixing, but we found this effect on the CO emission spectrum to be small.

As an error estimate for the mass of the CO we only quote the variation possible within these simple assumptions.  Obviously, the formation depends on model parameters not considered, such as clumps in both chemistry and density, large scale asymmetry, etc. Instead, we calculated a series of models with various amounts of core and $^{56}$Ni mixing to fit the observations.

\begin{figure}
\begin{center}
\includegraphics[angle=360,width=0.48\textwidth]{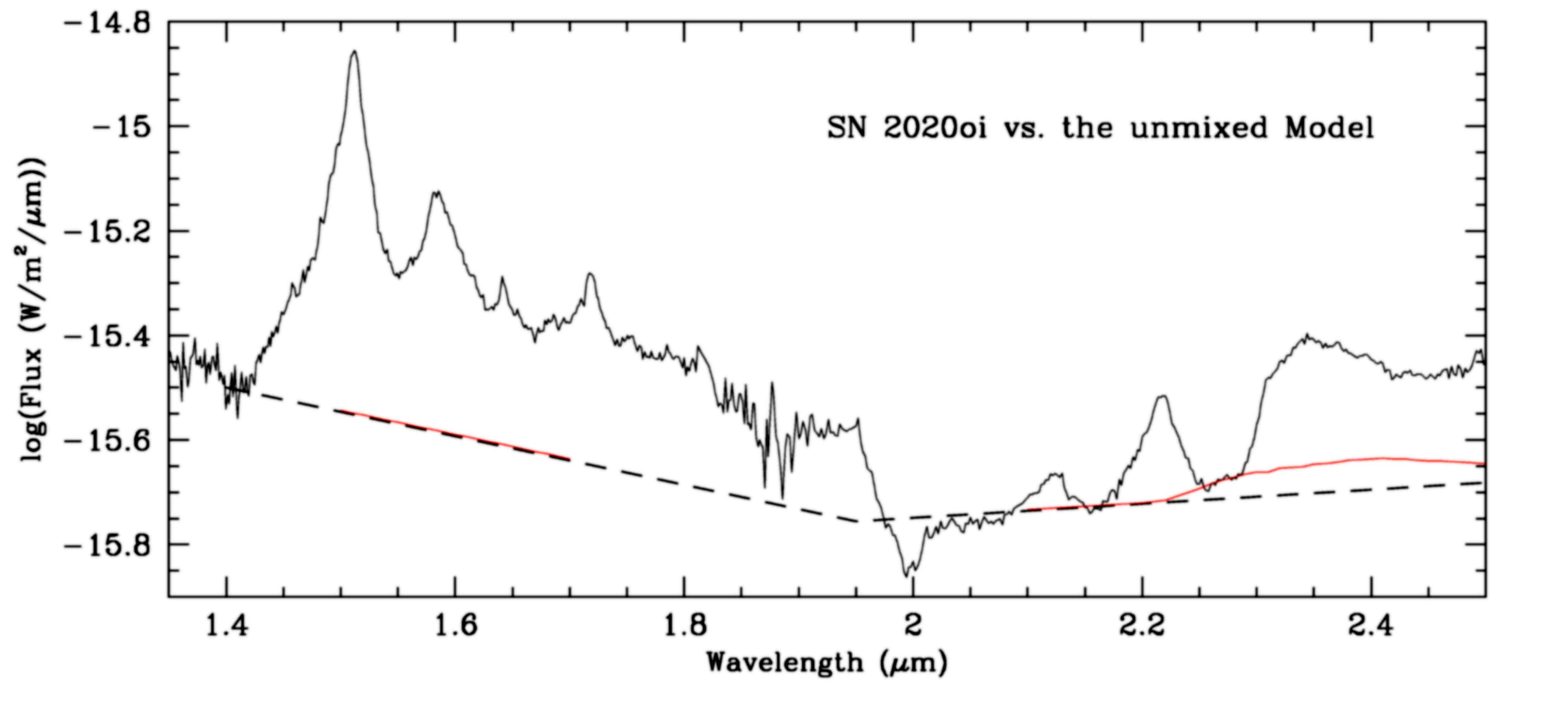}
 \includegraphics[angle=360,width=0.48\textwidth]{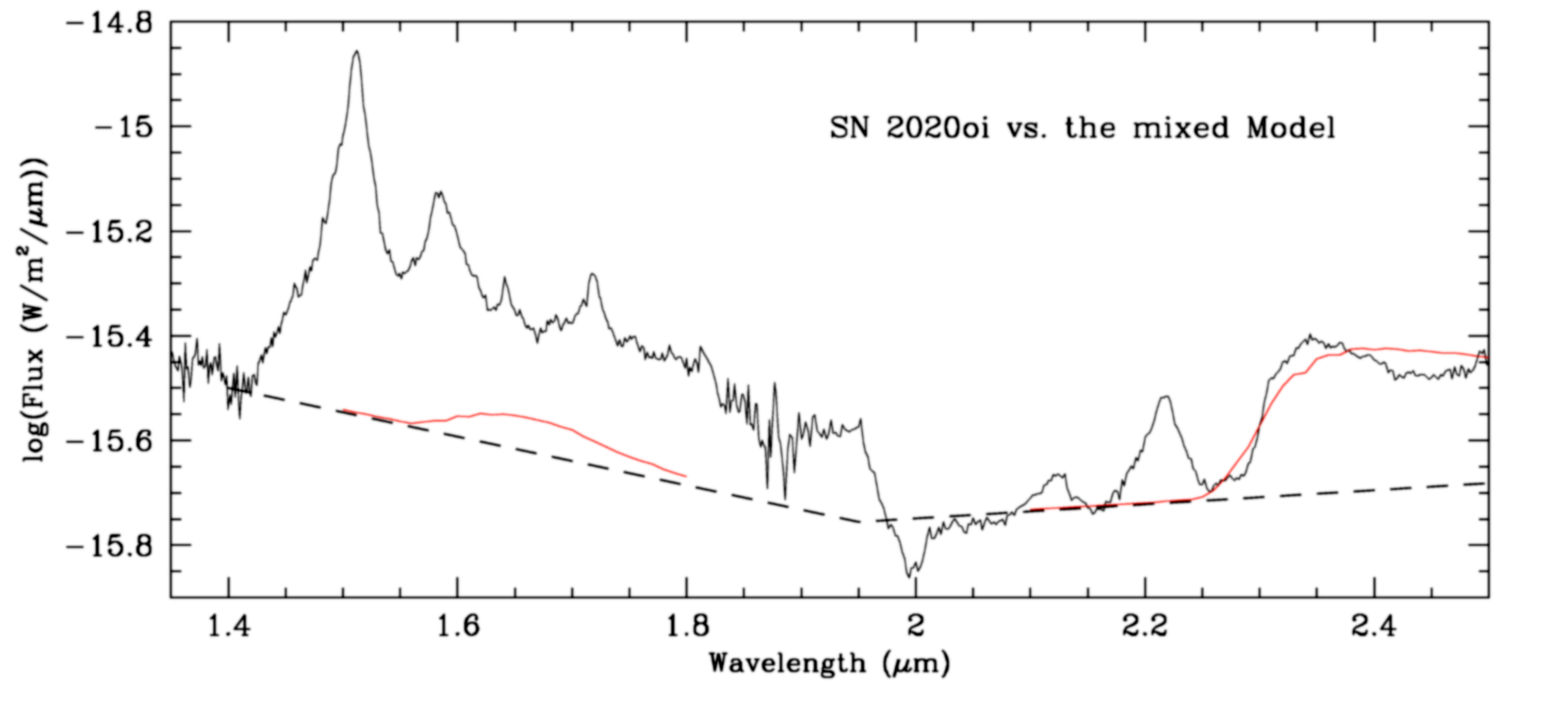}
\end{center}
\caption{Comparison between the CO contribution to the flux by time-dependent non-LTE models (red) and SN2020oi (black) in the spectral regions of the $1^{st}$ and $2^{nd}$ overtone of the  CO-bands (at 2.3-2.5 and 1.6-1.8 $\mu$m, respectively). Dashed lines are estimated continua. To produce a CO core early on, large scale mixing of the core is needed. See text for details.}
\label{COSTO}
\end{figure}

In Figures \ref{COST} $\&$  \ref{COSTO} we show the structure of the mixed model (up to $v \approx$ 7000 \kms), and the comparison between the observed and theoretical spectra of the mixed and unmixed model. Both models show some CO beginning about 50 days after the explosion. This phase coincides with the transition from the optically thick regime $\tau_{ros}$ to the optically thin regime, as it allows for effective cooling of the deeper, dense layers as a pre-requisite for molecule formation.

The early formation of CO is directly linked to low ejecta mass. CO formation would set in later for more massive envelopes because of the more effective trapping of thermal energy. Charged reactions are important for CO formation. In the inner hotter layers, a large fraction of the CO production occurs at the $`$photosphere' via $CO^{+}$ channel, as the photosphere recedes with time. However, even in our models, the rate $d{CO^+}/dt $ is larger than the expansion time  leading to the overall low ${CO^{+}}$ concentration seen in Figure \ref{COST}. Note that without extensive mixing, the CO formation time scales are much longer or the formation even suppressed and the CO forms well above the photosphere. As a result, the CO feature in our  unmixed model is both too weak and to broad (Fig.~\ref{COSTO}).

In the following we discuss the mixed model using the $1^{st}$ overtone of CO. About $0.5 - 1 \% $ of the total carbon is bound in molecules.  CO is important for the coupling between the radiation field and the plasma. However, only a small amount, $\approx 10^{-3} M_\odot$ is in the optically thin regime and contributes to the emission. The lower vibrational modes of $1^{st}$ CO-overtone only becomes optically thin at $ \approx 6000-8000$ \kms. This leads to a rather featureless synthetic CO-spectral profiles as observed. However, the overly broad blue wing to the first band-head (2.3 $\mu$m) and the suppression of the flux at the peak may indicate an excess of $CO$ formation at the time prior to the observation.

The $2^{nd}$ CO overtone (at $\sim$1.6-1.8 $\mu$m) is weak and does not conflict with the observations, which did not detect it. It may contribute to the flux at the $\approx 10 \% $ level, but atomic lines of allowed transitions still dominate the spectrum at day 60 until much later, namely the nebular phase. With mixing there is significant formation of SiO even some 2 months after the explosion which might be detected. As discussed in Section \ref{SnonLTECO}, layers with excess Si vs. C should show a strong feature if the Si core is exposed or Si/O compositions are mixed on macroscopic scale. With time, deeper layers become sufficiently cool for SiO to form. Combining near-IR CO and mid-IR SiO would provide a unique tool to distinguish various models and physical mechanisms. Temperatures are too high for early dust formation which, in SN2020oi, may favor the interpretation of pre-existing dust likely being heated by the SN radiation rather over the formation of new dust.

For SN2020oi low ejecta mass, extended mixing and explosion mechanisms which can $`$fill the inner void' are needed to match the observations. SN2020oi is an important representative of $`$low' mass end of SNe~Ic, with SN~1994I  being the only other representative. Our analysis of the CO feature shows that mixing of the inner region is needed from both its strength and profile. The spectral feature probes both the velocity and the time-domain. The light curves make only use of the time-domain and, thus, are less suitable to probe mixing and asymmetries. At early times, mixing and asymmetries will mostly change the shape of the light curves. To first order, the rise is dominated by the mass and energy of the ejecta whereas late-times are dominated by the amount of $^{56}Ni$. The  angular redistribution of photons is largest at early times. At late times, the radiation becomes isotropic. A full 3D-light curve analysis is beyond the scope of this paper. 

\subsubsection{Dust Emission in SN2020oi}

\begin{figure}
\includegraphics[scale=0.6,angle=0,width=8truecm]{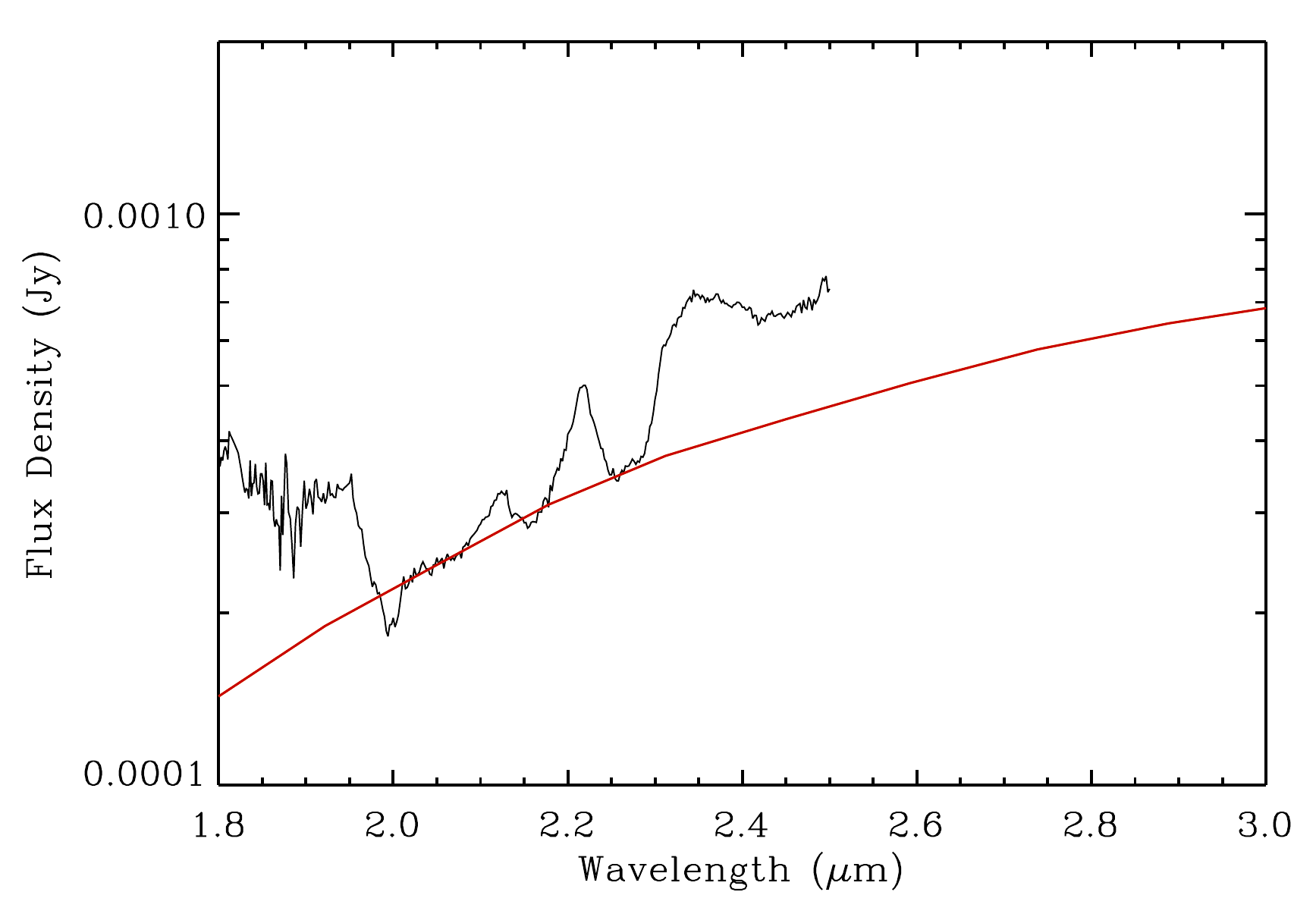}
\caption{ The spectrum of SN2020oi at 63 d superposed on a carbon dust model (in red) for the best-fitting dust temperature of $\sim$810 K. 
We performed the fitting only using the continuum between 2.00 and 2.32\mic. The best-fit model is plotted from 1.8 to 3\mic\ for display purpose.}
\label{Fdustmodel}
\end{figure}

As can be seen in Figure \ref{FnearIRspec} at day 63, the spectrum of SN2020oi exhibits a rising continuum flux density longward of 2.0 $\mu$m, as well as emission from CO longward of 2.3 $\mu$m. Clearly, the rising continuum is due to emission from dust. The dust continuum is fit with the Planck function [B$_{\nu}(T)$] multiplied by the absorption efficiency ($Q_{abs}$), where we assume carbon dust and the best-fit is shown in Figure \ref{Fdustmodel}. We use carbon dust because it condenses early, at temperature 1100 - 1700 K \citep{fedkin10}. The continuum we use is similar to that for the CO analysis (in Sections 4.1 - 4.3), but we add the third portion between 2.255 and 2.285 \mic\ in addition to the two parts between 2.01 and 2.08 \mic\ and between 2.155 and 2.17 \mic\ since Q$_{abs}$ of carbon dust has a curvature. Details of the fitting procedure using MPFIT \citep{markwardt09} and information about absorption coefficients are given in \cite{rho18}. The estimated dust temperature is 810$\pm$10 K, and the dust mass is 5.9$\pm$0.7 $\times$10$^{-5}$ M$_\odot$. There may be another lower-temperature dust component, but we cannot constrain it since the CO feature (see Figures \ref{FCOmodel} and \ref{COSTO}) continues at the end of the spectrum (at 2.5 \mic). Spectral coverage beyond 2.5 \mic\ is required to estimate a more accurate dust mass. Therefore, the dust mass of 5.9$\pm$0.7 $\times$10$^{-5}$ M$_\odot$ is a lower limit of the dust mass.

Two key questions about the emitting dust need to be addressed: (1) Is the rising continuum emitted by freshly formed dust in the SNe ejecta?  (2) Can dust form as early as at 63 days post explosion in SNe Type Ic? Dust formation in Type Ib SNe has been observed in the case of SN2006jc. SN SN2006jc showed increasing near-IR fluxe excess as early as 50 - 75 days based on photometry \citep{diCarlo08} and a rising continuum in optical spectra, indicating a dust temperature of 1700 K and dust mass of 6$\times$10$^{-6}$ M$_\odot$ \citep{smith08}. The dust emission at day 50 seen in SN2006jc was interpreted as due to either new dust in the ejecta or dust originating from the dense shell formed in the post-shock CSM \citep{smith08, pastorello08}. On day 20, AKARI observation of SN 2006jc detected near-IR to mid-IR emission which can be fitted with a two-temperature dust model with 800 K and 320 K and corresponding dust masses of 7$\times$10$^{-5}$ and 2.7$\times$10$^{-3}$ M$_\odot$, respectively \citep{sakon09}. The 800 K dust component is suggested to originate from freshly formed dust and the 320 K component is in the CSM dust.

\begin{figure*}
\includegraphics[scale=0.6,angle=0,width=4.35truecm,height=6.5truecm]{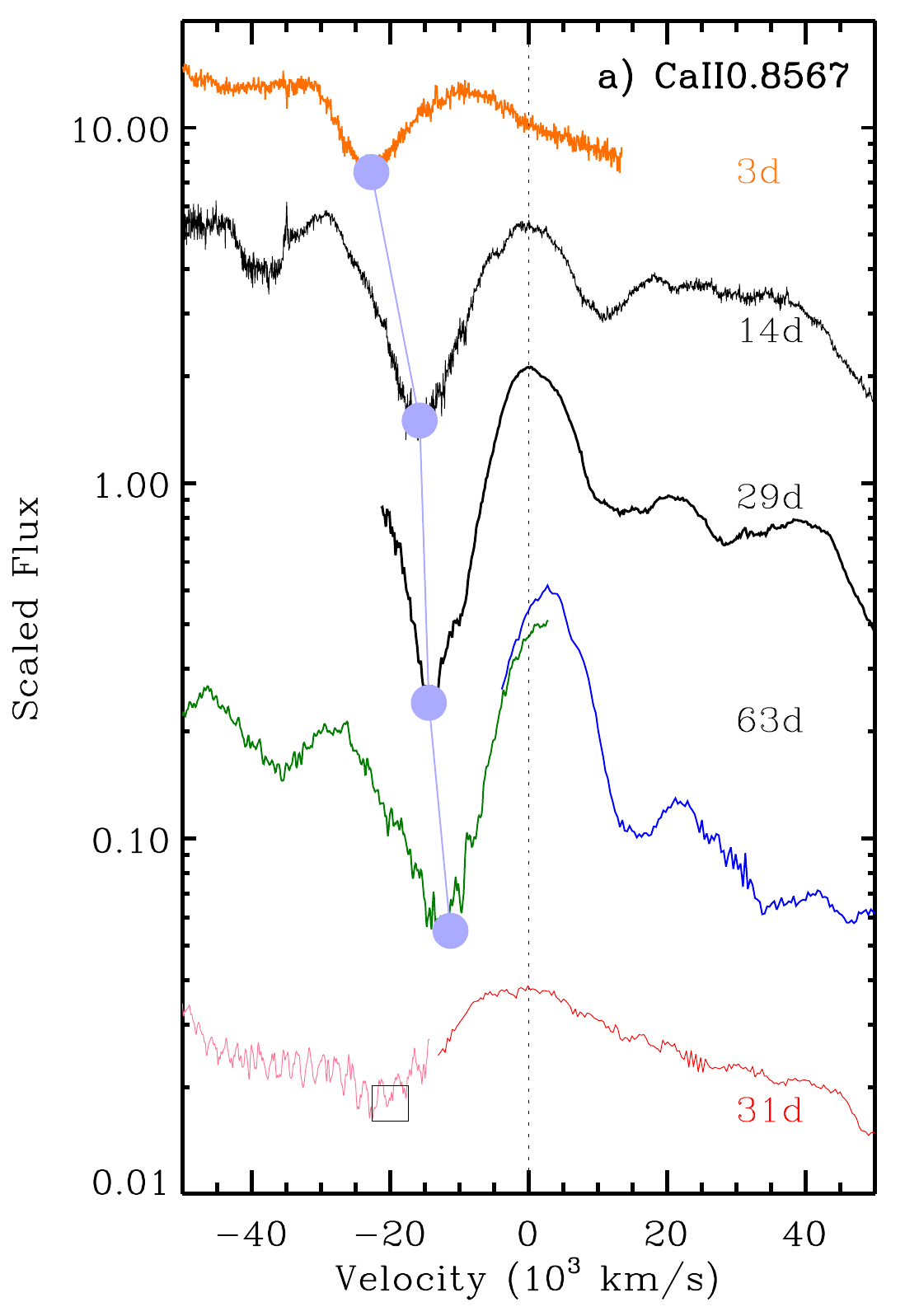}
\includegraphics[scale=0.6,angle=0,width=4.35truecm,height=6.5truecm]{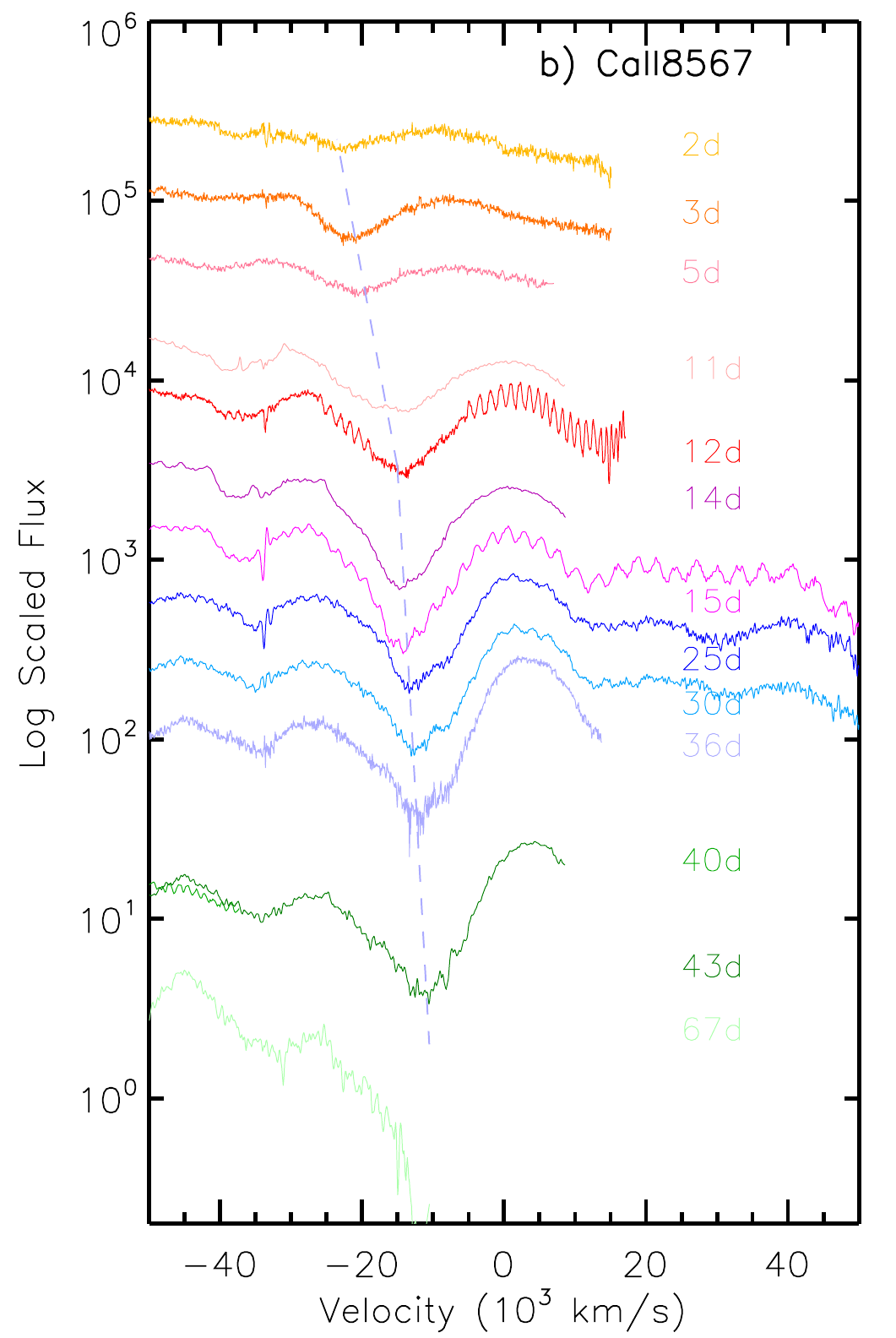}
\includegraphics[scale=0.6,angle=0,width=4.35truecm,height=6.5truecm]{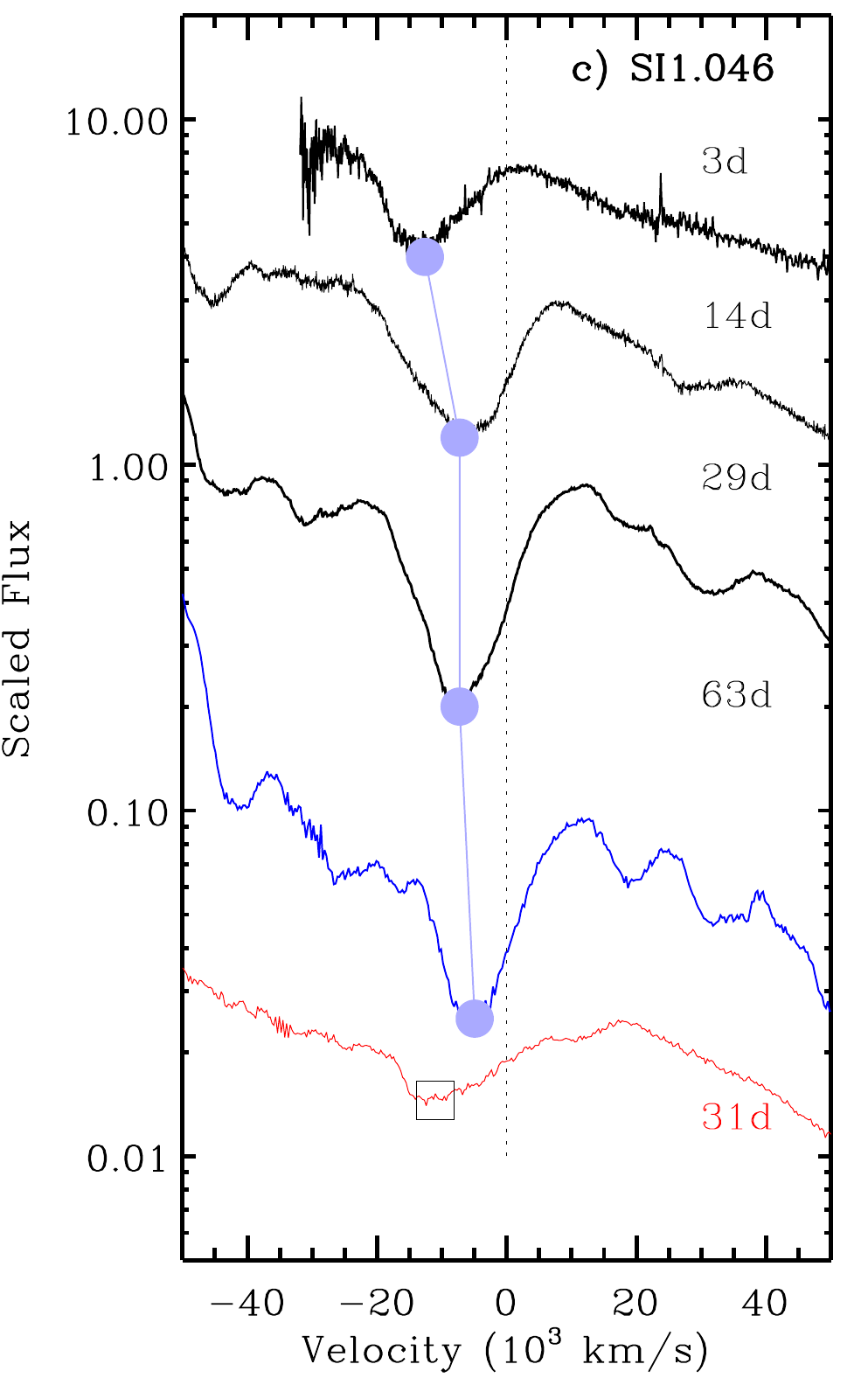}
\includegraphics[scale=0.6,angle=0,width=4.35truecm,height=6.5truecm]{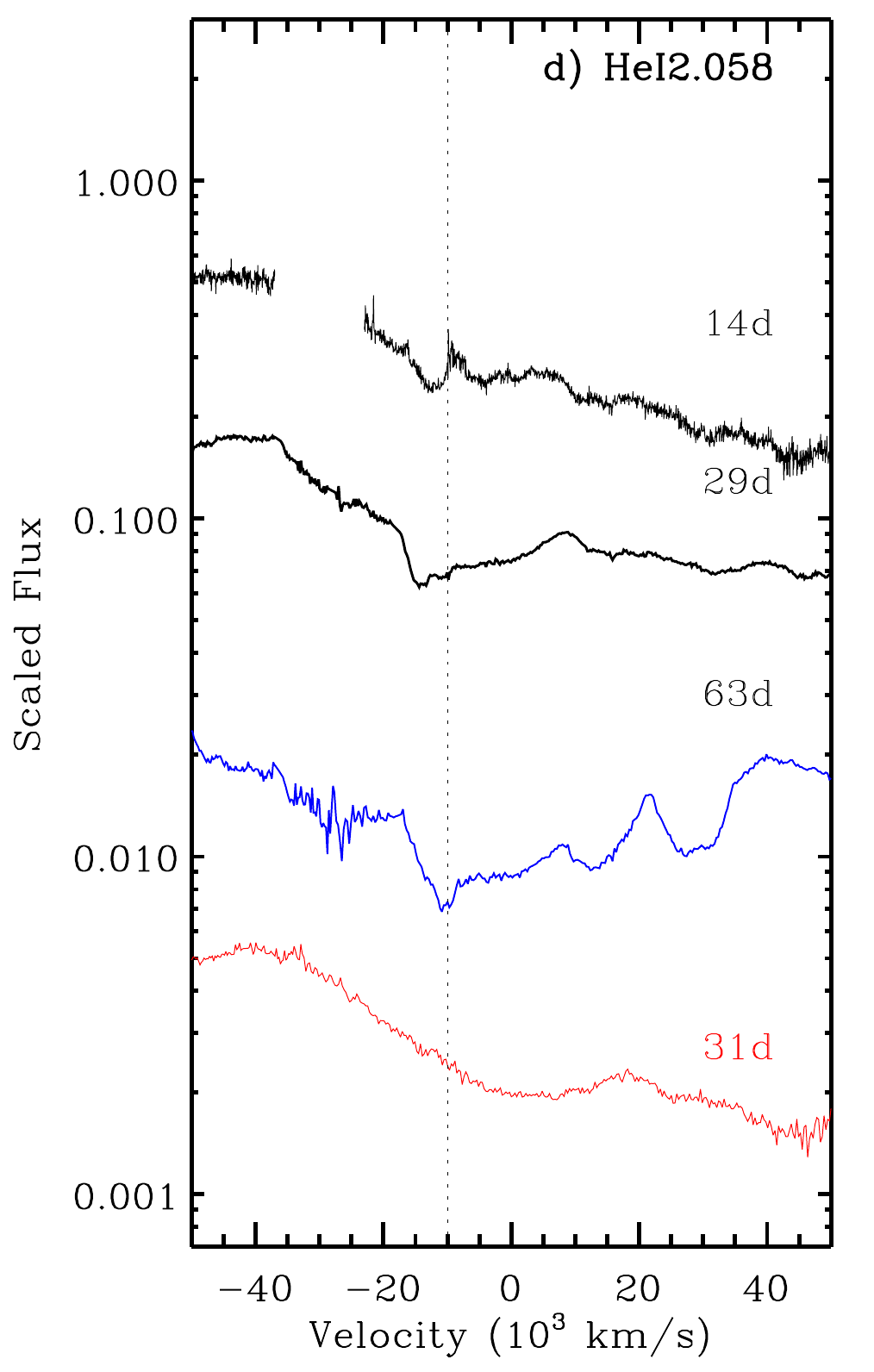}
\caption{\protect{(a) Velocity profiles and temporal evolution of (a) the \caii\ triplet (assumed intensity-weighted rest wavelength 0.8567 $\mu$m) in SN2020oi upper 4 curves and in SN2020bvc (bottom curve in red).  Color coding is as in Figure \ref{FnearIRspec} for near-IR spectra (panels a, c, d) and as in Figure \ref{Foptispec} for optical spectra (panel b). The \caii\ triplet in SN2020oi has both emission and absorption components; the velocities of absorption minima (in cyan) are marked in circles at the velocities 
-22,700,  -15,700,  -14,400,   -11,300 \kms\ for days 3, 14, 29 and 63, respectively.
The velocities are estimated using Gaussian fits.
(b) Same for \caii\ optical line in SN2020oi. Note the change in slope of the peak absorption velocity with time in both \caii\ lines after the maximum peak at day 12;  c) Same for  IR \si\ line at 1.046$\mu$m in both SNe. The absorption minima are at velocities of -12,700, -7,300, -7,300, and -5,000 \kms\ for days 3, 14, 29 and 63, respectively. (d) Same for potential He I line at 2.0587 $\mu$m in both SNe. The absorption minima are about -10,000 \kms\ (dotted line). The He I absorption feature is discussed in Section \ref{Shelium}.
}}
\label{Fvelocity1}
\end{figure*}

\noindent {\bf Freshly Formed Dust:}
The central region of CO in SN2020oi is optically thick so that radiation from CO does not efficiently cool the gas to low enough temperatures for the dust to form. The temperature in the CO forming region is $\sim$3000 K. In the outer layers (see Figures \ref{COST} and \ref{COSTO}) where CO can cool the gas efficiently, densities are too low ($\sim$4$\times$10$^4$ cm$^{-3}$) by many orders of magnitude for dust formation. Note that this conclusion  assumes a progenitor and fully mixed model as described earlier. Our dust-forming model based on \cite{dominik92} and \cite{dominik97}, which is an extension of the non-LTE CO modeling (Sections \ref{SnonLTECO} and \ref{SnonLTECO2}), shows that no prominent dust forms at day 63 in the Type Ic SN2020oi. However, chemically controlled dust models tend to result in earlier dust formation \citep{sluder18, sarangi15} and clumping of ejecta may change the timescale of dust formation.

The presence of newly formed dust either in ejecta or CSM interaction can be tested by other means. One possible signature of dust formation is a deviation in the light curves from those produced purely by the $^{56}$Co-to-$^{56}$Fe decay slope \citep{elmhamdi04}. An example is SN2006jc, which shows such a deviation around day 60, as shown in Figure \ref{FLCcomp1}. SN2020oi does not show a significant  deviation from the decay slope between days 20 and 60. Another signature of newly formed dust can come from line profiles. The red wings of emission lines can be suppressed due to extinction by newly condensing dust that obscures the receding gas. Dominant blue wings relative to red wings, or double-peaked line profiles due to condensing dust are often observed in hydrogen or helium lines \citep{elmhamdi04, smith08}, but SN2020oi did not display such line profiles. \cite{bevan19} have attributed the early-time ($\sim$70 d) line profile asymmetries in SN2005ip (Type IIn) to newly-formed ejecta dust rather than a CSM shell, although the dust mass at day 50 is only 10$^{-8}$ M$_\odot$. Atomic lines in the optical or near-IR spectra at day 63, which contains CO emission and a rising dust continuum (Figure \ref{FnearIRspec}) do not show any asymmetry or other unusual line shapes. Note, however, that spectra obtained after day 63 are too limited to reveal any temporal change, because the spectra lack hydrogen and helium lines.

A model of Type Ib SN2006jc by \cite{nozawa08} indicates that carbon dust formation can occur in Type Ib SNe as early as at 40 -- 60 days due to  the rapid decrease in the gas temperature in SNe Ib. This is consistent with the carbon dust we used in our dust fitting of SN2020oi above. The temperature at day 50 reaches 1000 -- 2000 K and carbon dust forms in the outer carbon layer. The condensation of carbon dust is followed by the condensation of silicate and oxide grains, up to 230 days; most of the dust forms at days 80 -- 180 \citep{nozawa08}. Compared to Type IIP SNe, Type Ib/Ic SNe have lost most of the hydrogen/helium envelopes before the explosion; thus, their ejected masses are smaller, and the expansion velocities are higher than in SNe IIP. This leads to a lower density of gas in the ejecta and more rapid cooling than in typical SNe II. The dust temperature and dust mass that we derive for SN2020oi are comparable to those in the dust formation model for SN SN2006jc of \citet{nozawa08}.

The equilibrium temperature of the dust $T_{\rm d}$ is determined in the usual way by assuming radiative balance between absorption and emission: \[ \frac{L_*}{4\pi{r}^2} \: \overline{Q}_{\rm abs}\pi{a}^2 = 4\pi{a}^2 \sigma T_{\rm d}^4 \: \langle{Q_{\rm abs}(T_{\rm d})}\rangle \:, \] where $L_*$ is the bolometric luminosity of the central object, $r$ is the distance from the SN that the dust has reached at time $t$, and $\sigma$ is the Stefan-Boltzmann constant. $\overline{Q}_{\rm abs}$ is the absorption efficiency of the dust, averaged over the spectral energy distribution of the heating source, and $\langle{Q_{\rm abs}(T_{\rm d})}\rangle$ is the Planck mean absorption efficiency of the dust. For carbon dust we take $\overline{Q}_{\rm abs}\simeq1$ and $\langle{Q_{\rm abs}(T_{\rm d})}\rangle=1.0\times10^{-4}\,a\,T_{\rm d}$, where $a$ is the grain radius in $\mu$m; this formulation for $\langle{Q_{\rm abs}(T_{\rm d})}\rangle$ is valid for $250<  T_{\rm d} \mbox{~(in~K)}< 1000$ and $a<0.3\,\mu$m \citep[see][for details]{tielens05book}. Thus
\begin{eqnarray}
T_{\rm d}  & = & \left ( \frac{L_*}{16{\pi}{r}^2} \: \frac{1}{10^{-4}a\sigma}\right )^{1/5} \nonumber \\
T_{\rm d} \mbox{~(in~K)}  & \simeq & 16.8 \: \left (\frac{L_*}{L_\odot} \right )^{1/5}
  \left ( \frac{r}{10^{17}\mbox{cm}} \right )^{-2/5}
    \left ( \frac{a}{1\,\mu\mbox{m}} \right )^{-1/5}  .\nonumber
\end{eqnarray}

We derive the luminosity of SN2020oi at day 63 using optical photometry to be 2.5$\times$10$^{40}$ \ergs, but this does not include the ultraviolet contribution. The luminosity from our light curve model gives $L_*=1.5\times10^{41}$~erg/s for the bolometric luminosity, which is consistent with the smallest luminosity of Type Ic SNe \citep{kumar18}; we use this value for $L_*$ in what follows.

As the ejecta are strongly decelerating (see Fig.~\ref{Fvelocity1}), we estimate the distance reached at time $t=63$~days by assuming for simplicity that the ejecta velocity decline with time as $v_{\rm ejecta}\propto{t}^{-\alpha}$, where $\alpha$ is a small positive constant. This is intended for convenience only and clearly not as a physical model. We integrate $v_{\rm{ejecta}}$ over time, using the times and velocity values for the  8567\AA\ \caii\ line and the 1.046\mic\ [S\,I] line in the caption to Figure~\ref{Fvelocity1}. We find that $\alpha\simeq0.22$ for the \caii\ line, and $\alpha\simeq0.29$ for the [S\,I] line. In this way we find that $r\simeq2.6\times10^{17}$~cm from the \caii\ line, and $r\simeq9.3\times10^{16}$~cm from the [S\,I] line. 
These values of $r$ lead to $T_{\rm d}\simeq$ 625 (395)~K and 945 (595)~K respectively for a grain size of 0.1 (1) \mic. Based on SN dust formation models \citep{sluder18, sarangi15}, the peak of grain sizes is close to 0.1\mic\ although $\sim$1\mic\ size grains still exist. These values are in line with the value deduced ($T_{\rm d}\sim$800~K) from the IR excess.

These considerations point to the fact that the dust contributing to the IR excess of SN2020oi in Figure~\ref{Fdustmodel} arises from dust condensing in the SN ejecta.

\noindent {\bf CSM dust:} Other possibilities to explain the rising continuum seen in SN2020oi are that the emission originates in the pre-existing CSM by the radiative heating of pre-existing CSM dust by SNe or by newly formed dust in the swept-up CSM dense knots. The newly formed dust can result from the CSM interaction when the reverse shock hits the dense CSM shell or knots. In the case of SN IIn 2005ip, \cite{fox10} suggested that the warmer dust with temperature 900-1100 K and mass $\sim$5$\times$10$^{-4}$ M$_\odot$, originates from newly formed dust in the ejecta, or possibly the cool, dense shell, and is continuously heated by the interaction of the ejecta with the pre-existing CSM. A cooler dust-component with a temperature of $\sim$300 K was also observed in SN2005ip, which they suggested originated from heated dust.

For the Type IIn SN2010jl, \citet{andrews11} estimated the temperature of its pre-existing CSM dust shell to be $\sim$750 K on day 90, which is comparable to the dust temperature we estimate for SN2020oi on day 63. \cite{gall14} suggested that the formation of dust between days 40 and 240 occurs in its dense circumstellar medium of SN2010jl. However, the theoretical model by \cite{sarangi18}, which included the heating of the post-shock gas, indicated that dust formation could only commence after day $\sim$380 after the radiation from the shock weakens for the case of a Type IIn SN. 
We conclude that the observed dust in SN2020oi may be pre-existing, radiatively heated dust from a circumstellar shell.

\noindent {\bf IR echo:} An alternative interpretation of the dust emission is that it is due to an infrared echo \citep[see][]{bode80, bode80snII}. The SN outburst heats the pre-existing dust (the dust shell of the progenitor or interstellar dust in the vicinity of the SN) emits IR in moving features. {\it Spitzer} detected an IR echo in SN2004et with a temperature of 115$\pm$15 K from the ISM dust of the host galaxy \citep{kotak09}. The dust knots producing the echo in the young SNR Cas A show a similar temperature, $\sim$150 K \citep{dwek08}.

\cite{graham86} shows how to discriminate between dust condensation and an IR echo: from the color temperature of the dust to determining the distance, and hence the speed of recession, of the dust from the SN. Light-travel-time effects result in the bulk of the observed IR dust emission arising at the vertex of a paraboloid of revolution with focus at the SN, from which the vertex recedes at speed $c$/2, with $c$ being the light speed \citep[][]{graham86, dwek08}. The grain temperature at the vertex may in principle have any value up to the grain vaporization temperature \citep[e.g.,][]{graham86}, find the initial dust temperature for SN 1982e to be $\sim$1300 K. The dust temperature of IR echo  depends on the SN burst luminosity and properties of pre-existing dust which produce diverse spectral energy distributions \citep{dwek08}. The dust temperature of 810 K observed in SN2020oi is therefore not inconsistent with an infrared echo.

In summary, the fact that the observed dust temperature is consistent with the equilibrium dust temperature from the SN raises the possibility that the rising continuum can come from newly formed dust either in the ejecta or in circumstellar knots. However, heated dust from a circumstellar shell or an IR echo is still a plausible explanation for the rising continuum. Distinguishing between dust freshly formed in the SN ejecta, an IR echo, and dust formed in a pre-existing CSM shell requires a higher sampling of dust emission with time. For heated dust, one would observe the dust temperature increases with time. More frequent sampling of IR spectra between day $\sim$20 and $\sim$63 may be able to determine the changes in the dust temperature unambiguously. This, combined with a pathway and diagnostics of CO and SiO features from NIR/MIR observations as described in the Appendix are crucial for advancing understanding of molecule formation/destruction and dust evolution in ccSNe, which will be investigated in greater detail and to later (fainter) stages of evolution during the era of JWST.

\begin{figure}[!ht]
\includegraphics[scale=0.8,angle=0,width=8.5truecm,height=6.5truecm]{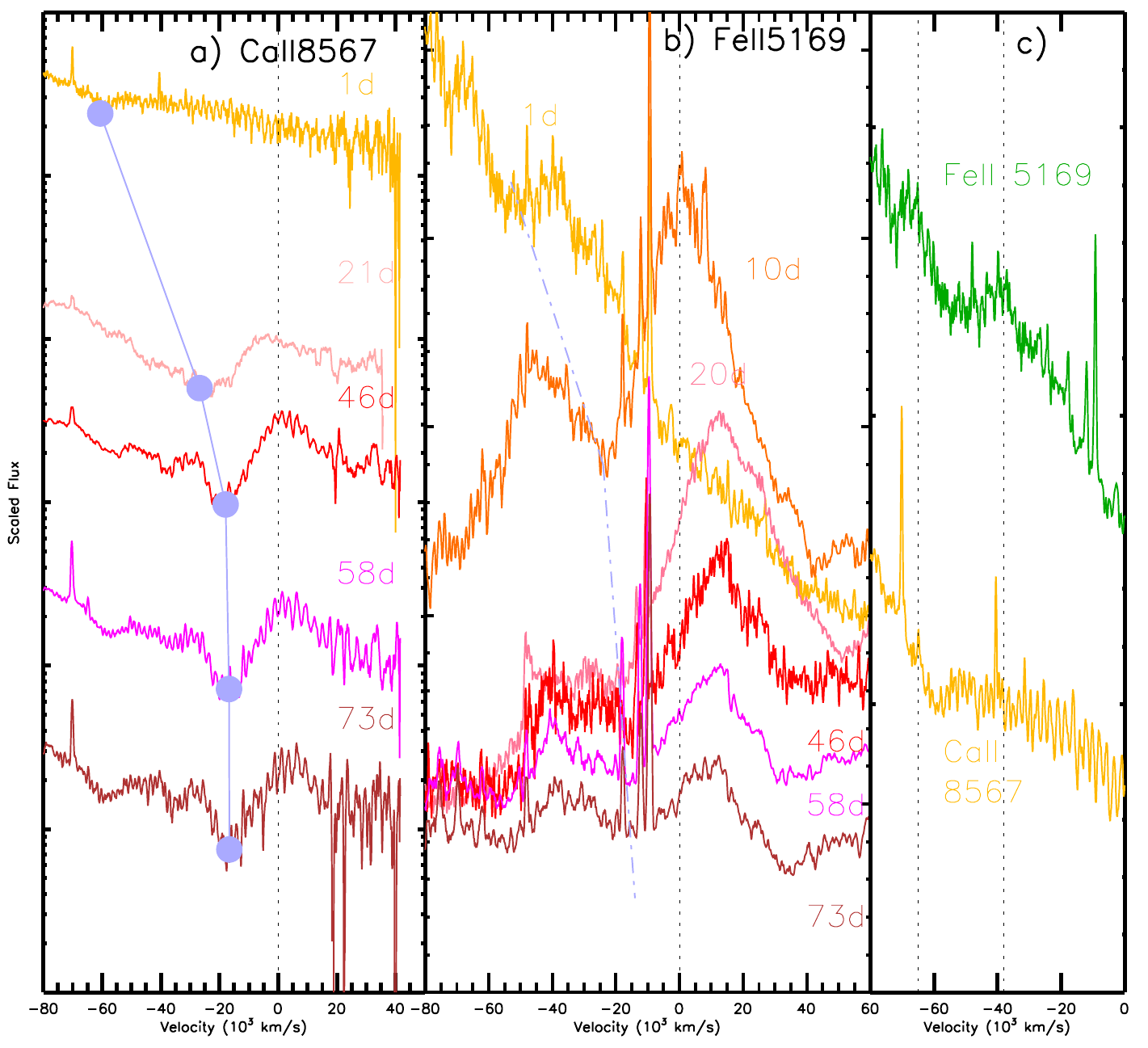}
\caption{Velocity profiles and temporal evolution of IR \caii\ line (panel a) at 8567 \AA\ and \feii\ at 5169 \AA\ (panel b) in SN2020bvc. 
Color coding (panels a and b) is as in Figure \ref{Foptispec}. The absorption minima of \caii\ are at velocities -60,600, -26,800, -17,900, -16,800
and -16,700 \kms\ for days 1, 21, 46, 58, 73, respectively.
The temporal changes of the \feii\ 5169 \AA\ velocity are shown in Figure \ref{fig:bvc_Fe_Vel}. Panel (c); Zoomed version of velocity profiles of the 8567 \AA\ \caii\ (yellow) and \feii\ at 5169\AA\ (green) showing absorption feature at similar velocity. The dotted lines are at velocities of -65,000 and -38,000 \kms. 
}
\label{Fvelocity2}
\end{figure}

\subsection{Spectral Evolution of Type Ic and Ic-BL}
\label{Svelocity}

\subsubsection{Velocity Profiles of Type Ic SN2020oi}
\label{Svelocitysnoi}

Many spectral lines, including multiple atomic line features of \feii\ and \siii, appeared in the optical band (Figure \ref{Foptispec}). It is often difficult to determine if the spectra features are due to different atomic lines or due to temporal velocity changes of the same line. Therefore, we use the strong absorption feature produced by the IR \caii\ triplet at 0.8567 $\mu$m, which appears both in the optical and NIR spectra (see Figures \ref{Foptispec} and \ref{FnearIRspec}) as a guide. Figures \ref{Fvelocity1}a and \ref{Fvelocity1}b show the velocity profiles and temporal evolution of the \caii\ triplet. We also show the velocity profiles of \si\ at 1.046 $\mu$m and temporal changes in Figure \ref{Fvelocity1}c. 

We estimated velocities at the absorption dip (at minimum intensity) by fitting multiple-Gaussian components as needed; the components include both absorption and emission features. The velocity of the \caii\ (\si) peak absorption changes from -22,700 to -15,700 \kms\ (-12,700 to -7,300 \kms), a change of 7,000 (5,400) \kms\ between the explosion and day 14, but between day 14 and day 63 the velocity changes only by 4,400 (2,300) \kms\, using near-IR spectra complimented by the optical spectrum at day 3 in Figures \ref{Fvelocity1}a and \ref{Fvelocity1}c.  The optical spectra in Figure \ref{Fvelocity1}b confirm this trend. Since the optical spectra have a denser time-sampling, they further constrain the steepness of velocity change to be maximum at day 12, which corresponds to the brightness peak (V$_{max}$). 

\subsubsection{Type Ic-BL SN2020bvc: \caii\ and \feii\ Velocities}

SN Ic-BL are classified by their broad lines and high expansion velocities \citep[i.e., 15,000 to 25,000 \kms\ at maximum light;][]{modjaz16} compared to those of normal SN Ic. SNe Ic-BL are the only supernova type associated with GRBs \citep{woosley2006sn_grb, modjaz11-rev,cano17_obs_guide}, indicating that their jet may be the origins of the high ejecta velocities \citep{barnes18}. When their jets fail to break out of the progenitors' envelopes, they can release sufficient energy, to produce the high expansion velocities characteristic of SNe Ic-BL \citep{lazzati12}. However, not all SN Ic-BL are associated with observed GRBs. It is an open question as to whether there are different progenitor/explosion scenarios, or if the GRBs are sometimes simply missed due to their being off-axis, or undetected due to sensitivity limitations \citep[see][and references therein]{modjaz16}.

We have used the IR \caii\ triplet absorption to determine the expansion velocities of SN2020bvc, as we have done for SN2020oi. SN2020bvc shows a higher velocity change between days 1 and 20 than at later times. The first optical spectrum (day 1) shows an absorption centered at $\sim$6850 \AA\ (see Figure \ref{Foptispec}) which corresponds to an expansion velocity of $\sim$60,000 \kms\ for the \caii\ triplets at 8567 \AA\ (see Figure \ref{Fvelocity2}c). The temporal evolution of the triplet is shown in Figure \ref{Fvelocity2}a. The change in velocity between day 1 and 20 is $\sim$34,000 \kms, while from day 20 to 73 it is $\sim$10,000 \kms. Such high expansion velocities in SN2020bvc have also been reported by \cite{izzo20} and \cite{ho20}, using the \caii\ triplet absorption feature in the same FLOYDS spectrum (available in TNS) on day 1. Significant velocity changes in eight optical spectra taken during the pre-max stage are also shown by \cite{ho20}, who report the change of 32,000 \kms\ in expansion velocity between day 0.76 and 12.5. Their results are consistent with our findings. The separation between the absorption minimum and the emission peak of \caii\ triplet in SN2020bvc is about 13,000 -- 15,000 \kms\ at late times, which is comparable to that in the Type Ic SN2020oi. The blue-shifted lines indicate high velocities, up to 60,000 \kms\ for SN2020bvc and 20,000 \kms\ for SN2020oi, and the expansion velocity rapidly declines before the optical maximum. After the optical maximum (e.g., V$_{max}$), the velocity slowly declines in both SNe.

\begin{figure}[ht!]
\includegraphics[scale=0.35]{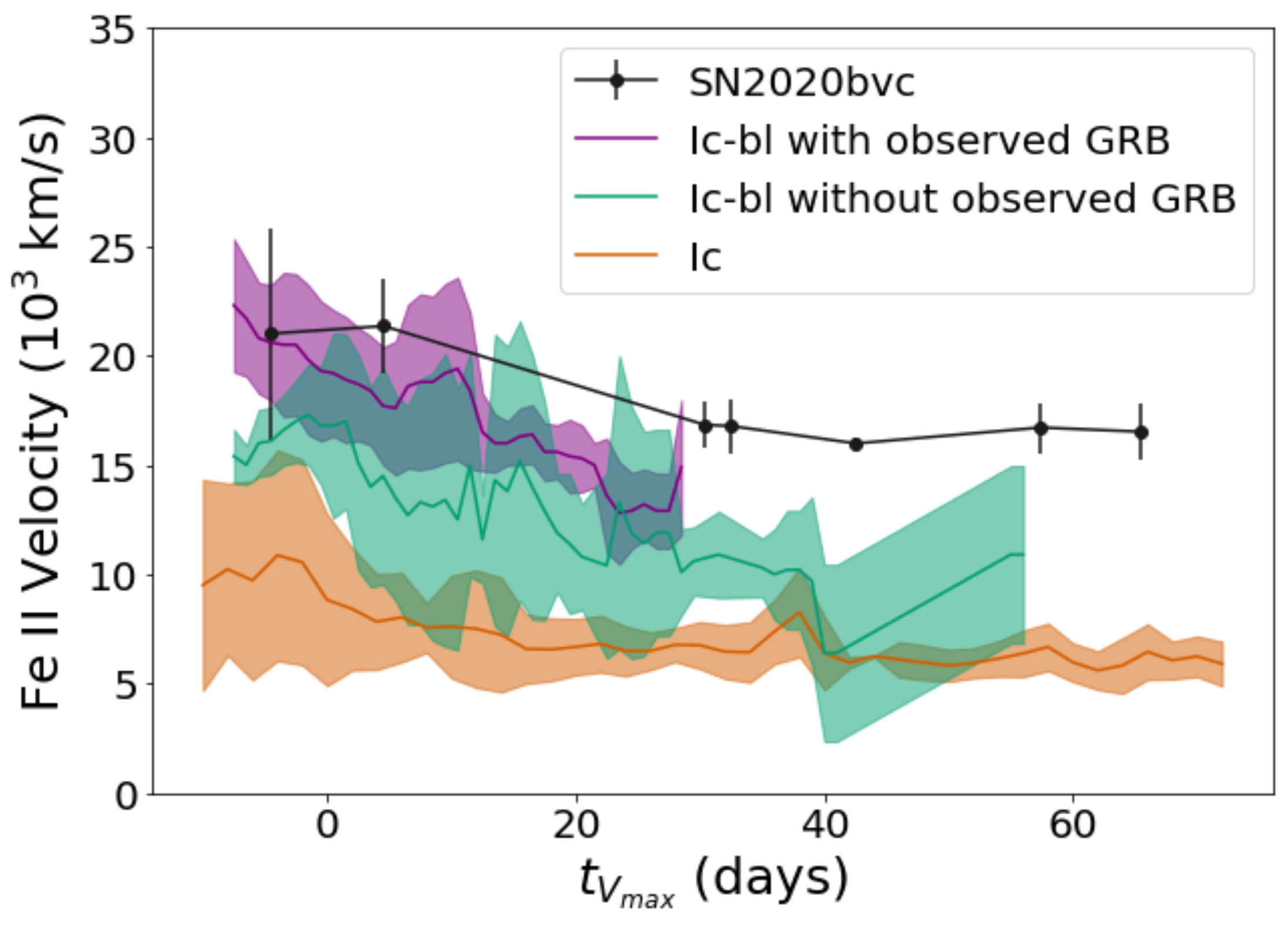}
    \caption{\feii\ 5169$\AA$ absorption velocity in SN2020bvc (black) compared to the weighted mean velocities of SNe Ic (red), SNe Ic-BL with GRBs (purple), and SNe Ic-BL without observed GRBs (green), from \cite{modjaz16}, measured in the same way as for SN2020bvc. Time, $t_{V_{max}}$ (= t$_0$ - 16.3) is measured relative to date of max in the $V$-band. The high velocities of SN~2020bvc are more consistent with those of SNe Ic-BL associated with GRBs, indicating that the on-axis-unobserved GRB model may be more appropriate for SN2020bvc.}
    \label{fig:bvc_Fe_Vel}
\end{figure}

The temporal changes of the velocity profiles from the absorption minima of 5169 \AA\ \feii\ are shown in Figure \ref{Fvelocity2}b. The optical spectrum at 10d includes this \feii\ line. The evolution of the profile is similar to that of the \caii\ feature. We measured the velocity of the blended \feii\ 5169 \AA\ absorption line of SN 2020bvc using the method developed by \cite{modjaz16}, which uses emcee \citep{foreman-Mackey13}, a Monte Carlo Markov Chain sampler, and accounts for both blueshift and broadening of the lines. The high velocities derived from \feii\ lines shown in Figure \ref{fig:bvc_Fe_Vel} are in good agreement with those from the \caii\ triplet in Figure \ref{Fvelocity2}a. Figure \ref{fig:bvc_Fe_Vel} shows the temporal changes in the \feii\ 5169\AA\ absorption velocities in SN2020bvc, and compares them with a large sample of SNe Ic-BL \citep{modjaz16}. The high velocities of SN2020bvc are in good agreement with those SNe Ic-BL associated with GRBs than those with Type Ic-BL without associated GRBs. If SNe Ic-BL without associated GRBs are off-axis GRBs, our results indicate that the on-axis-unobserved GRB scenario is more appropriate for SN~2020bvc, which is supported by the work by \cite{ho2020sn2020bvc}. In contrast, \cite{izzo20} suggest an off-axis GRB for SN2020bvc because of lack of early X-ray detection. Based on the analysis of the expansion velocity in  SN2020bvc, we favor an on-axis-unobserved GRB over an off-axis GRB scenario.

\begin{figure}
\includegraphics[scale=0.6,angle=0,width=8.5truecm]{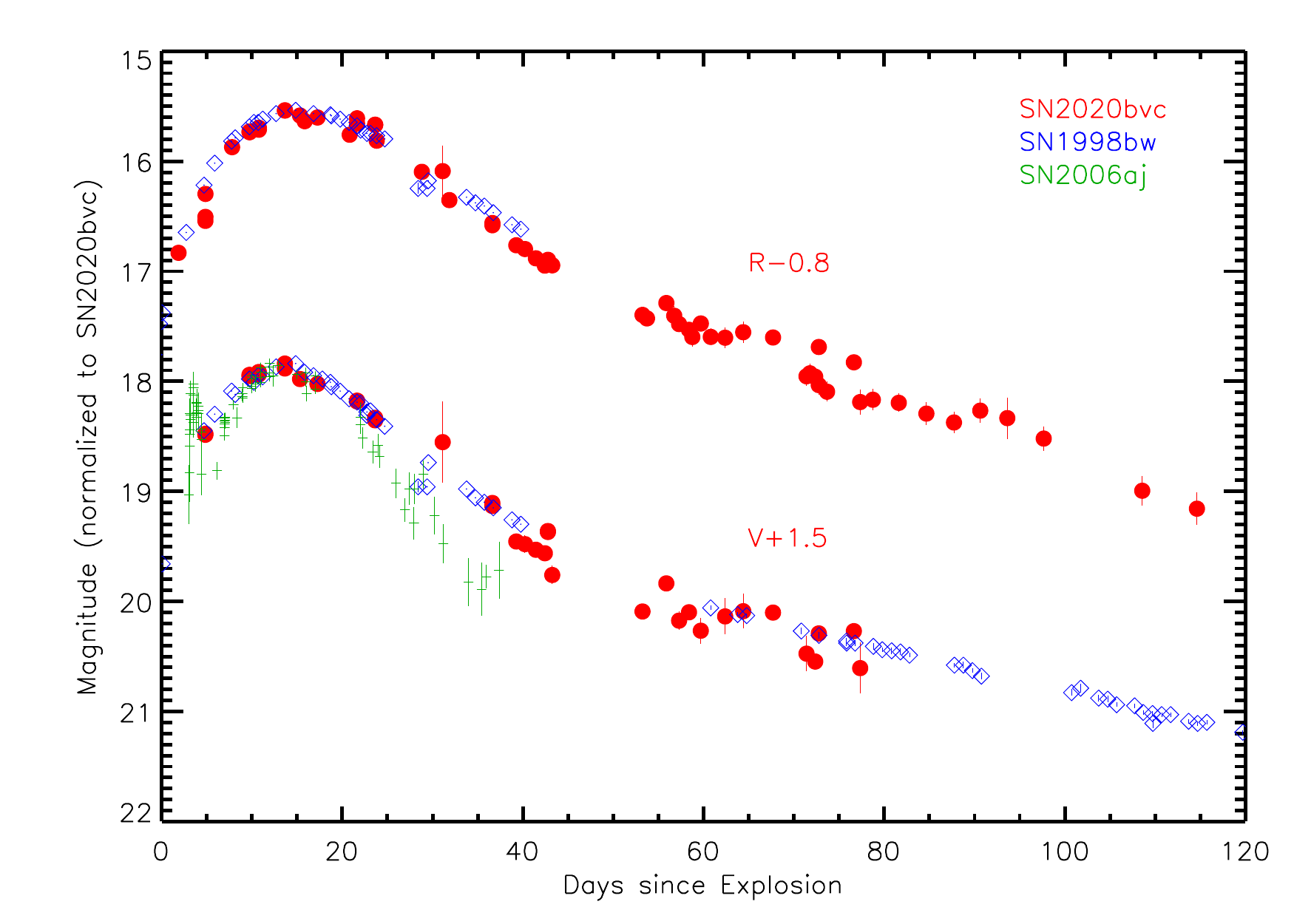}
\caption{Optical light curve of SN2020bvc in $V$ and $R$ bands compared with SN1998bw and SN2006aj.
The light curves of the two comparison SNe are scaled to match those of SN2020bvc at its peaks.}
\label{FLCcomp2}
\end{figure}

\subsection{SN2020bvc Light Curves and Their Implication}
\label{Sdoublepeak}

The light curves of SN 2020bvc in the V and R bands are compared with SN1998bw and SN2006aj in Figure \ref{FLCcomp2} and their progenitor and explosion parameters are compared in Table \ref{Tprogenitor}. The light curves of SN 2020bvc at $V$ and $R$ have early bright peaks as shown in Figure \ref{FLCmodel2} \citep[see][]{ho20}. Such early first peaks are not usually found in SN Ic-BL, (SN 2006aj shows a similar first peak in V). The overall SN parameters (energy, ejecta mass, and nickel mass) of SN 2020bvc are comparable to those of SN 1998bw and the light curves of SN2020bvc are similar to those of SN1998bw. The light curve of SN2006aj shows a more rapid decline following the peak than either SN2020bvc or SN1998bw.

Our best-fit model parameters for SN2020bvc ($E_\mathrm{K} = 1.1\times10^{52}$~erg, $M_\mathrm{ej} = 6.36~M_\odot$, $M_\mathrm{Ni} = 0.4~M_\odot$) are comparable  to SN 1998bw, in general \citep[$M_\mathrm{ej}\sim$6.8, $M_\mathrm{Ni}\sim$0.4, $E_\mathrm{K}\sim2\times10^{52}$ erg; e.g.,][see Table \ref{Tprogenitor}]{cano13}. The nickel mass largely determines the peak brightness and does not sensitively depend on the ejecta mass and energy, assuming that there is no other power source such as a magnetar \citep{maeda07}. The energy and ejecta mass, together with the nickel mass determined from the peak luminosity, give the width of the light curve. A degeneracy exists between $E_\mathrm{K}$, and $M_\mathrm{ej}$, and thus our model is not a unique solution. For example, the light curve of SN 2020bvc might also be explained by a smaller ejecta mass and energy for the same nickel mass.

In a test model, we have applied the SN parameters (see Table \ref{Tprogenitor}) derived by \cite{ho20}, who give a smaller explosion energy and ejecta and nickel masses (i.e., $E_\mathrm{exp} = 3\times10^{51}~\mathrm{erg}$, $M_\mathrm{ej} = 1.0~M_\odot$, and $M_\mathrm{Ni} = 0.11 ~M_\odot$). 
We find that in all bands and all times the light resulting from Ho's parameters is much fainter than the observed light curves. 
The Ni mass of 0.11 ~M$_\odot$ is too small to explain the observed optical brightness where the difference is up to 1.5 magnitudes. The ejecta mass of 1.0~$M_\odot$ and the energy of 3$\times10^{51}~\mathrm{erg}$ result in too narrow light curves to explain the broad light curve.

There are only about a dozen published Stripped SNe (i.e., SNe of Types IIb, Ib, Ic, and Ic-BL) that show double-peak light curves \citep[see Figure 4][]{modjaz19}. As we have modeled the light curves of SN2020bvc, the interaction with an extended CSM shell around the progenitor \citep[e.g.,][]{piro15} can reproduce the first peak, which is caused by shock cooling emission of the CSM  heated by the SN shock. Our estimated CSM mass $M_\mathrm{CSM}$ = 0.1 $M_\odot$ within the region of $R_\mathrm{CSM}$ = 10$^{14}~\mathrm{cm}$ implies that the progenitor experienced a mass eruption with $\dot{M} \sim 1.0~(v_\mathrm{w}/100 \mathrm{[km~s^{-1}]})~M_\odot~\mathrm{yr^{-1}}$ shortly before the SN explosion (i.e., $t <$ 1.0~yr) with the CSM material in a confined region ($\sim$10$^{14}$ cm). This high mass-loss rate is difficult to reconcile with the previously suggested scenarios for pre-SN mass loss from stripped-envelope SN progenitors \citep[e.g.,][]{fuller18, aguilera18}. A luminous blue-variable (LBV) star's present-day mass-loss rate is typically about 10$^{-3}$ M$_\odot$ yr$^{-1}$. However, during a certain time-period of Eta Carinae, the mass-loss rate was much higher, at about 1 M$_\odot$ yr$^{-1}$ \citep{smith03, smith07}, which is approximately the same order of magnitude as the inferred high mass-loss rate in SN2020bvc.

There are two other possible scenarios to explain the bright first peak. It might result from an expansion of the progenitor's outer region during the pre-SN stage due to an energy injection from the inner convective layers \citep{fuller18}. \cite{ho20} suggest that the first peak is due to shock-cooling emission \citep{nakar14} from extended low-mass material (mass $\mathrm{M}$ $<$ 10$^{-2}$ M$_\odot$ at radius $\mathrm{R}$ $>$ 10$^{12}$ $\mathrm{cm}$). Alternatively, the first peak may be caused by the interaction with a binary companion, similar to the case of SNe Ia \citep[][]{kasen10}. A more precise determination of SN parameters would require an exploration of the full parameter space, which is outside of this paper's scope and will be the subject of future work. SN2020bvc is a valuable addition to the sample of SNe with double peaks.

\begin{figure}[ht!]
    \includegraphics[scale=0.23]{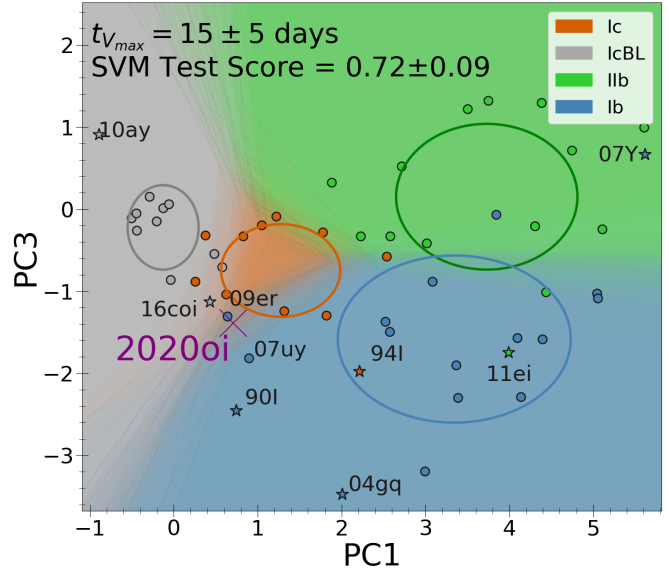}
    \caption{Classification of SN2020oi (purple 'X') is shown in the phase space presented in \cite{williamson19}. PC1 (Principle Component 1) and PC3 are Principle Component decompositions from the mean spectra of many Type Ic SNe over 0-5 days after t$_{V(max)}$ and 5-15 days, respectively. The colored regions are determined by a support vector machine trained on a principal component decomposition of a stripped-envelope supernova dataset. Ellipses mark the regions within one standard deviation for each subtype. SNe that are two standard deviation outliers are marked with stars and labeled. SN2020oi is located between the SN Ic (red) and SN Ib (blue) regions.}
    \label{Fpca}
\end{figure}

\subsection{SN2020oi Classification and Helium}
\label{Shelium}

SN~2020oi is classified as Ic based on the absence of hydrogen features in its spectra and the lack of strong helium lines typical of a SN Ib. The distinction between Type Ic and Type Ib SNe is based on the presence of helium. However, helium may be hidden in SN Ib, and the difference may be to what extent radioactive nickel is sufficiently mixed into the He layers \citep{wheeler87}. In order to compare SN2020oi to the SN Ic population to understand how ``typical'' it is, we compare its spectra at multiple phases with the SN Ic mean spectral templates produced by \cite{liu16}. 

\begin{figure}
\includegraphics[scale=0.6,angle=0,width=8.5truecm,height=15truecm]{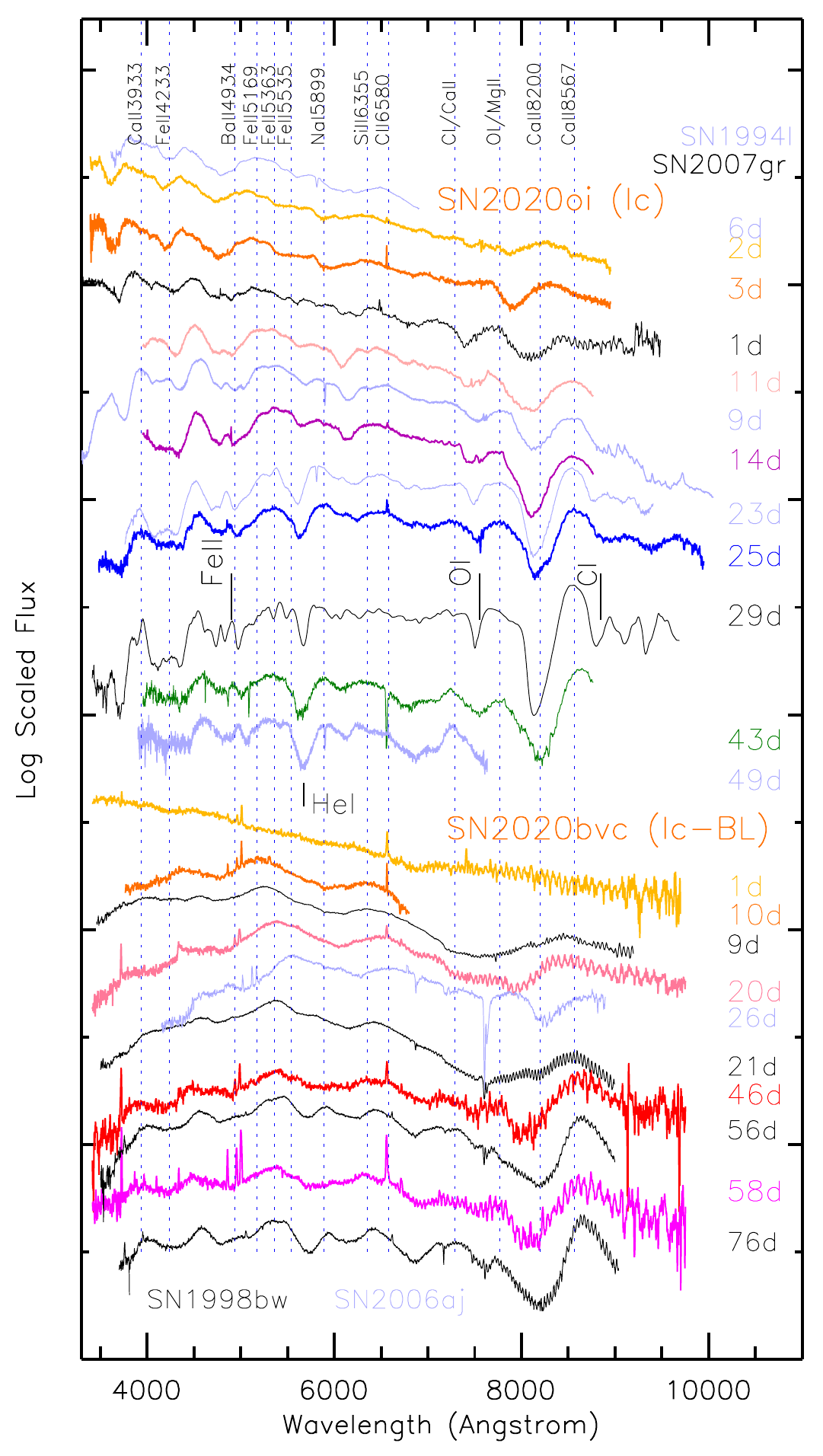}
\caption{
Upper half: optical spectra of SN2020oi compared with SN2007gr (in black) and SN1994I (in light blue). Lower half: optical spectra of SN2020bvc are with SN1998bw (in black) and SN2006aj (in light blue). The colors of SN2020oi and SN2020bvc are the same as those in Figure \ref{Foptispec}. The numbers on the right side are approximately the days since the explosion. The locations of optical \hei, \feii, and \oi\ and \ci\ lines are marked in the middle. and the order of the spectra is in the numbers.
}
\label{Foptispeccomp}
\end{figure}

In the SN2020oi spectra, we examined absorption features that could be consistent with \hei\ 5875 $\AA$ and 1.0831 \mic. However, these lines are commonly contaminated with \nai\ D and \si, respectively. The IR \hei\ line at 2.0587 \mic\ is a much more precise indicator of He abundance in SNe ejecta \citep{chugai1990supernova, patat01, modjaz09}. With the assumption that the spectral features near the wavelength of this line are indeed produced by \hei, we present velocity profiles of the line in Figure ~\ref{Fvelocity1}d. The profiles are consistent with an absorption line at an expansion velocity $\approx$10,000 \kms, which is typical of the \caii\ and \si, as shown in Figure ~\ref{Fvelocity1}, although the putative \hei\ absorption trough is not as strong as in bona fide SNe Ib \cite[e.g., SN~2008D; see Figure 14 of][]{modjaz09}. Figure ~\ref{Fvelocity1}d hints temporal changes of the velocities of \hei. The optical \hei\ 5875 $\AA$ line profile also is consistent, with a similar expansion velocity after day 12 in Figure \ref{Foptispec} (this He line is marked on the spectrum on day 49 in Figure \ref{Foptispeccomp}.)

To classify SN2020oi we have applied the principal component and support vector machine method \citep[see][for details]{williamson19}. The results are shown in Figure \ref{Fpca}. We find that SN2020oi is consistent with the SN Ic population at the $1\sigma$ level, but that it is near the boundary between the SN Ic (red) and SN Ib (blue) regions. In particular, two of the top five spectral matches (matches are nearest neighbor spectra in the Principle Component Analysis (PCA) phase space) are the peculiar SNe Ib SN2007uy and SN2009er. These two SNe have weaker He lines than the typical SNe Ib \citep{modjaz14} along with broader features at higher velocities than the typical Ib.

It is challenging to confidently identify the lines (and infer abundances) in stripped-envelope supernovae without full radiative transfer simulations because of the large line shifts caused by high expansion velocities. Multiple groups have produced synthetic models of SN Ic spectra, including ones with small amounts of He (i.e., $<0.1~M_{\odot}$) that are consistent with observations \citep[][Williamson et al.\,2020, in prep.]{dessart12, hachinger12}. However, so far, the synthetic spectral models of SNe have been limited to optical spectra. Expanding the wavelength coverage to the near-IR should allow most of the observedatomic lines. However including molecule formation, and possibly including a dust model, would be required to reproduce the CO features, and the rising K-band continuum such as is seen in SN2020oi at day 63 (see Figure \ref{FnearIRspec}). Efforts in at least some of those directions will be the focus of future research.

\subsection{Spectral Evolution and Comparison with Other SNe}

\subsubsection{ Type Ic SN2020oi}

In Figure \ref{Foptispeccomp} the spectra of SN2020oi are compared with two other Type Ic SNe, SN2007gr and SN1994I, and those of SN2020bvc with two Ic-BL SNe, SN1998bw and SN2006aj. Here the day numbers for each spectrum at right in the figure are days after the maximum light \citep[for example, in B band for SN2007gr;][]{hunter09}. To make the number of days compared to those SN2020oi, we added 11.5 days from the B$_{max}$ day, which was an approximation.

Overall the spectra of SN2020oi are similar to those of SN2007gr, but at later epochs ($>$ 25 days), SN2020oi has weaker lines than SN2007, including those of Fe around $\sim$4800 \AA), \oi\ (at 7774 \AA) and \ci\ (at 8727 \AA), marked on the spectra of day 29 in Figure \ref{Foptispeccomp}. The progenitor parameters derived from the light curves also show that SN2020oi has an ejecta mass of 0.71 M$_\odot$, lower than the 2-3 M$_\odot$ for SN2007gr (see Table \ref{Tprogenitor}), although the Ni masses of both SNe are $\sim$0.07 M$_\odot$. Strong  Fe lines in optical band are suggested to be due to Fe clumping \citep{mazzali10}. 

The four SN1994I spectra shown in Figure \ref{Foptispeccomp} are remarkably similar to those of SN2020oi except for a slight difference in velocity shifts. However, after day 40, the velocities are comparable. The spectra between SN2020oi at day 25 are very similar to that of SN1994I at day 23 except for the strength of the Na I D line caused by different amounts of interstellar gas on their sightlines. This confirms the finding of similarity between SN2020oi and SN1994I from the light curves in Section \ref{SLCsnoi} and the CO emission analysis in Section \ref{SCOdust}. SN1994I has commonly been considered a standard example of Type Ic.  SN2020oi is the second, such standard example. We encourage the development of more comprehensive hydrodynamical models in order to create a rich set of model spectra for future studies of elemental abundances and mixing.

\subsubsection{Type IC-BL SN2020bvc}

Figure \ref{Foptispeccomp} shows that the spectra of SN2020bvc are similar to those of another Ic-BL SN1998bw. In particular the spectrum at day 20 of SN2020bvc is nearly identical to that of SN1998bw at day 21, including the peaks of the \feii\ and \caii\ absorption features. SN1998bw shows a $\sim$35,000 \kms\ blueshifted line, similar to what is observed in SN2020bvc. In SN1998bw, at the intermediate phase between photospheric and fully nebular phase, the expansion velocities ($\sim$10$^4$ \kms) remained exceptionally high compared to those of other recorded ccSNe at similar phases \citep{patat01}. The velocity change in SN2020bvc is also $\sim$10,000 \kms, again comparable to those in SN1998bw. However, after day 20, while the spectral shape of SN2020bvc remains unchanged, SN1998bw shows noticeable changes in its spectra; some lines become more prominent while the widths of the main lines become narrower. The line broadening remains in SN2020bvc the same (about $\sim$20,000 \kms) until our final spectrum at day 73. In SN1998bw, more lines appeared, including nebular lines, which became noticeable from day 51 (see spectra at day 56 and 76 in Figure \ref{Foptispeccomp}). The spectrum of SN1998bw became more similar to typical Type Ic at later times, but this is not the case for SN2020bvc, as shown in Figure \ref{Foptispeccomp}. The properties of the explosion and the progenitor derived for SN2020bvc and SN199bw are comparable; both have an explosion energy of 10$^{52}$ erg. There are some additional differences between the two SNe, though. Some He weak lines in SN1998bw are identified \citep{patat01}, while the presence of He lines in SN2020bvc is unclear. Most importantly, SN2020bvc does not have an identified $\gamma$-ray counterpart so far, whereas SN1998bw is coincident with GRB 980425. According to \cite{woosley99} its $\gamma$-ray emission is due to shock breakout and relativistic shock deceleration in circumstellar material in a highly asymmetric explosion.

The spectral comparison between SN2020bvc (day 20) and SN2006aj (day 26) in Figure \ref{Foptispeccomp} shows that they have similar broadened line patterns (the widths of the lines are comparable). Interestingly, the blueshifts in SN2006aj are much smaller than those of SN2020bvc. Optical spectra of SN2006aj do not show high-velocity lines such as the ones in SN2020bvc \citep{modjaz06, waxman07}.

The light curve and spectral comparison described above show that SN2020bvc is more similar to SN1998bw than SN2006aj. Thus, our results suggest that the explosion and progenitor parameters of SN2020bvc and SN1998bw are different from SN2006aj (Table \ref{Tprogenitor}) with larger explosion energies, progenitor stars, and Ni masses.

The temporal behaviors of the spectral shape and line profiles in SN2020bvc are very similar to those of Ic-BL SN1998bw. Our findings that SN2020bvc and SN1998bw have very similar light curves, explosion parameters, and high expansion velocities, while the nearby SN1998bw had an observed GRB of very low luminosity, are consistent with the suggestion by \cite{ho20} that SN2020bvc may also have had an associated GRB  which was not found by GRB satellites.

\section*{Conclusion}

1. We have presented near-infrared and optical observations of the Type Ic SN2020oi in the galaxy M100 and the broad-lined Type Ic SN 2020bvc in UGC 09379, using Gemini, Las Cumbres Observatory, SOAR, and other ground-based telescopes. The light curves of SN2020oi cover $\sim$100 days after the explosion and for 80 days after the explosion for SN2020bvc. We present an analysis of 13 optical spectra and four near-IR spectra of SN2020oi, and eight optical spectra and one near-IR spectrum of SN2020bvc. These two explosions provide a rare opportunity to compare their temporal and spectral evolutions of Ic and Ic-BL. The spectra of both SN2020oi and SN2020bvc exhibit dominant emission lines from \feii, \siii, \caii, \ci, \si\, and \sii.

2. The light curves of SN2020oi show a gradual increase over the first ten days since the explosion and a rapid decrease for 25 days. At later times the light curves show slight decreases. Calculations performed using the one-dimensional multi-group radiation hydrodynamics code STELLA imply that SN2020oi has a canonical explosion energy, a Ni mass of 0.07 M$_\odot$, and an ejecta mass of 0.7 M$_\odot$, and is remarkably similar to SN 1994I. 

3. In the Type Ic SN2020oi, CO emission is detected at day 63. The CO emission profile is featureless, presumably due to the emission occurring over a wide velocity range. The CO emission appears to be in transition from the optically thick to the optically thin regime, the latter phase allowing more effective cooling of the deeper, dense layers. The CO mass is greater than 10$^{-3}$ M$_\odot$, and the CO temperature is 3000 - 3300 K; a fully mixed model is required to reproduce the observed CO features. 

4. The near-IR spectrum of SN2020oi at day 63 also reveals a rising continuum in the K band, which must be due to emission by heated dust, at temperature of $\sim$800 K and a mass of 10$^{-5}$ M$_\odot$. This the first detection of dust in a Type Ic SN.   We explore scenarios for creating dust in the SN ejecta or in the circumstellar medium (CSM), heating pre-existing CSM dust or emission via an infrared echo as explanations of the rising continuum.
The fact that the observed dust temperature is consistent with the equilibrium dust temperature from the SN raises the possibility that the rising continuum can come from newly formed dust either in the ejecta or in circumstellar knots. However, heated dust from a circumstellar shell or an IR echo is still a plausible explanation for the rising continuum.

5. A new STELLA SN model consisting of a helium-poor CO star of $M = 8.26~M_\odot$, corresponding to an initial mass of 40 - 50~$M_\odot$, explosion energy of $E_\mathrm{exp} = 12\times10^{51}$~erg and a radioactive nickel mass of 0.4 $M_\odot$ produces a reasonably good fit for the main peak and the light curves of SN2020bvc. The Ni in the SN is assumed to be uniformly mixed. A model with the progenitor surrounded by massive circumstellar matter can reproduce the first-peak of the light curves. The light curve and explosion parameters are similar to SN 1998bw.

6. Temporal changes of the blueshifted absorption lines, which are most noticeable in the near IR \caii\ triplet  and \siii\ lines, are observed in both SN2020oi and SN2020bvc. These blueshifted absorptions are at high velocities, up to 60,000 \kms\ for SN2020bvc and 20,000 \kms\ for SN2020oi during pre-maximum. The expansion velocities thereafter decrease to $\sim$6000--10,000 \kms\ in both SNe. The overall shape of SN2020bvc spectrum after day 20 remains surprisingly unchanged up to day 73. The temporal behavior of SN2020bvc in terms of spectral shape and line profiles are very similar to those of Ic-BL SN1998bw.

7. We show a potential helium absorption in the SN2020oi spectra. However, our PCA analysis, compared with many other Type Ic SNe, shows that SN2020oi is still consistent with the SN Ic population.

\acknowledgements
We thank an anonymous referee for helpful comments and suggestions, which helped to improve the paper. JR thanks Nathan Smith for the helpful discussion on Eta Carina and Bill Reach on the dust grain temperature. 
This paper is based in part on observations obtained at the international Gemini Observatory, a program of NSF OIR Lab, which is managed by the Association of Universities for Research in Astronomy (AURA) under a cooperative agreement with the National Science Foundation on behalf of the Gemini Observatory partnership: the National Science Foundation (United States), National Research Council (Canada), Agencia Nacional de Investigaci\'{o}n y Desarrollo (Chile), Ministerio de Ciencia, Tecnolog\'{i}a e Innovaci\'{o}n (Argentina), Minist\'{e}rio da Ci\^{e}ncia, Tecnologia, Inova\c{c}\~{o}es e Comunica\c{c}\~{o}es (Brazil), and Korea Astronomy and Space Science Institute (Republic of Korea), and based on in part observations obtained at the Southern Astrophysical Research (SOAR) telescope, which is a joint project of the Minist\'{e}rio da Ci\^{e}ncia, Tecnologia e Inova\c{c}\~{o}es (MCTI/LNA) do Brasil, the US National Science Foundation’s NOIRLab, the University of North Carolina at Chapel Hill (UNC), and Michigan State University (MSU). This work makes use of observations obtained with the Las Cumbres Observatory network.

JR acknowledges support from NASA ADAP grant (80NSSC20K0449) and various Guest Observer Programs for the study of SN dust. DAH, CM, JB, and DH are supported by NASA grants 80NSSC19k1639 NSF AST-1911225 and AST-1911151. SV is supported by NSF AST-1813176. DPKB is supported by a CSIR Emeritus Scientist grant-in-aid, which is being hosted by the Physical Research Laboratory, Ahmedabad. PH acknowledges support by the National Science Foundation (NSF) grant (AST-1008962). SCY is supported by the National Research Foundation of Korea (NRF) grant (NRF- 2019R1A2C2010885). LG was funded by the European Union's Horizon 2020 research and innovation programme under the Marie Sk\l{}odowska-Curie grant agreement No. 839090. This work has been partially supported by the Spanish grant PGC2018-095317-B-C21 within the European Funds for Regional Development (FEDER). HA acknowledges support from the Basic Science Research Program through the National Research Foundation of Korea (NRF) funded by the Ministry of Science, ICT \& Future Planning (NRF-2017R1C1B2004566). MM and the SNYU group are supported by the NSF CAREER award AST-1352405 and by a Humboldt Faculty Fellowship. JV and his group at Konkoly Observatory is supported by the project $`$Transient Astrophysical Objects'' GINOP 2.3.2-15-2016-00033 of the National Research, Development and Innovation Office (NKFIH), Hungary, funded by the European Union and partly by the KEP-7/2018 grant of the Hungarian Academy of Sciences.  KV and LK received support from the NKFIH/OTKA grants KH-130526 and  K-131508. KV also acknowledges the partial support from the Lend\"ulet Program of the Hungarian Academy of Sciences, project No. LP2018-7/2019. We thank participating observers on the UW APO ZTF follow-up team, including Brigitta Spi{\H o}cz, Eric Bellm, Zach Golkhou, Keaton Bell, and James Davenport. MLG acknowledges support from the DiRAC Institute in the Department of Astronomy at the University of Washington. The DiRAC Institute is supported through generous gifts from the Charles and Lisa Simonyi Fund for Arts and Sciences, and the Washington Research Foundation.

\software{Figaro \citep{shortridge92}, Gemini IRAF Package, lcogtsnpipe \citep{valenti16},
PyZOGY \citep{guevel17}, IRAF \citep{tody86, tody93},
FLOYDS pipeline \citep{valenti14}, PyDIS \citep{davenport18}, STELLA \citep{blinnikov00, blinnikov06},
MPFIT \citep{markwardt09}, Spextool \citep{cushing04}}

\bibliography{msrefsall}
\appendix
\section{Diagnostics and Sensitivities of Molecular Bands}

In this Appendix, we demonstrate the diagnostics and their sense of observational signatures for conditions and abundances found in SNe~Ic, i.e., C/O/Si-rich mixtures. We assume a thermal radiation field and neglect time-dependence of the formation because it will change the flux but hardly the opacity profiles. This allows linking results from SN2020oi to many future SNe of ccSNe with ground-based near-IR and mid-IR observations such as JWST.

\begin{figure*}[ht]
\begin{center}$
\begin{array}{cc}
\includegraphics[angle=360,width=0.46\textwidth, height=0.2\textheight]{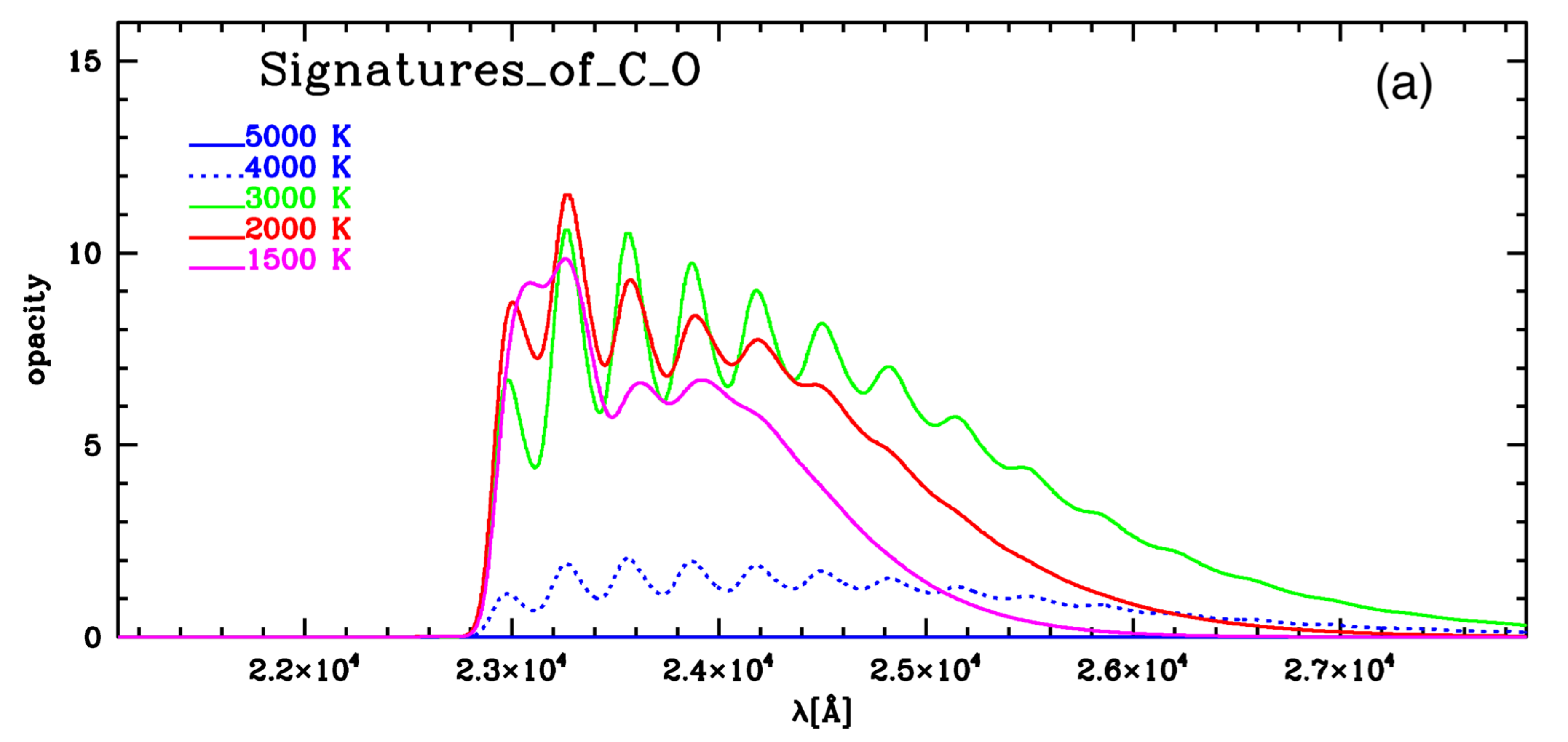} &
\includegraphics[angle=360,width=0.46\textwidth, height=0.2\textheight]{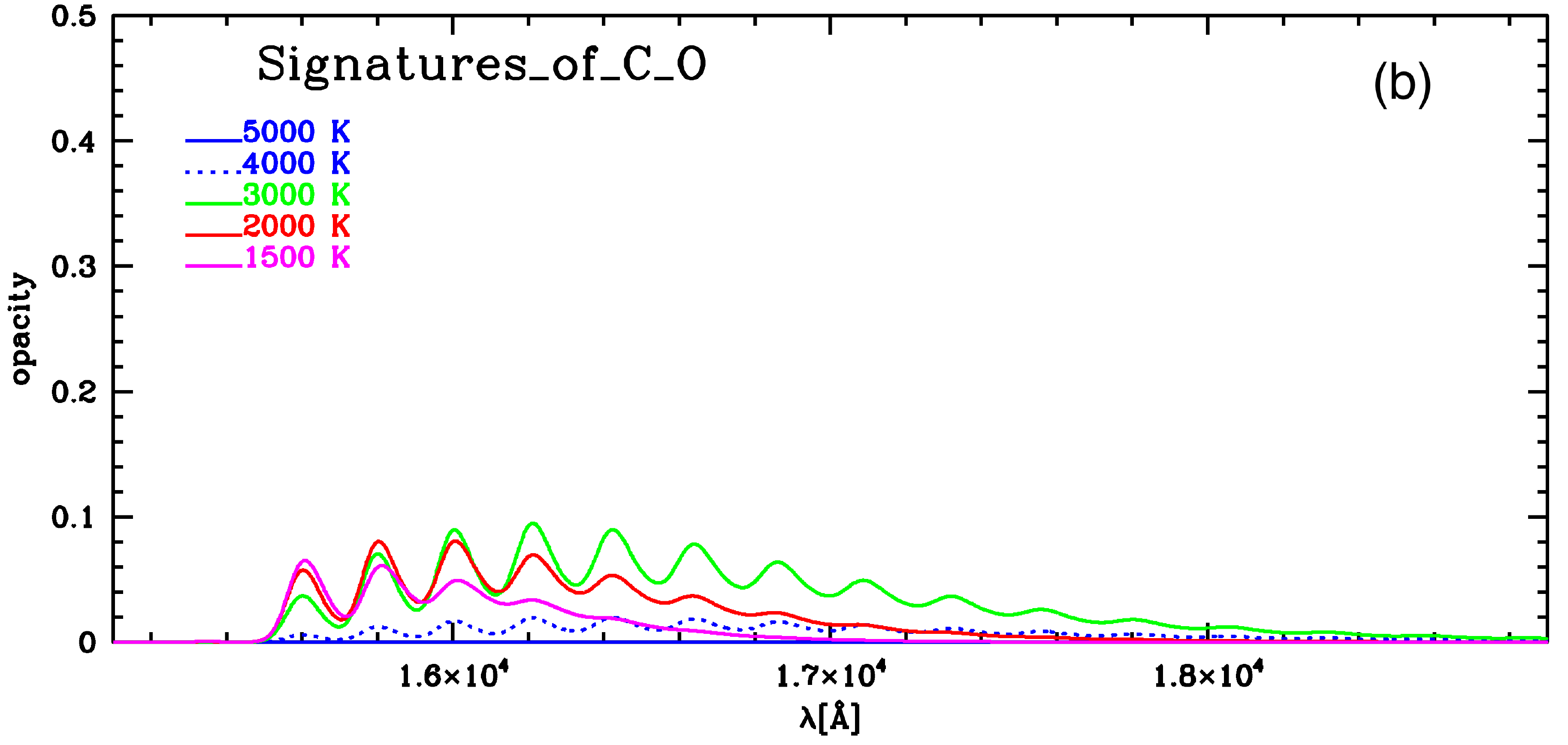} \\

\includegraphics[angle=360,width=0.46\textwidth, height=0.2\textheight]{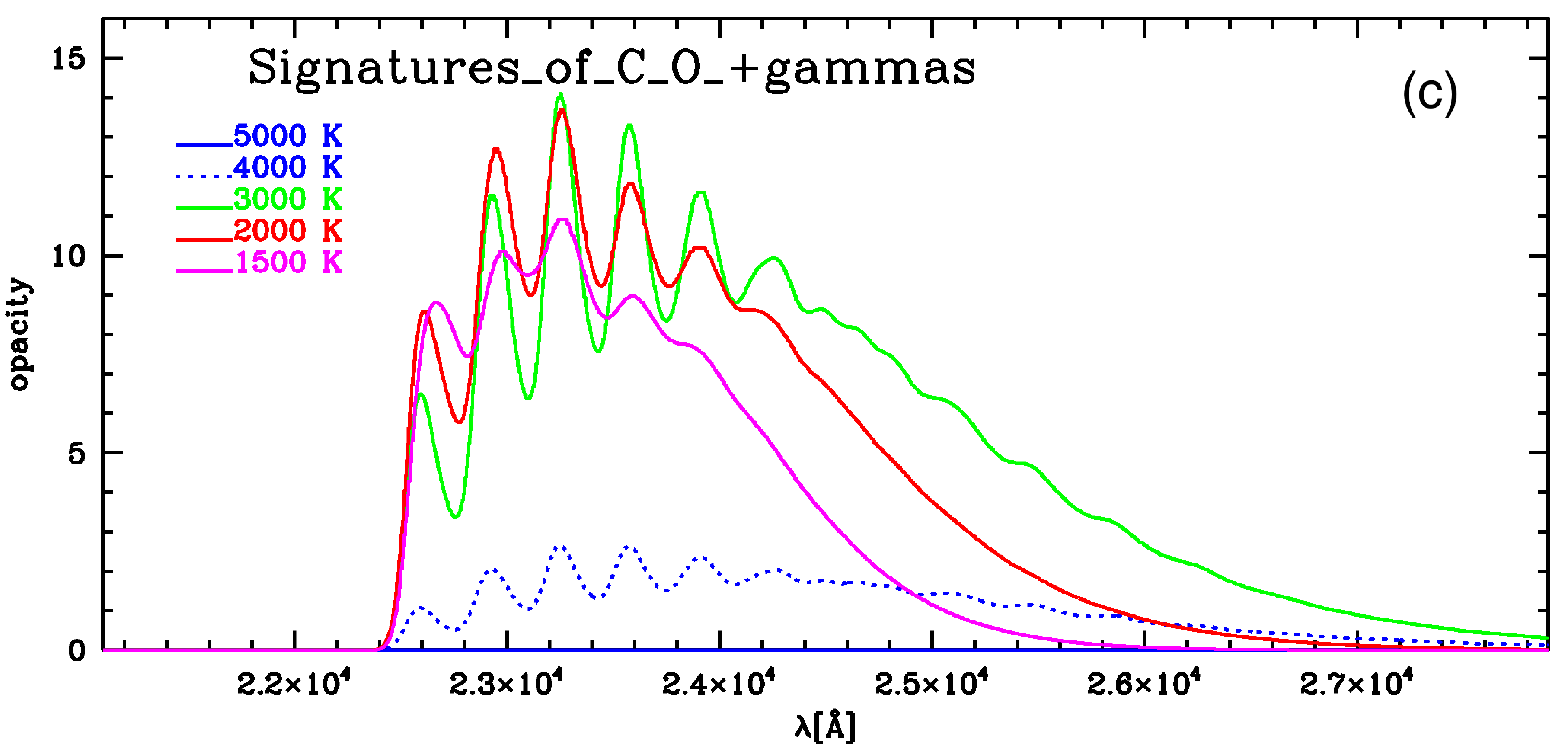} &
\includegraphics[angle=360,width=0.46\textwidth, height=0.2\textheight]{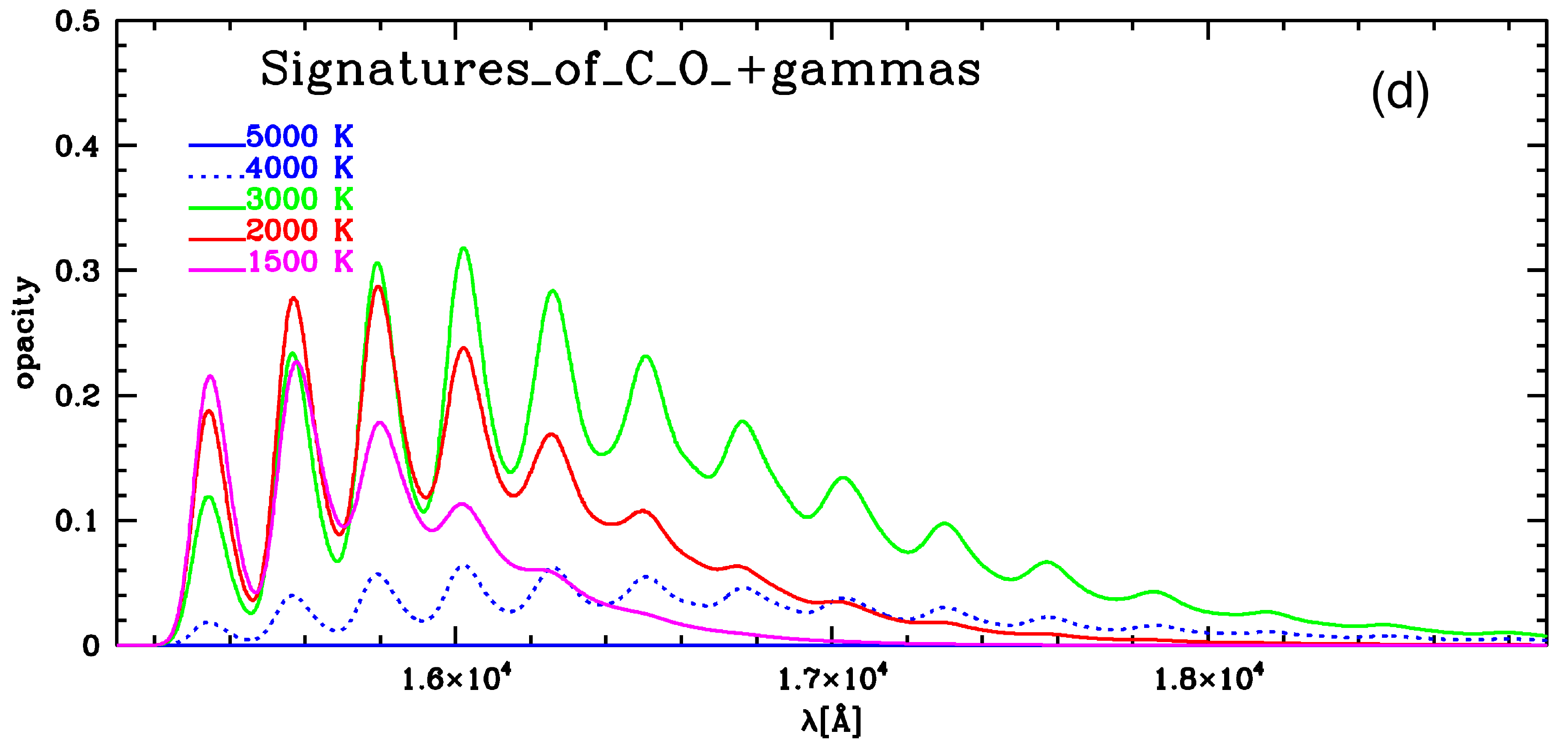} \\

\includegraphics[angle=360,width=0.46\textwidth, height=0.2\textheight]{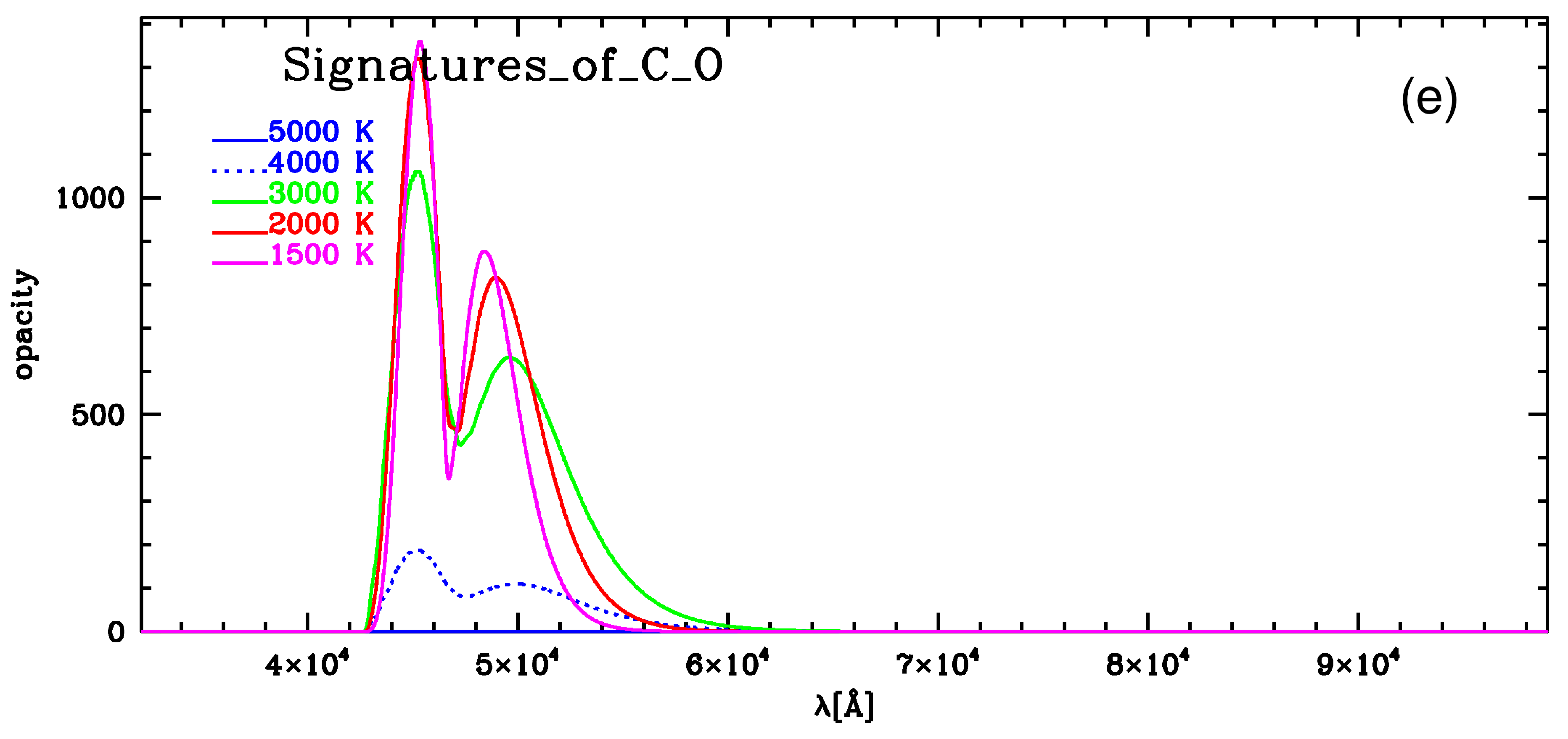} &
\includegraphics[angle=360,width=0.46\textwidth, height=0.2\textheight]{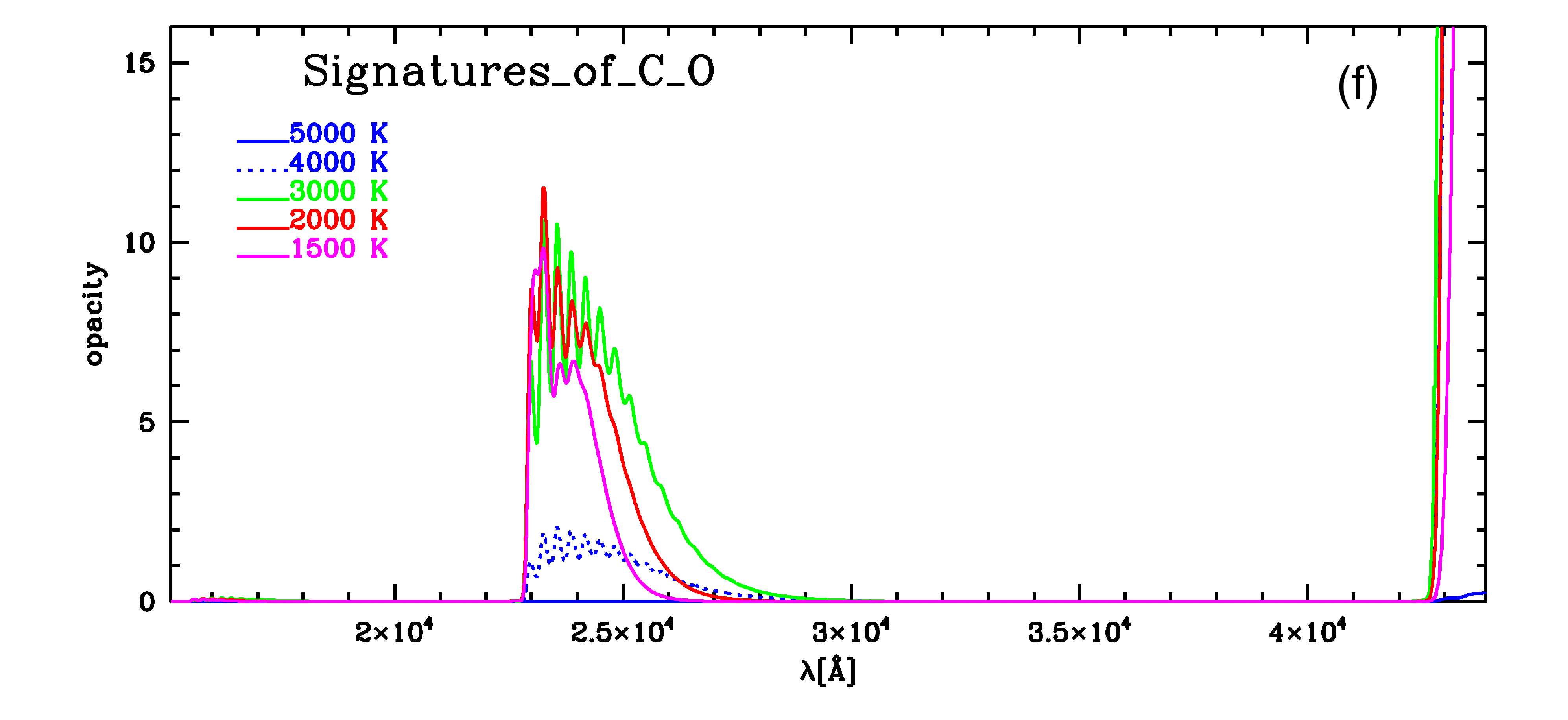} \\

\includegraphics[angle=360,width=0.46\textwidth, height=0.2\textheight]{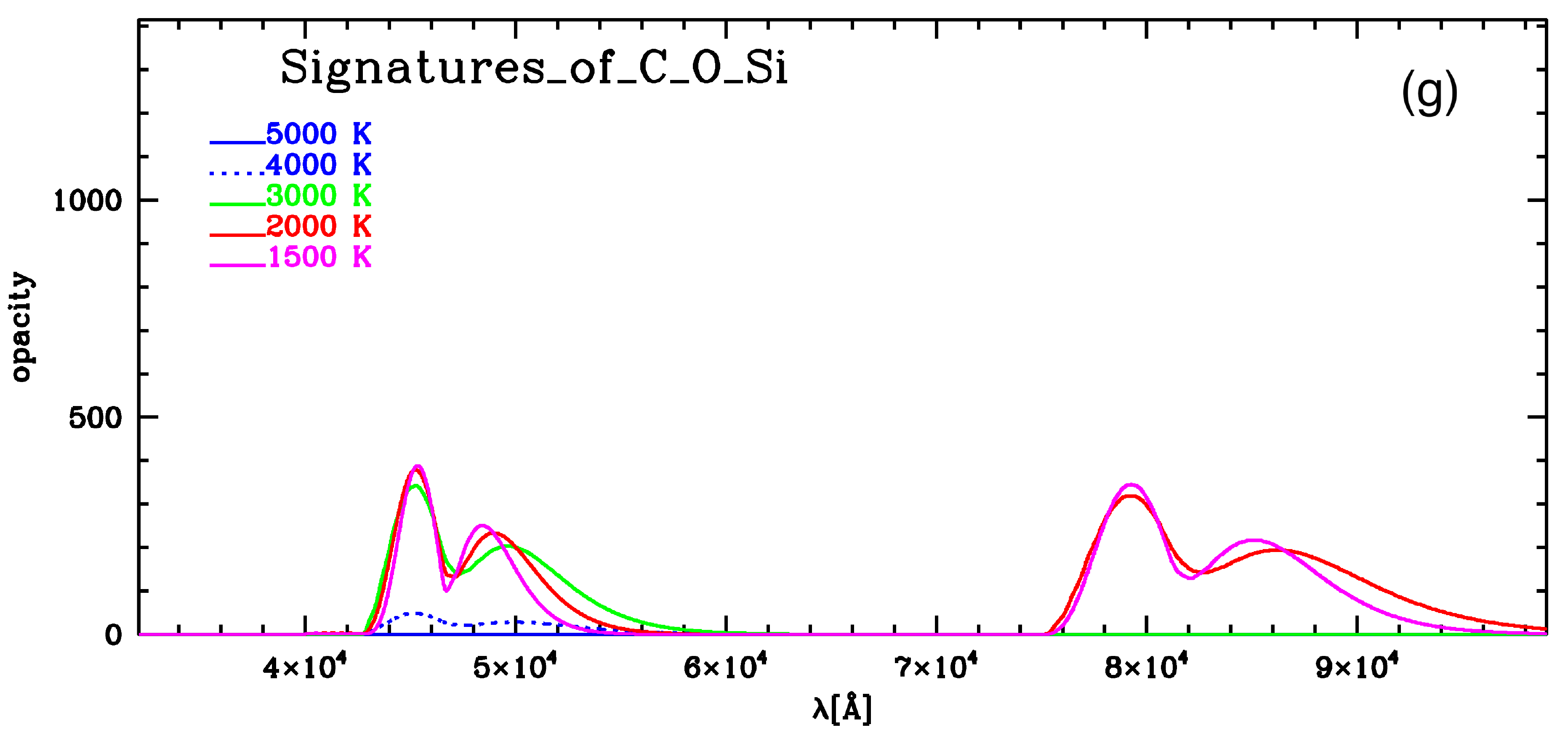} &
\includegraphics[angle=360,width=0.46\textwidth, height=0.2\textheight]{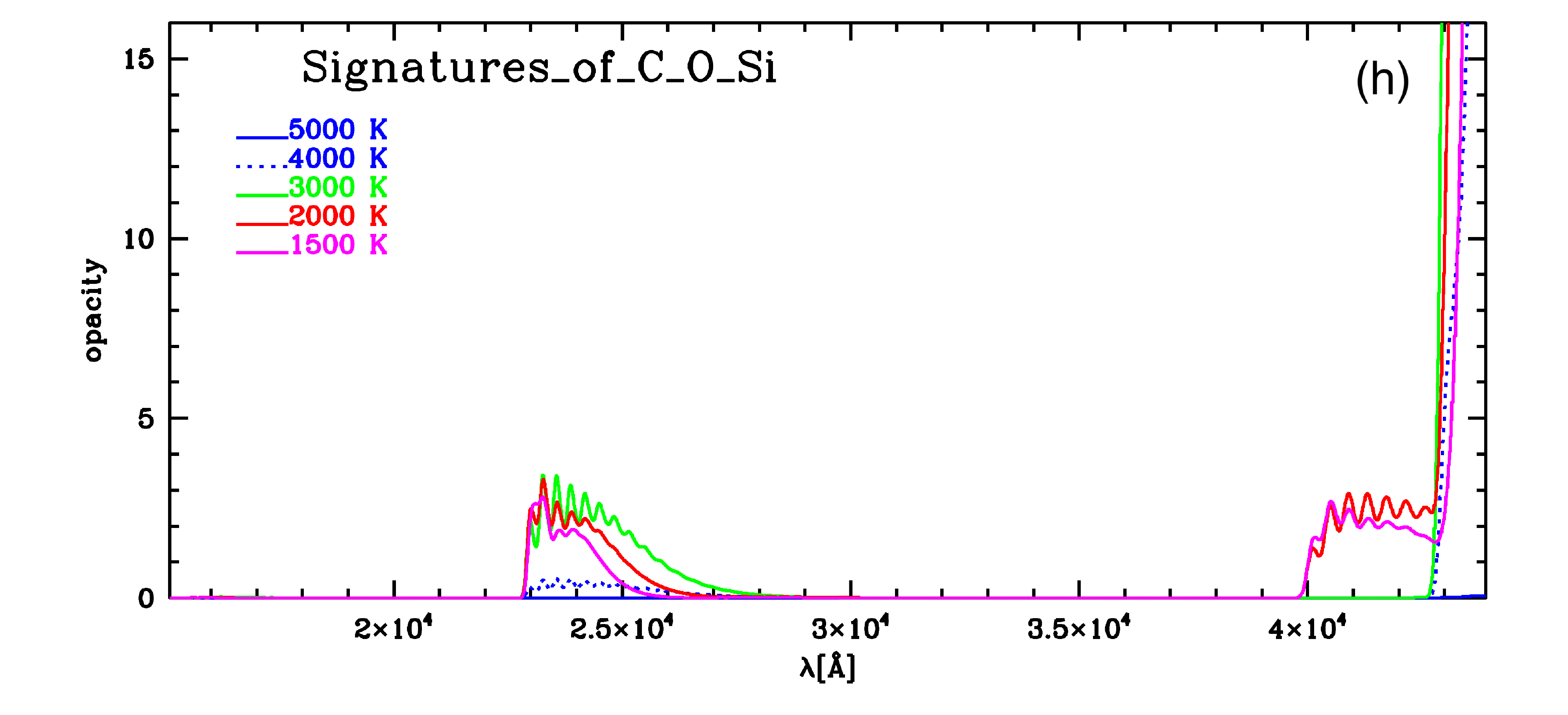} \\
\end{array} $
\caption{Near and mid-IR (1.4 - 10 \mic) spectral signatures of the fundamental, 1$^{st}$ and 2$^{nd}$ vibrational bands of CO, CO+ and SiO for C/O layer {\it (a - f)} and C/O/Si-rich layer with equal C and Si abundances {\it (g - h)}, and for a velocity dispersion of 1,000 km s$^{-1}$. All simulations include CO, CO+ and SiO. In each panel the temperature-dependence of the spectral features is shown. {\it (a - d):} The effect of non-thermal excitation on the 1$^{st}$ {\it (a,c)} and 2$^{nd}$ overtone {\it (b,d)} of CO and CO+ without {\it (a,b)} and with {\it (c,d)} non-thermal excitation. {\it (e - h):} Spectral signatures due to the effect of the abundances. The fundamental CO {\it (e)} and 1$^{st}$ overtone CO {\it (f)} bands dominate for CO-rich layers but, in addition, the fundamental SiO {\it (g)} and 1$^{st}$ overtone SiO {\it (h)} bands start to emerge for C/O/Si layers (the second overtone of SiO at $\sim$28000 \AA\ is too faint to appear). Non-thermal excitation and ionization have been neglected in all models but for those in {\it (c,d)} we assumed $CO^{+}/CO=1$. Note that non-thermal excitation with long recombination and charge-exchange time scales will produce a strong $2^{nd}$ CO overtone in the NIR and, for temperature $\leq$2000K, the $1^{st}$ SiO-overtone produces a $`$wing’ on the short wavelength edge of the fundamental CO-band {\it (h)} comparable in strength to the 1st overtone of CO. For more details see the text.}
\label{COSE}
\end{center}
\end{figure*}

 For the dependence of the density and velocity dispersion, we refer to studies cited in Section \ref{SnonLTECO}. In Figure \ref{COSE}, we demonstrate the sensitivity of the temperature, abundances for (C/O/Si)-rich compositions with abundance ratios of (1:1:0) and (1:2:1), a particle abundance of $5 \times 10^{11} cm^{-3}$ and a velocity dispersion of $1000$ \kms\ typically found in SN~Ic envelopes. The influence of non-thermal ionization by radioactive decay is parameterized by the $^{+}CO/CO$ ratio.

 In the near-IR and for temperatures  $T$ below $5000K$, the opacity of the $1^{st}$ CO and $CO^+$ overtones decrease rapidly with T and, under SN Ic conditions, produce optically thick features. Below $T \leq 2000 K$, the first and second vibrational modes dominate (Fig. \ref{COSE}). Note that second vibrational mode is strong because of its statistical weight.

 Non-thermal excitation has three signatures in the Near-IR: a) $CO^+$ adds a blue-shifted emission component, b) the peak of the emission rises to the $3^{rd}$ component in the spectra even at low T, and c) the appearance of a strong $2^{nd}$ overtone. b) and c) can be used to detect or constrain the amount of mixing in SN~Ic. Note that c) requires good signal to noise spectra but it is not sensitive to inhomogeneities and large asphericity effects expected in SNe~Ib/c (see above). Moreover, ground-based near-IR spectra allow for significantly better S/N in the H-band compared to the K-band.  In practice, a problem is during the semi-transparent phase is line-blending by allowed transitions which are produced in the region inside the CO forming region but dominate the spectra until the nebular phase \footnote{by pattern recognition algorithms, the signatures may be recovered, though}. Moreover, the wide wavelength range spans both the J to the H-band range making pattern recognition of the different modes problematic for ground-based observations.

The MIR molecule spectra are dominated by the fundamental CO-band.  The large opacity allows for detailed studies of the onset of CO-formation be expected to become be optically thick in SNe~Ibc with exception of the high-overtones. Thus, MIR observations should cover a wide wavelength range. The change of the opacity profile can be understood in the same way as the overtone bands.

For $T \leq 2000K$ one may expect to see the fundamental band of $SiO $ if the photosphere has receded to the corresponding layers in a massive stellar explosion, or if large asymmetric explosions `squeezed-out' Si as described above. The appearance of an SiO band is a clear indication of low T at the inner layers but, its non-appearance does not imply the lack of large or small scale mixing. Note that the fundamental SiO bands may have been observed in the SN1998S with the Spitzer Space Telescope though, unfortunately, the spectra have been hampered by the order overlap and possibly calibration issues at the red wing \citep{gerardy02}.

At low T, the $1^{st}$ SiO overtone is clearly present in compositions with more O than C. It is a probe for large vs. small-scale mixing processes at work. In our example (Fig. \ref{COSE}, lower right panel), its strength is comparable to the $1^{st}$ CO-overtone which allows the analysis based on its profile. However, it provides a sensitive temperature indicator for at values close to the dust formation temperature.

\end{document}